\documentclass{aa}  
\usepackage{graphicx}
\usepackage{xcolor}
\usepackage{txfonts}

\usepackage{subfig}
\usepackage{pdflscape}

\newcommand{\kms }{km~s$^{-1}$}

\newcommand{ \Msun } {M$_{\odot}$}

\newcommand{ \Msunyr } {M$_{\odot}$~yr$^{-1}$}

\newcommand{ \Ha} {H$\alpha$}
\newcommand{ \Hb} {H$\beta$}
\newcommand{ \Hg} {H$\gamma$}
\newcommand{ \OIII} {[O\,\textsc{iii}]}
\newcommand{ \NII} {[N\,\textsc{ii}]}
\newcommand{ \SII} {[S\,\textsc{ii}]}
\newcommand{ \OII} {[O\,\textsc{ii}]}
\newcommand{ \OI} {[O\,\textsc{i}]}
\newcommand{ \SIII} {[S\,\textsc{iii}]}

\newcommand{ \cigale}{{\tt CIGALE}}
\newcommand{ \emcee}{{\tt emcee}}
\newcommand{ \gaia}{\textit{Gaia}}

\newcommand{ \micron}{$\mu$m }

\newcommand{\clump}{c}

\usepackage[colorlinks=true, citecolor=blue, linkcolor=violet]{hyperref}
\begin{document} 

   \title{GA-NIFS: JWST/NIRSpec IFS view of the $z\sim3.5$ galaxy GS5001 and \\ its close environment at the core of a large-scale overdensity}
  
  \titlerunning{JWST/NIRSpec view of GS5001}

   \author{Isabella  Lamperti \inst{\ref{iCAB},\ref{iUNIFI},\ref{iOAA}}\fnmsep\thanks{E-mail: isabella.lamperti@unifi.it} \and
   Santiago Arribas\inst{\ref{iCAB}}  \and
   Michele Perna\inst{\ref{iCAB}}  \and
   Bruno Rodr\'iguez Del Pino\inst{\ref{iCAB}} \and
   Chiara Circosta\inst{\ref{iESAC}, \ref{iUCL}} \and
   Pablo~G.~P\'erez-Gonz\'alez\inst{\ref{iCAB}} \and
   Andrew~J.~Bunker\inst{\ref{iOxford}} \and
   Stefano Carniani\inst{\ref{iSNS}} \and
   Stéphane Charlot\inst{\ref{iSorbonne}} \and
   Francesco D'Eugenio\inst{\ref{iKav},\ref{iCav}} \and
   Roberto Maiolino\inst{\ref{iKav},\ref{iCav}, \ref{iUCL}} \and
   Hannah~\"Ubler\inst{\ref{iKav},\ref{iCav}} \and
   Chris J.~Willott\inst{\ref{iNRC}} \and
   Elena~Bertola\inst{\ref{iOAA}} \and
   Torsten~Böker\inst{\ref{iESA}} \and
   Giovanni~Cresci\inst{\ref{iOAA}} \and
   Mirko~Curti\inst{\ref{iESO}} \and
   Gareth C. Jones\inst{\ref{iOxford}} \and
   Nimisha~Kumari\inst{\ref{iAURA}} \and
   Eleonora Parlanti\inst{\ref{iSNS}} \and
   Jan Scholtz\inst{\ref{iKav},\ref{iCav}} \and
   Giacomo Venturi\inst{\ref{iSNS}}}
      
\authorrunning{I.  Lamperti et al.}

\institute{
Centro de Astrobiología (CAB),  CSIC-INTA,  Ctra. de Ajalvir Km. 4, 28850 Torrejón de Ardoz, Madrid, Spain\label{iCAB}
\and Dipartimento di Fisica e Astronomia, Università di Firenze, Via G. Sansone 1, 50019, Sesto F.no (Firenze), Italy\label{iUNIFI}
\and INAF - Osservatorio Astrofisico di Arcetri, largo E. Fermi 5, 50127 Firenze, Italy\label{iOAA}
\and European Space Agency (ESA), European Space Astronomy Centre (ESAC), Camino Bajo del Castillo s/n, 28692 Villanueva de la Cañada, Madrid, Spain\label{iESAC}
\and Department of Physics and Astronomy, University College London, Gower Street, London WC1E 6BT, UK\label{iUCL}
\and  University of Oxford, Department of Physics, Denys Wilkinson Building, Keble Road, Oxford OX13RH, United Kingdom\label{iOxford}
\and Scuola Normale Superiore, Piazza dei Cavalieri 7, I-56126 Pisa, Italy\label{iSNS}
\and Sorbonne Universit\'e, CNRS, UMR 7095, Institut d'Astrophysique de Paris, 98 bis bd Arago, 75014 Paris, France\label{iSorbonne}
\and Kavli Institute for Cosmology, University of Cambridge, Madingley Road, Cambridge, CB3 0HA, UK\label{iKav}
\and Cavendish Laboratory - Astrophysics Group, University of Cambridge, 19 JJ Thomson Avenue, Cambridge, CB3 0HE, UK\label{iCav}
\and NRC Herzberg, 5071 West Saanich Rd, Victoria, BC V9E 2E7, Canada\label{iNRC}
\and European Space Agency, c/o STScI, 3700 San Martin Drive, Baltimore, MD 21218, USA\label{iESA}
\and European Southern Observatory, Karl-Schwarzschild-Strasse 2, 85748 Garching, Germany\label{iESO}
\and AURA for the European Space Agency, Space Telescope Science Institute, Baltimore, MD, USA\label{iAURA}
}


 
  \abstract
  {
  We present JWST Near-Infrared Spectrograph (NIRSpec) observations in integral field spectroscopic (IFS) mode of the galaxy GS5001 at redshift z=3.47,  the central  
  member of a candidate protocluster in the GOODS-S field.
  The data cover a field of view (FoV) of $4\arcsec\times4\arcsec$ ($\sim30\times30$~kpc$^2$) and were obtained as part of the Galaxy Assembly with NIRSpec IFS (GA-NIFS) GTO programme.
The observations include both high (R$\sim$2700) and low (R$\sim$100) spectral resolution data, spanning the rest-frame wavelength ranges 3700-6780~\AA\ and 1300-11850~\AA, respectively. These observations enable the detection and mapping of the main optical emission lines from \OII$\lambda \lambda3726,29$  to \SIII$\lambda9531$. 
We analysed the spatially resolved ionised gas kinematics and interstellar medium properties, including obscuration, gas metallicity, excitation, ionisation parameter, and electron density.
In addition to the main galaxy (GS5001), the NIRSpec FoV covers a close companion in the south, 
with three sub-structures with velocities blueshifted by  $\sim-150$~\kms\ with respect to GS5001, and another source in the north redshifted by $\sim200$~\kms. 
Optical line ratio diagnostics indicate star formation ionisation and electron densities of $\sim 500$~cm$^{-3}$ across all sources in the FoV.
The gas-phase metallicity in the main galaxy is 12+log(O/H) $= 8.45\pm0.04$, and slightly lower in the companions (12+log(O/H)$ = 8.34-8.42$), consistent with the mass-metallicity relation at $z\sim3$.
We find peculiar line ratios (high $\log$\NII/\Ha $=[-0.45, -0.3]$, low $\log$\OIII/\Hb $= [0.06, 0.10]$) in the northern part of  GS5001. These could be attributed to either higher metallicity, or to shocks resulting from the interaction of the main galaxy with the northern source.
We identify a spatially resolved outflow in the main galaxy, traced by a broad symmetric component in \Ha\ and \OIII, with an extension of about 3~kpc. We find maximum outflow velocities of $\sim400$~\kms, an outflow mass of $(1.7\pm0.4)\times 10^8$~\Msun, a mass outflow rate of $23\pm5$ \Msunyr, and a mass loading factor of 0.23. These properties are compatible with star formation being the driver of the outflow.
Our analysis of these JWST NIRSpec IFS data therefore provides valuable, unprecedented insights into the interplay between star formation, galactic outflows, and interactions in the core of a $z\sim 3.5$  candidate protocluster.
 }

   {}
   {}
   {}
   {}
   {}


	\maketitle
%

\section{Introduction}
\label{sec:intro}

Understanding the complex processes governing galaxy formation and evolution in the early Universe requires detailed observations of their interstellar medium (ISM) properties and kinematics.
The ISM's physical properties and kinematics provide essential clues to the mechanisms driving galaxy evolution, including star formation, gas inflows and outflows, and the chemical enrichment \citep[e.g.][]{Naab2017, Cresci2018, Maiolino2019}.
Over the past few decades, several studies have investigated how the ISM properties evolve with cosmic time, focussing for instance on the ionisation parameter, electron density and temperature, and metallicity \citep[e.g.][]{Nakajima2014, Kaasinen2017, Kaasinen2018, Davies2021, Curti2024,  Reddy2023b, Reddy2023c, Sanders2016a, Sanders2021, Sanders2024}.

One of the most significant advancements in this field has been the advent of spatially resolved studies of galaxies at redshifts of $z=1-3$. 
These studies have highlighted the importance of resolving the inner structure of the ISM properties in order to understand galaxy evolution \citep[e.g.][]{Cresci2010, Troncoso2014, Stott2016,  Curti2020b_KLEVER, Gillman2022}.
Spatially resolved observations of optical emission lines have also allowed the study of the internal kinematics of galaxies up to high redshift \citep[$z\sim1-3$, e.g.][]{Wisnioski2015, Wisnioski2019, Harrison2017a, FoersterSchreiber2018,  vanHoudt2021, Genzel2023, Uebler2024a}.
These studies have also identified and characterised outflows powered by active galactic nuclei (AGN) and star-bursts \citep[e.g.][]{Genzel2014, ForsterSchreiber2014, ForsterSchreiber2019}.

The ability to resolve the internal structure of galaxies is crucial for understanding how they evolve over time and how the conditions within their ISM change. 
By mapping the distribution of star formation, one can determine where stars are forming within a galaxy, and hence whether star formation is concentrated in the nucleus or more spread out across the disc \citep[e.g.][]{GonzalezDelgado2016, Spindler2018}.  
 Mapping gradients of gas-phase metallicity allows us to study the flows of gas within galaxies and where the enrichment of metals of the ISM takes place \citep[e.g.][]{Cresci2010, Maiolino2019}. 
Moreover, the internal kinematics of galaxies allow us to assess whether galaxies at high redshift were already settled into well-ordered rotating discs or if they exhibited more chaotic, turbulent motions.
Additionally, kinematic information, gas-phase metallicity, and star formation distributions are essential for understanding the role of inflows and outflows in regulating the growth of galaxies across cosmic time \citep[e.g.][]{Naab2017, Nelson2019, FoersterSchreiber2020, Scharre2024}. 

At cosmic noon ($z=1-2$), many of these studies have been performed using ground-based facilities with near-infrared integral field spectroscopy (IFS) instruments.
However, ground-based observations are limited in their ability to probe beyond a redshift of $z\sim 2.5$ for emission lines like \Ha, and up to $z\sim 3.5$ for \Hb\ and \OIII. These transitions are key to infer the ISM chemical and kinematic properties.
The IFS capabilities of the NIRSpec instrument \citep{Jakobsen2022, Boker2022} on board the \textit{James Webb} Space Telescope (JWST) enable one for the first time to map the optical lines at $z>3$ with a spatial resolution of $\sim 0.1$".
Several works have demonstrated the capability of JWST/NIRSpec integral field units to study the ISM properties and gas kinematics in star-forming galaxies and AGN host galaxies at $z>3$ \citep[e.g.][]{Wylezalek2022, Uebler2023,  Uebler2024c, Uebler2024, Perna2023, Arribas2024, Parlanti2024,
Vayner2024,  Wang2024arxiv, Roy2024arxiv, Saxena2024arxiv}.

\begin{figure*}[!]
\centering 
\includegraphics[width=0.28\textwidth]{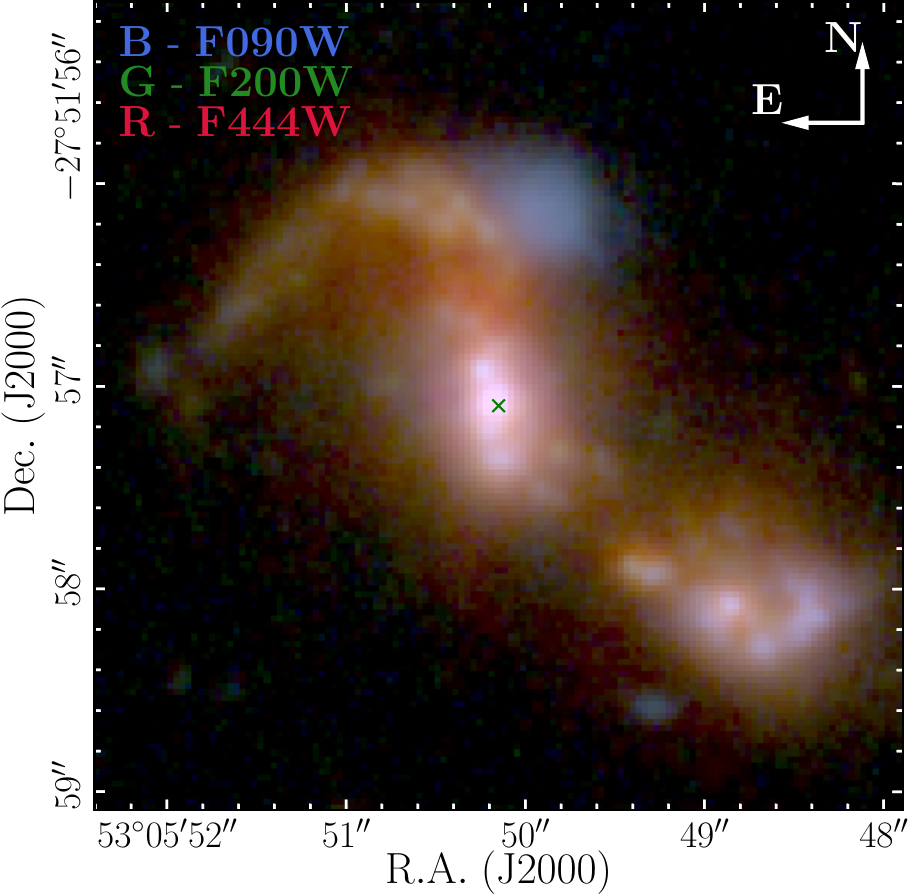}
\includegraphics[width=0.7\textwidth]{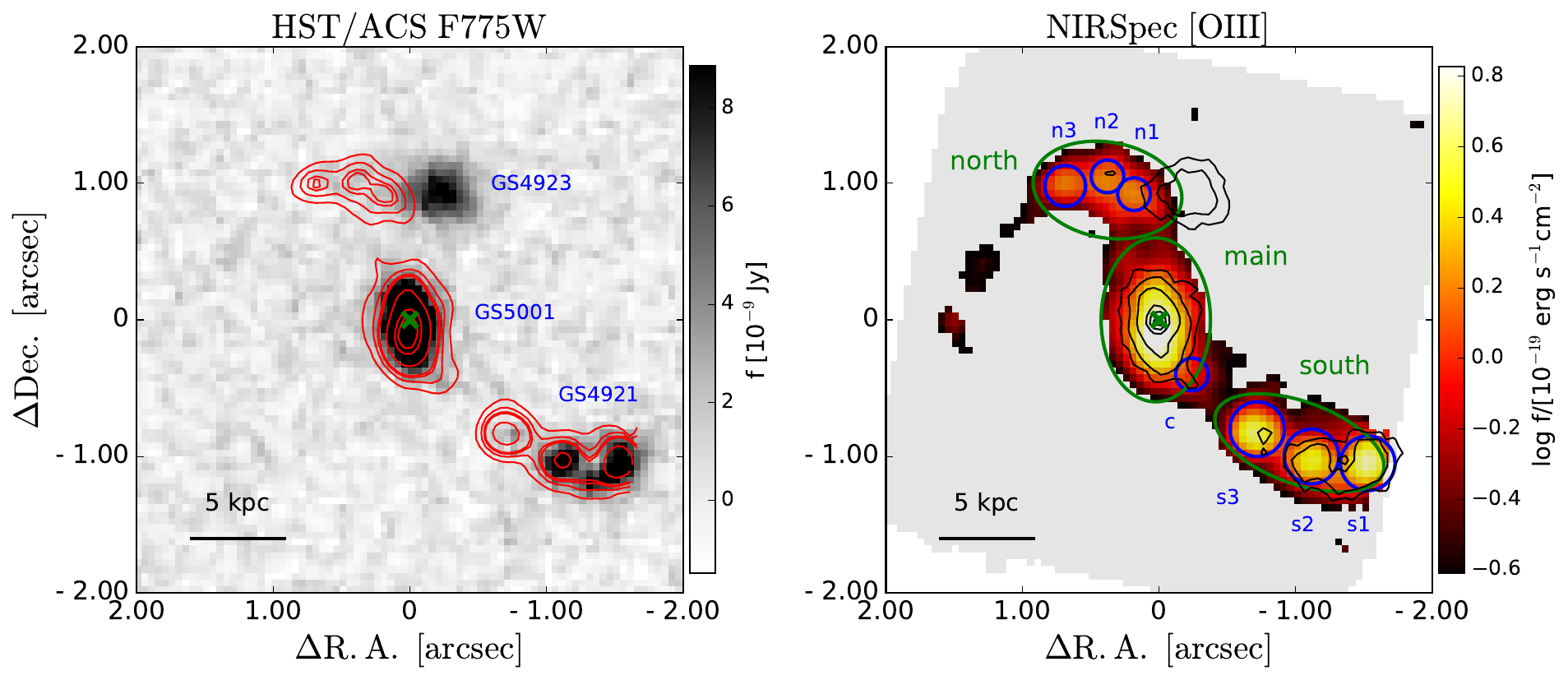}

\caption{Identification of galaxies and structures around GS5001. \textit{Left:} NIRCam three-colour image of GS5001 (F090W filter in blue, F200W filter in green and F444W filter in red).  
\textit{Middle:} HST/ACS F775W band image with contours of the NIRSpec \OIII\ emission in red. In blue are the names of the sources identified in CANDELS. \textit{Right:} NIRSpec \OIII\ flux map with contours of the HST ACS/F775W emission in black. Blue and green regions mark the different components identified in the \OIII\ map, from which we extracted the integrated spectra reported in Fig.~\ref{fig:int_spectrum} and Fig.~\ref{fig:int_spectra} in the appendix. 
 The green crosses show the centre of the continuum emission in the HST/WFC3 F125W filter, used for astrometric registration.}
\label{fig:HST-NIRSpec_maps}
\end{figure*}

In this work, we present JWST/NIRSpec IFS observations of the galaxy GS5001 and its close companions. 
GS5001 \citep[also known as Candels-5001 and as ID 4417 in the GOODS-MUSIC catalogue,][]{Grazian2006} is a Lyman-break selected galaxy in the GOODS-S field at z=3.473 \citep{Maiolino2008} at co-ordinates R.A. 03h32m23.35s, Dec. $-$27$^{\circ}$51'57.13'' (J2000).
Two other galaxies detected in the HST/UV images lie in a close projected separation (1-2$\arcsec$, 7-15~kpc): GS4921 to the south (ID 4414 in the GOODS-MUSIC catalogue) and GS4923 to the north (see central panel in Fig.~\ref{fig:HST-NIRSpec_maps}). GS4921 was confirmed to be at a similar redshift ($z=3.471$), based on VLT/SINFONI observations \citep{Maiolino2008}. GS4923 has a photometric redshift of 0.2 \citep{Perez-Gonzalez2005}, but some \OIII\ emission at a similar redshift of 3.47 was tentatively identified at its location using VLT/SINFONI data \citep{Ginolfi2017}.

GS5001 is the central member of a system classified as a candidate protocluster, as it lies in a large-scale overdensity of galaxies \citep{Franck2016a}. 
\citet{Ginolfi2017} detect a large reservoir of molecular gas, traced by CO(4--3), which extends up to 40~kpc around this system.
Additionally, they detect several CO(4--3)-emitting systems within a radius of 250 kpc, mostly distributed along the same northeast-southwest direction as the CO(4--3) reservoir (see their Fig. 9); they suggest that these systems are tracing the inner and densest regions of a large-scale accreting stream feeding the central massive galaxy GS5001.

GS5001 has a star formation rate (SFR) in the range 150-240 \Msunyr\ \citep[][]{Troncoso2014, Pacifici2016, Perez-Gonzalez2005, Guo2013}, derived from spectral energy distribution (SED) fitting of the UV to FIR photometry, and a stellar mass in the range $M_{\star}= 1-4\times 10^{10}$~\Msun\ \citep[][]{Maiolino2008, Perez-Gonzalez2005, Guo2013, Santini2015, Pacifici2016}. 
The southern companion, GS4921, has SFR = $35-70$~\Msunyr\ and stellar mass $M_{\star}= 0.5-2.3\times 10^{10}$~\Msun\ \citep[][]{Maiolino2008, Perez-Gonzalez2005, Guo2013}. 
All values reported here have been homogenised to a \citet{Chabrier2003} IMF.
GS5001 is also detected in the X-ray, although with a small number of counts (20 counts), and has an aperture corrected X-ray luminosity of $L_{0.5-7\,\mathrm{keV}}=9.5\times 10^{42}$ erg s$^{-1}$ \citep{Fiore2012,Luo2017}. 
Despite the high X-ray luminosity, it is not clear whether this object hosts an AGN, because its very high SFR could be responsible for most of the X-ray flux \citep{Fiore2012}.

The goal of this work is the detailed characterisation of the physical and kinematic properties of GS5001 and its close environment.
Other recent studies characterising dense environments in the early universe using JWST/NIRSpec IFS include those by \citet[][LBQS 0302-0019; a dual AGN with multiple companions at $z\sim3.3$]{Perna2023}, \citet[][GS4891; a galaxy group at $z\sim3.7$]{RodriguezDelPino2024},  \citet[][HZ10; a galaxy group at $z\sim5.7$]{Jones2024barXiv}, \citet[][HFLS3; an overdensity at $z\sim6.3$]{Jones2024}, and \citet[][SPT0311-58; a protocluster core at $z\sim6.9$]{Arribas2024}.
This paper is organised as follows. Section~\ref{sec:obs_datared} presents the JWST/NIRSpec observations and the data reduction. In Section~\ref{sec:system_description} we describe the different components of the GS5001 system.
Section~\ref{sec:line_fit} describes the method used for the emission line fitting.
In Section~\ref{sec:results} we present the results. First, we show the ionised gas kinematics (Sec.~\ref{sec:kinematics}) and then the properties of the ISM (attenuation, source of excitation, metallicity, ionisation parameter, electron density, Sec.~\ref{sec:ISM_prop}). The Discussion is in Section~\ref{sec:discussion} and the Conclusions in Section~\ref{sec:conclusions}.

Throughout this work, we assume a cosmological model with $\Omega_{\lambda} = 0.7$, $\Omega_{\text{M}}= 0.3$, and $H_0 = 70$ km s$^{-1}$ Mpc$^{-1}$, 
which results in a scale of 7.34~kpc/arcsec at $z =3.47$.
In this work, we assume a \citet{Chabrier2003} initial mass function (IMF).

\section{Observations and data reduction}
\label{sec:obs_datared}

\subsection{JWST/NIRSpec observations }
\label{sec:obs}
We present observations of GS5001 obtained on December 10 2022 as part of the Galaxy Assembly with NIRSpec Integral Field Spectroscopy (GA-NIFS\footnote{\href{https://ga-nifs.github.io}{https://ga-nifs.github.io}}) GTO programme (PIs: S.~Arribas, R.~Maiolino), contained within programme \#1216 (PI: K.~Isaak).
The main goal of GA-NIFS is to study the internal structure and the close environment of a sample of 55 galaxies and AGN at $z=2-11$ \citep{Perna2023a}.

GS5001 was observed with NIRSpec IFS \citep{Boker2022, Rigby2023} both with the high-resolution ($R\sim2700$) and with the low-resolution ($R\sim100$) mode. 
The R2700 observations use the grating/filter pair G235H/F170LP, covering the wavelength range $1.66-3.16$~\micron\ with a spectral resolution R$\sim1900-3500$, corresponding to $\sigma \sim 35-70$~\kms  \citep{Jakobsen2022}.
The R100 PRISM/CLEAR data covers the range $0.6-5.3$~\micron\ with R$\sim30-300$, corresponding to $\sigma \sim 300-4000$~\kms  \citep{Jakobsen2022}.
Both observations were taken using an IRS$^2$ detector readout pattern \citep[NRSIRS2 for R2700 and NRSIRS2RAPID for R100;][]{Rauscher2017}, which significantly reduces the detector noise with respect to the standard procedure.
 The R2700 observations have an exposure time of 4.1 hours, 
while the R100 observations an exposure time of 1.1 hours.
 We use an 8-point medium cycling dither pattern, which provides a total observed field of view (FoV) of $\sim4"\times4"$ ($\sim30\times30$~kpc$^2$ at the redshift of the target).

\subsection{Data reduction}
\label{sec:dara_red}

The raw data were reduced with the JWST calibration pipeline version 1.8.2, using the CRDS context file {\it jwst\_1068.pmap}. 
The default reduction pipeline was modified to improve the data quality, as is explained in detail in \citet{Perna2023} and \citet{DEugenio2023arXiv}.
We build the final cubes with a spaxel size of 0.05" using the `drizzle' method.

For the subtraction of the background emission from the R100 data cube, we extract spectra from spaxels in regions of the FoV where no galaxy emission is present 
and calculate a median spectrum. Then, we subtract this median spectrum from the data cube.
The background subtraction is not needed for the R2700 data, as the background contribution is not significant and it is accounted for during the line fitting procedure. 

We note that the noise given in the data cube (`ERR' extension) is underestimated compared to the actual noise in the data. Therefore, we re-scale the `error' vector in each spaxel by a factor of (median 1.6) to match the noise estimated from the standard deviation of the continuum in spectral regions free of line emission \citep[following the method used by e.g.][]{Uebler2023, RodriguezDelPino2024}. 
 Specifically, we selected the continuum spectral regions $4700-4800$~\AA, $5050-5200$~\AA, and $6000-6400$~\AA\ (rest-frame wavelength).

\begin{figure*}[!t]
\centering 
\includegraphics[width=0.95\textwidth]{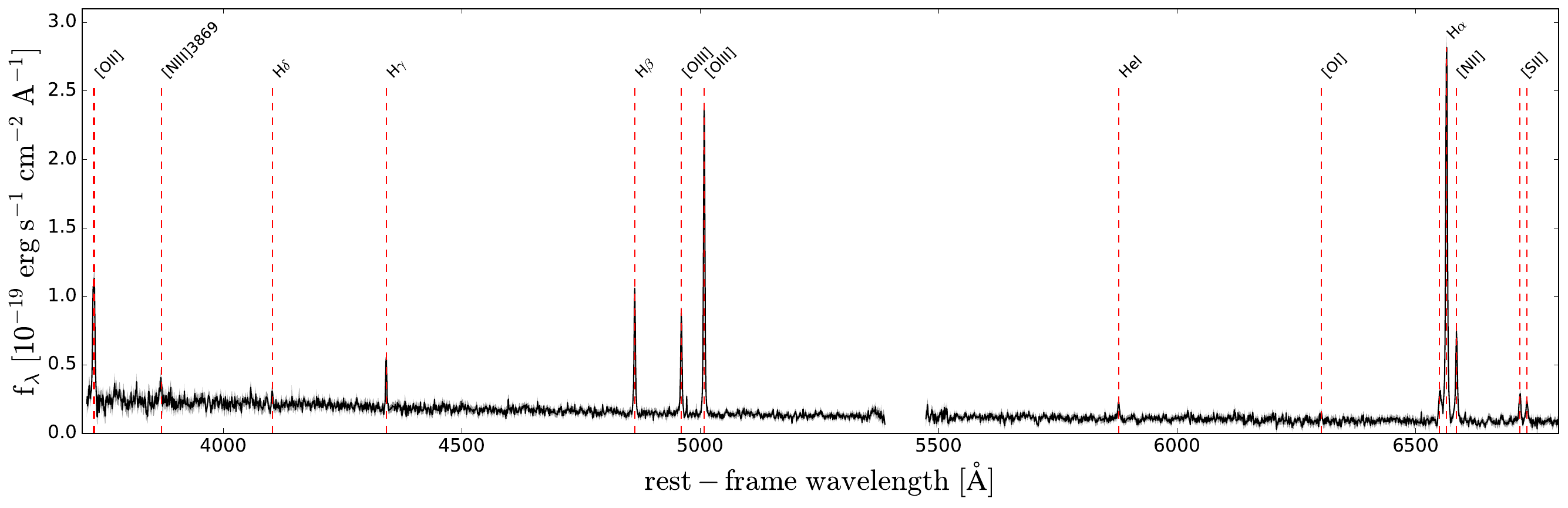}
\includegraphics[width=0.95\textwidth]
{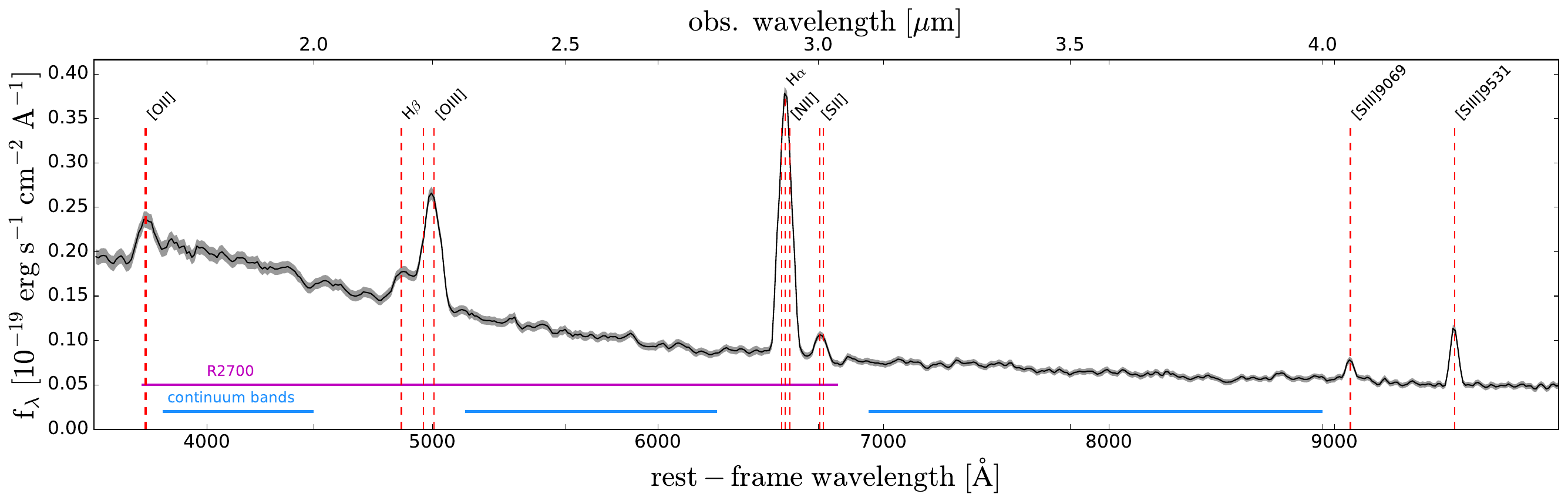}
\caption{Average rest-frame spectrum of the main galaxy central $3\times3$ spaxels ($0.15\arcsec \times0.15\arcsec$), from the high-resolution ($R\sim2700$, upper) and low-resolution ($R\sim100$, lower) NIRSpec data. The aperture was chosen as it is representative of the aperture used for the  spaxel-by-spaxel fit. For  R100, only the portion of the spectrum with the lines of interest is shown. Vertical dashed lines mark the position of the detected emission lines.  The shaded grey area shows the error on the fluxes. The magenta line in the R100 panel shows the wavelength range covered by the R2700 data.  The light blue lines show the wavelength ranges used to create the continuum maps shown in Fig.~\ref{fig:cont_maps}.}
\label{fig:int_spectrum_spaxel}
\end{figure*}

Fig.~\ref{fig:int_spectrum_spaxel} shows the R2700 and R100 spectra extracted from the central $3\times3$ spaxels of the main galaxy.
This figure illustrates the wavelength range spanned by the two spectra, the main emission lines and the quality of the data.

\begin{figure*}[!t]
\centering 
\includegraphics[width=0.98\textwidth]{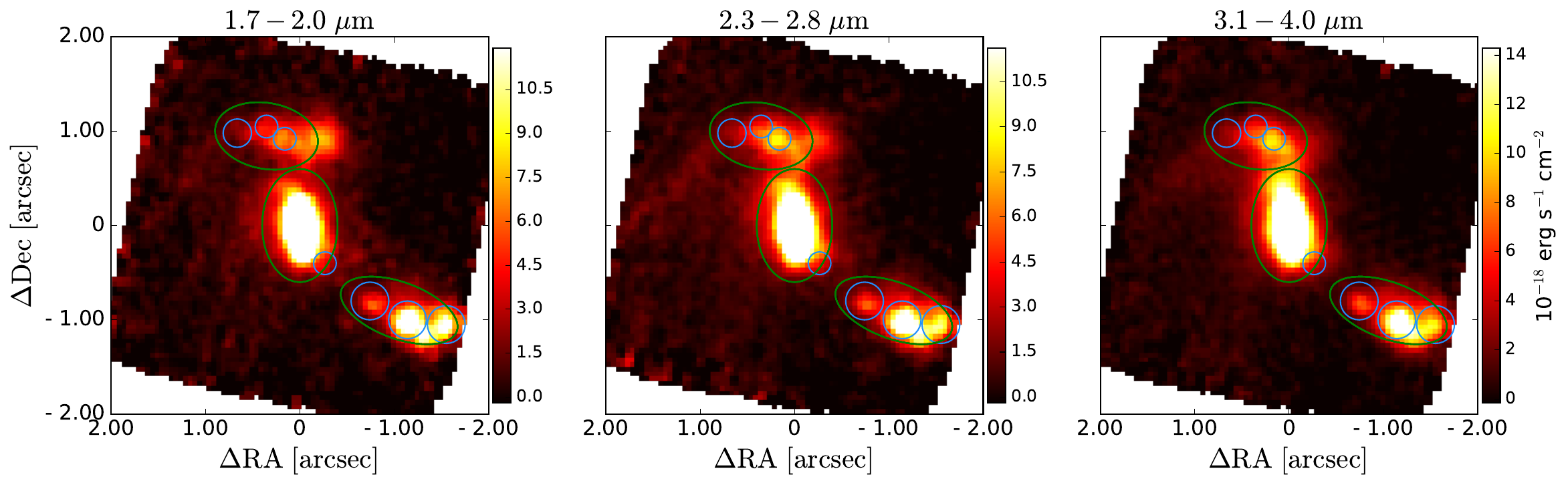}
\caption{Continuum maps obtained collapsing the R100 data cube over three different (observed) wavelength ranges, as is indicated above the panels. The wavelength ranges are also shown in the R100 spectrum  in Fig.~\ref{fig:int_spectrum_spaxel}. The green ellipses indicate the three main components (`main', `south' and `north'), the light blue circles indicate the identified sub-clumps.}
\label{fig:cont_maps}
\end{figure*}

\subsection{Ancillary data}
In addition to the NIRSpec data, we also consider other observations of GS5001. In particular, we use images from HST and JWST/NIRCam, and ALMA observations of the CO(4--3) emission line.

The HST/ACS WFC F755W and HST/WFC3 F125W images of GS5001 have been downloaded from the Rainbow Cosmological Survey database\footnote{\href{https://arcoirix.cab.inta-csic.es/Rainbow_Database/}{https://arcoirix.cab.inta-csic.es/Rainbow$\_$Database/} }. The astrometric alignment is explained in the following section.
We also download publicly available JWST/NIRCam images (F090W, F200W, F444W) of GS5001 from the multi-object JWST Deep Extragalactic Survey \citep[JADES;][]{Eisenstein2023arxiv, Rieke2023}. The astrometric alignment of the NIRCam images is described in \citet{Rieke2023}.

In this work, we also use ALMA Band 3 observations of the CO(4--3) emission line at rest-frequency 461.04~GHz (project ID 2012.1.00423.S, P.I. T.~Nagao), which have been presented in \citet{Ginolfi2017}.
We obtain the calibrated measurement sets (MS) from the European ALMA Regional Centre (ARC). We create a data cube of the CO(4--3) spectral window using the ALMA data reduction software {\tt CASA} v6.2.1 \citep{CASA}.
 We use the CASA task {\tt tclean} applying natural weighting to generate a data cube with synthesised beam FWHM $1.9 \times 1.3$ arcsec$^2$.
The original spectral resolution of the observations is $\sim2.5$~\kms and we binned the spectral channels to reach a resolution of $\sim10$~\kms.
The CO(4--3) intensity map (moment 0) is produced by integrating the continuum subtracted data cube in the spectral channels within $\pm800$~\kms\ from the CO(4--3) emission frequency at $z = 3.47$.

\subsection{Astrometry}
\label{sec:astrometry}

We register the astrometry of the NIRSpec data by matching the central position of the target GS5001 in the NIRSpec data cube and in the HST/WFC3 F125W band image, after aligning the HST image to the \gaia\ co-ordinates frame. 
To derive the absolute astrometry of the HST image, we used the two stars present in a $90''\times90''$ FoV around GS5001 and calculated the mean offset of their position from the co-ordinates reported in the \gaia\ DR3 catalogue \citep{Gaia2016, Gaia2023j}. 
The derived offset (co-ord(HST)-co-ord(\gaia)) is $\Delta \text{R.A.}= -0.1969 \pm0.0430$ arcsec and $\Delta \text{Dec.}= -0.27479\pm0.0081$ arcsec. The uncertainties, calculated from the standard deviation of the offsets of the two stars, are 43mas in R.A. and 8mas in Dec.

Then, we align the position of the peak emission of our target (GS5001) in the HST/WFC3 125W band, derived with a 2D Gaussian fit, and in an image created by collapsing the NIRSpec R100 data cube over a similar wavelength range ($1.1- 1.4\mu$m observed wavelength). 
In this way, we derive a correction of $\Delta \text{R.A.} = -0.1291$ arcsec and $\Delta \text{Dec.} =-0.1554$ arcsec for the astrometry of the NIRSpec data cubes. This offset is consistent with the expected accuracy of JWST pointing without a dedicated target acquisition procedure \citep{Boeker2023}.
We check that the positions of the other sources in the FoV are well aligned in the HST and NIRSpec image, so no rotation term is considered. 

\section{Description of the GS5001 system}
\label{sec:system_description}
Figure~\ref{fig:HST-NIRSpec_maps} shows the map of the integrated \OIII\ flux of the system (right panel) obtained from the emission line fit  of the R2700 NIRSpec data cube (see Sec.~\ref{sec:line_fit}), together with the HST/ACS F755W image (central panel) and a NIRCam three-colour image (left panel; F090W filter in blue, F200W filter in green and F444W filter in red). 
The ionised gas emission extends over more than 20~kpc and is divided in different structures: a central galaxy, and two other components, one extending to the north-east and one to the south-west.

In the south component, we identify three sub-structures (s1, s2, s3), two of which are well detected in the HST/ACS image (s1 and s2). These two sub-structures were previously identified in the CANDELS catalogue as the source GS4921 \citep[][see Fig.~\ref{fig:HST-NIRSpec_maps}]{Guo2013}.
In the north, the \OIII\ emission is concentrated in a region to the east of the UV-bright galaxy detected in the HST/ACS image, with CANDELS identifier 4923. This galaxy has a photometric redshift of $z=0.2$ \citep{Perez-Gonzalez2005}.  Given the lack of emission lines in the NIRSpec data cubes,  it is indeed compatible with being a low-redshift foreground galaxy. 
The \OIII\ emission in the north shows three peaks (n1, n2, n3) and a long tail elongated towards the south-east.
 The  \OIII\ emission detected in the SINFONI data to the north of GS5001, with a similar redshift of $z\sim3.47$ and originally attributed to GS4923 \citep{Ginolfi2017}, is likely coming from the north companion (see Fig.~\ref{fig:HST-NIRSpec_maps}).
 We identify an additional clump (`\clump') to the south-west of the main source,  which is clearly visible in the velocity channels $[130-150]$~\kms\ (see Fig.~\ref{fig:clump_SW} in the appendix). 
We shall discuss the nature of this clump in Section~\ref{disc:dyn}. 
We also detect diffuse ionised gas emission in the regions in between the three components (called `main', `south' and `north' from now on).

To investigate the nature of the north and south components, we construct continuum maps by collapsing the R100 data cube in three wavelength ranges that are free from strong line emission: 1.7--2.08~$\mu$m, 2.3--2.8~\micron and 3.1--4.0~\micron (observed wavelength). We could not use the NIRCam images for this purpose because they are contaminated by strong emission lines (\OII, \OIII, \SIII).
The south component and its three sub-clumps are detected in the three continuum bands (see Fig.~\ref{fig:cont_maps}). The source s3 is fainter than the other two, but it is well detected in continuum. The north component is also detected in the three continuum bands. The continuum is strongest where the n1 peak is located, and fainter in the position of the n3 peak.

The presence of extended \OIII\ and continuum emission suggests that GS5001 is in a complex environment, with two close companions and evidence of gravitational interactions. The nature of each substructure in the three systems is further investigated in the next sections.

\section{Analysis: Emission line fitting}
\label{sec:line_fit}
To derive the properties of the ionised gas in this system, we use two approaches. First, we run a spaxel-by-spaxel fit, which allows us to derive spatially resolved maps of the gas physical conditions and kinematics. Since some emission lines are faint, we also extract aperture spectra from different regions of the system to increase the S/N and derive more robust measurements of their physical properties. The apertures used to extract the spectra are shown in the right panel of Fig.~\ref{fig:HST-NIRSpec_maps}. In particular, we extract the spectra of the `main' galaxy and the two `north' and `south' components (green ellipses in Fig.~\ref{fig:HST-NIRSpec_maps}). Additionally, we extract spectra of the `sub-structures' identified as \OIII\ peaks (n1, n2, n3, s1, s2, s3) and the clump `\clump' (blue circles in Fig.~\ref{fig:HST-NIRSpec_maps}).

\subsection{R2700 data}
In this section, we describe the method used to derive the emission line properties by fitting the R2700 data cube. 
We first measure the continuum level by taking the mean of the flux in  two continuum spectral windows free from emission lines on the two sides of the  \Ha\ line. 
In the cases where the continuum is detected (i.e. the mean flux is larger than zero),  we model it by interpolating the flux across the entire wavelength range, excluding the spectral regions within $\pm10$~\AA\  from the position of the emission lines. We interpolate the flux using a Savitzky-Golay filter \citep{Savitzky1964} and first-order polynomials.
If the continuum is not detected, we do not apply this interpolation.
 
After subtracting the continuum, we fit the following emission lines: \Ha, \Hb, \Hg, \OIII$\lambda \lambda$4959,5007,  \NII$\lambda \lambda$6548,6583,  \SII$\lambda \lambda$6716,6731,  \OII$\lambda \lambda$3726,3729, \OI$\lambda$6300.
We fit all these lines simultaneously using a combination of Gaussian profiles. We force the kinematic parameters (velocity centroid and line velocity width) of all lines to be the same. 
We fix the flux ratio of the \NII$\lambda$6583/\NII$\lambda$6548 and \OIII$\lambda$5007/\OIII$\lambda$4959 lines to the theoretical value 2.99 \citep{Osterbrock2006}.
The relative flux of the \OII\ and \SII\ doublets is allowed to vary between 0.384 < \OII$\lambda$3729/ \OII$\lambda$3726 < 1.456 and 0.438 < \SII$\lambda$6716/\SII$\lambda$6731 < 1.448, which correspond to the theoretical limits for densities in the range $1-10^5$~cm$^{-3}$ \citep{Sanders2016a}.
To retrieve the intrinsic line width, we convolve the Gaussian profiles with the line spread function at the corresponding wavelength during the fitting procedure, using the resolving power as a function of wavelength presented in \citet{Jakobsen2022}\footnote{Downloaded from \href{https://jwst-docs.stsci.edu/jwst-near-infrared-spectrograph/nirspec-instrumentation/nirspec-dispersers-and-filters}{https://jwst-docs.stsci.edu/jwst-near-infrared-spectrograph/nirspec-instrumentation/nirspec-dispersers-and-filters}.}.

We consider two models: one with a single Gaussian and one with two Gaussian components (narrow+broad) for each emission line.  For the one-component model,  we constrain the width of the line to be $20$~\kms $ < \sigma< 300$~\kms. We set the lower limit due to the spectral resolution, and the higher value to guide the fit and avoid fitting noise.
For the two-component model, we constrain the width of the narrow component to be $20$~\kms $ <\sigma_{n}< 120$~\kms\ and the width of the broad component to be $140$~\kms < $\sigma_{b}< 300$~\kms.
We use as initial guess for the redshift the value $z=3.473$ \citep{Maiolino2008}.

\noindent - \textit{Spaxel-by-spaxel spectral fit}:\\
We  fit the emission lines across the FoV of the data cube to derive spatially resolved properties.
In order to increase the S/N,  we calculate the average of the spectrum in the selected spaxel and in the eight adjacent spaxels ($3\times3$).
We fit the emission lines using the Python routine `scipy.optimize.curve$\_$fit', which uses non-linear least-squares to fit a function to data.  
We fit the data cube with the one-component and two-component models, and then we select the number of components to use in each spaxel based on the Bayesian Information Criterion \citep[BIC,][]{Schwarz1978}.
We select the two-component model if it causes an improvement in the BIC larger than 20$\%$.

In Fig.~\ref{fig:spectrum_fit_example} in the appendix 
we show the spectrum of one spaxel together with its best-fit model as an example.
Figure~\ref{fig:Ha_map} shows the maps of the \Ha\ flux and kinematics obtained with the best-fit model.  
 For comparison, we also report in Fig.~Fig.~\ref{fig:Ha_map_1comp} the appendix 
 the velocity maps obtained with the one-component Gaussian fit, showing no significant differences from those in Fig.~\ref{fig:Ha_map}.
 The S/N in each line is calculated by taking the ratio of the amplitude of the best-fit model over the noise, estimated by taking the standard deviation of a close spectral region free of line emission. 

\noindent - \textit{Spatially integrated spectral fit:}\\
We also fit the integrated R2700 spectra of the different components of the system identified in Figure~\ref{fig:HST-NIRSpec_maps}. 
 Specifically, we extract integrated spectra of the main galaxy, the south component and its three sub-structures in the south (s1, s2, s3), the north component  and its three sub-structures  (n1, n2, n3). The positions and sizes of the apertures are reported in Table~\ref{tab:integrated_prop} and shown in Figure~\ref{fig:HST-NIRSpec_maps}.
We use the same methodology (same model and parameter constraints) as for the spatially resolved fit. In this case, we fitted the spectra using the Markov chain Monte Carlo (MCMC) programme \emcee\ \citep{Foreman-Mackey2013}, which allows us to robustly estimate the uncertainties. The MCMC approach was not used for the spatially resolved fit because it is too computationally expensive.
 We assume a Gaussian likelihood and use uniform priors within the velocity ranges specified above. 
We derive the best-fit parameters by taking the median values from the marginal posterior distributions. The uncertainties are given as the average of the differences between the median and the 16th and 84th percentiles of the posterior distributions.
To decide whether a two-component model is needed, we use the BIC, 
and require that BIC(2comp) is smaller than the BIC(1comp) by $> 20\%$. The two-component model is needed only for the spectrum of the main galaxy. 

he integrated spectrum of the main galaxy together with the best-fit model is shown in Fig.~\ref{fig:int_spectrum}. The integrated spectra of the other components (south, north, s1, s2, s3, n1, n2, n3, \clump) are shown in Fig.~\ref{fig:int_spectra} in the appendix. 
In Table~\ref{tab:integrated_prop} we report physical properties derived for the different regions, while the integrated fluxes measured from the integrated spectra are reported in Table~\ref{tab:integrated_fluxes}.

\subsection{R100 data}
We fit the R100 data cube to extract the \SIII$\lambda \lambda 9069,9531$  line fluxes, since this wavelength range is not covered by the R2700 data (see Fig.~\ref{fig:int_spectrum_spaxel}). These fluxes, together with the \SII $\lambda \lambda 6717, 6731$ fluxes, will be used to estimate the ionisation parameter, as is described in Sec.~\ref{sec:ionization}.
We use the same methodology as for the fit of the R2700 data cube.
In this case, we do not convolve the line width with the line spread function of the NIRSpec PRISM,  as we note that in some regions the emission lines are narrower than the pre-flight estimates of the spectral resolution curve reported in \citet{Jakobsen2022}.
This does not affect our results as we are only interested in extracting the line fluxes from the R100 data and not the kinematics.

The continuum is modelled and subtracted before the emission line fit using the same method described above for the R2700 data.
Given the low spectral resolution, we just employ one Gaussian to model each line. 
We model simultaneously the two \SIII\ lines, using a single Gaussian component for each line with the same width (in velocity space). 
We also fit the \Ha, \NII\ and \SII\ complex in a separate fit, forcing the line width to be the same for all the lines. The spectral resolution is too low to resolve the \SII\ doublet, but for the purpose of deriving the ionisation parameter, we only need the sum of the fluxes of the two lines. 
 As specified above,  we use the \SIII\ fluxes  together with the \SII\ fluxes to estimate the ionisation parameter. For this aim, we use \SIII\ and \SII\ fluxes derived from the same cube (R100) to avoid potential discrepancies in the flux calibration of the R100 and R2700 observations \citep[e.g.][]{Arribas2024}.
 Table~\ref{tab:integrated_fluxes_R100} reports the measured fluxes of the \SII\ and \SIII\ doublets derived from the R100 data cube for the different regions.

\section{Results}
\label{sec:results}

\begin{figure*}[!]
\centering 
\includegraphics[width=0.99\textwidth]{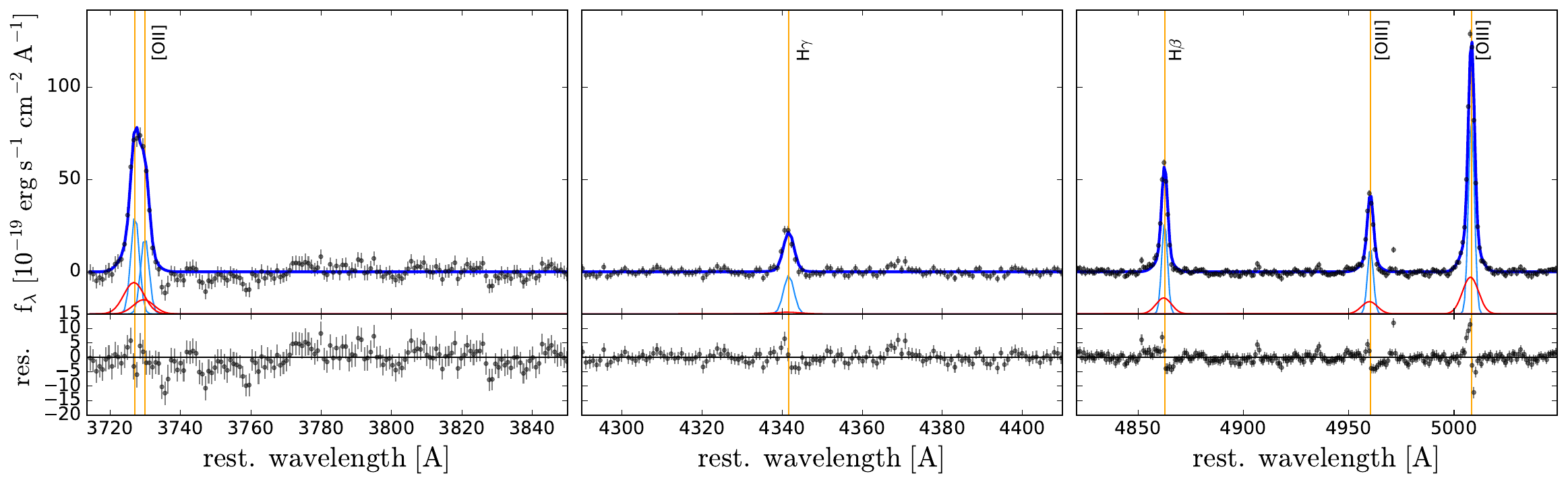}
\includegraphics[width=0.99\textwidth]{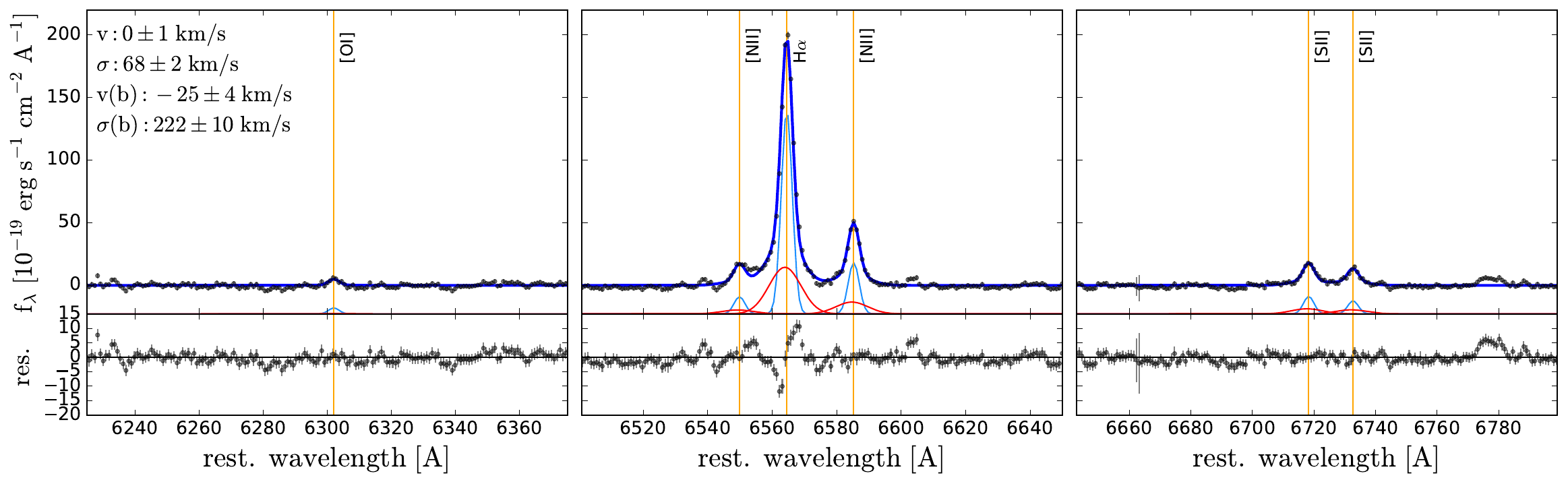}
\caption{Integrated spectrum of GS5001 (`main' source in Fig.~\ref{fig:HST-NIRSpec_maps}). Data are shown in black, the total best fit in blue, the narrow and broad components are shown in light blue and red, respectively, shifted vertically for visual purposes.  
The vertical lines mark the wavelength positions of the emission lines at the systemic redshift of the source ($z=3.4705)$. The fitting residuals are shown in the bottom panel.}
\label{fig:int_spectrum}
\end{figure*}

\begin{table*}
\centering
\caption{Physical properties of the different components of the system.}
\setlength{\tabcolsep}{3pt}
\begin{tabular}{lcccccccccccc}
\hline

region & $\Delta$RA  & $\Delta$Dec  & apert. & angle & SFR  & A$_V$ & 12+log(O/H) & log U &  n$_e$ & $v$ & $\sigma$  \\ 
   &   [arcsec] & [arcsec] & [arcsec]  & [degrees] &[\Msun\ yr$^{-1}$]     &  [mag.] & &  & [cm$^{-3}$]  & [\kms] &  [\kms]\\
 (1) & (2)  & (3) & (4) &(5) & (6) & (7) & (8) &  (9) & (10) & (11) & (12) \\ 
  \hline \hline

\hline


main$^{*}$ & 0 &  0 & 0.25$\times$0.45 & 0 & 180$\pm$1  & 1.54$\pm$0.07 & 8.45$\pm$0.04 &  --2.9$\pm$0.1 &  540$^{+120}_{-100}$& --1$\pm$1 & 93$\pm$1 \\
main narrow &  &  & & & 100$\pm$2 & 1.37$\pm$0.14 & 8.45$\pm$0.04  & - & - & 0$\pm$1 & 68$\pm$2  \\
main broad &  & & & &-  & 2.29$\pm$0.53 & 8.47$\pm$0.07 & - & - & --25$\pm$4 & 222$\pm$10  \\

south & --1.29 &  --1.00 & 0.55$\times$0.30 & -20 & 76$\pm$1  & 1.31$\pm$0.09 & 8.39$\pm$0.04 & --3.1$\pm$0.1 & 750$^{+220}_{-180}$& --133$\pm$1 & 80$\pm$1\\
north & 0.40 &  0.95 & 0.45$\times$0.25 & -10 & 106$\pm$1  & 2.44$\pm$0.16 & 8.39$\pm$0.06  & --2.8$\pm$0.1 & 150$^{+130}_{-90}$& 191$\pm$1 & 100$\pm$2\\
\hline
s1 & --1.75 &  --1.05 & 0.20$\times$0.20 & 0 & 12$\pm$1  & 0.63$\pm$0.20 & 8.36$\pm$0.07  & --2.7$\pm$0.2 & - & --161$\pm$2 & 68$\pm$2  \\
s2 & --1.29 &  --1.00 & 0.20$\times$0.20& 0 & 19$\pm$1  & 1.34$\pm$0.13 & 8.41$\pm$0.04  & --3.2$\pm$0.1 & 630$^{+210}_{-180}$& --154$\pm$1 & 77$\pm$1 \\
s3 & --0.84 &  --0.80 & 0.20$\times$0.20 & 0 & 30$\pm$1  & 1.75$\pm$0.10 & 8.37$\pm$0.05 & --2.7$\pm$0.2 & 200$^{+130}_{-100}$& --82$\pm$1 & 77$\pm$1 \\
n1 & 0.18 &  0.92 & 0.12$\times$0.12& 0 & 12$\pm$1  & 1.82$\pm$0.16 & 8.42$\pm$0.05 & --2.9$\pm$0.2& 260$^{+200}_{-150}$ & 211$\pm$1 & 80$\pm$2 \\
n2 & 0.40 &  1.05 & 0.12$\times$0.12 & 0 & 13$\pm$1  & 2.43$\pm$0.22 & 8.36$\pm$0.07 & --3.0$\pm$0.2 & 510$^{+320}_{-220}$& 217$\pm$1 & 73$\pm$2 \\
n3 & 0.75 &  0.98 & 0.15$\times$0.15 & 0 & 27$\pm$1  & 3.20$\pm$0.48 & 8.34$\pm$0.11  &  --3.1$\pm$0.1 & -& 180$\pm$3 & 104$\pm$3 \\
\clump & --0.30 &  --0.40 & 0.12$\times$0.12& 0 & 6$\pm$1  & 1.75$\pm$0.25 & 8.44$\pm$0.07  &--3.2$\pm$0.2 & -& 67$\pm$2 & 73$\pm$3 & \\

\hline


\hline
\end{tabular} 
\label{tab:integrated_prop}
\tablefoot{
Properties of the different components identified in Fig.~\ref{fig:HST-NIRSpec_maps}. 
(1) Name of the component.
(2) and (3) Distance in R.A. and Dec. from the central (main) source  at co-ordinates R.A. 03h32m23.3497s, Dec. --27$^{\circ}$ 51'57.13''.
(4) Semi-axes of the ellipse used as extraction aperture.
(5) Angle of the major axis of the ellipse used as extraction aperture measured east of north (anticlockwise).
(6) SFR derived from the dust-corrected \Ha\ luminosity using the relation from \citet{Kennicutt2012}, assuming a \citet{Chabrier2003} IMF. 
(7) Extinction derived from the Balmer decrement, assuming a \citet{Cardelli1989} attenuation law.
(8) Metallicity calculated from the R3, O3O2 and R23 calibrations from \citet{Curti2020}.
(9) Ionisation parameter inferred using the S2S3 calibration from \citet{Diaz2000}.
(10) Electron density estimated from the simultaneous fit of the \SII\ and \OII\ line ratio, using the formulas from \citet{Sanders2016a}, assuming the typical electron temperature of HII regions $T_e=10^4$~K. 
(11) Velocity with respect to the redshift of the narrow component of the main target  ($z= 3.4705\pm 0.0001$).
(12) Velocity dispersion.\\
$^{*}$ For the main target, we show the properties inferred from the fit with one component, and the properties derived separately from the narrow and broad component of the two components fit.}
\end{table*}

\subsection{Kinematic properties}
\label{sec:kinematics}

\subsubsection{Velocity maps}
\begin{figure*}[!t]
\centering 
\includegraphics[width=0.99\textwidth]{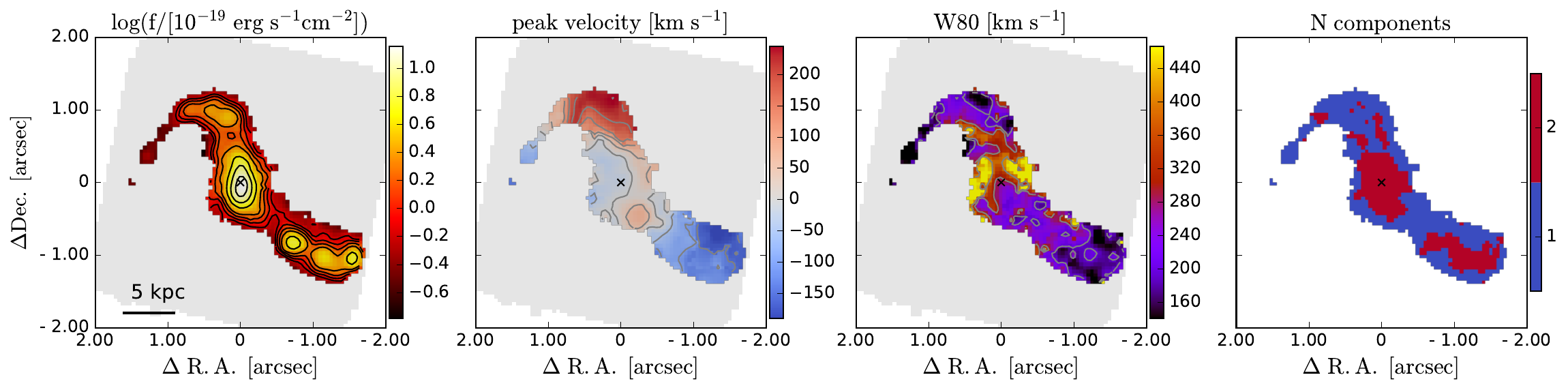}
\caption{Maps of the observed \Ha\ emission  obtained from the emission line fit with the best-model (with one or two Gaussian components). The number of components used for the fit in each spaxel is indicated in the right panel. Only spaxels with peak S/N> 3.5 in the \Ha\ line are shown.
From left to right: integrated observed flux (not corrected for obscuration), velocity at the peak of the emission,  line width W80 (width encompassing 80\% of the flux, derived from the total line profile), number of Gaussian components used in the fit. Contours in the flux map show the [20, 40, 60, 80, 90] percentiles; contours in the velocity map start at --150~\kms and increase every 50~\kms; contours in the W80 map are  at [200, 300, 400]~\kms.}
\label{fig:Ha_map}
\end{figure*}

Figure~\ref{fig:Ha_map} shows the maps of the \Ha\ line derived from the fit with one or two Gaussian components. The optimal number of Gaussian components to be used in each spaxel was selected based on a BIC threshold. Only spaxels with peak S/N> 3.5 in the \Ha\ line are shown. The maps show the total observed line flux (not corrected for obscuration), the velocity at the peak of the line,  and the line width W80 (width encompassing 80\% of the total line emission,  calculated as the difference between the 10th and 90th percentiles velocities).
The zero velocity is determined based on the redshift of the central target, derived from the integrated spectrum ($z=3.4705 \pm 0.0001$).

The southern clumps s1, s2, and s3 have velocities blueshifted by $-160, -153, -80$~\kms, respectively, with respect to the central galaxy (see Table~\ref{tab:integrated_prop}).
The northern clumps  n1, n2, and n3 instead are redshifted by $213, 218, 182$~\kms, respectively, and the tail shows a velocity gradient going from $\sim100$~\kms\ close to the north galaxy to $-100$~\kms\ towards the SE direction.

In the central galaxy, we see a gradient with the velocity increasing from --30~\kms\ in the north-east to 50~\kms\ in the south-west. If we interpret this gradient as rotation, the kinematic major axis is roughly along the NE-SW direction. 
We note that the direction of this velocity gradient (kinematic major axis)  is not aligned with the photometric major axis, which is roughly in the north-south direction (shift of $\sim 40^{\circ}$ between photometric and kinematic major axes), as commonly observed in interacting systems and mergers at low-redshift \citep[e.g.][]{Barrera-Ballesteros2015, Perna2022}.
The peak velocity increases at the location of the clump `\clump', which indeed has a velocity redshifted by $\sim140$~\kms.
If we were to exclude this clump, the velocity gradient could be more aligned with the E-W direction, instead of NE-SW. However, the low S/N prevents us from separating the  clump `\clump' from the rest of the galaxy, so, the direction of the kinematic axis remains uncertain.

We note that the velocities of the companions do not follow the rotation of the central target. The NE part of the central target has blueshifted velocities, while the north companion has redshifted velocities. Similarly, the SW part of the central target shows redshifted velocities while the south companions have bluer velocities.

The W80 map  shows values in the range $140-450$~\kms (corresponding to $\sigma \sim 55-175$~\kms), with lower values in the companions ($140-250$~\kms, see third panel in Fig.~\ref{fig:Ha_map}).
This map shows two regions with enhanced values in the main target (W80 > 400~\kms). 
They lie at the edge of the emission, roughly aligned with the photometric minor axis.
These regions could indicate the presence of an outflow. Another possibility is that these increases in W80 are caused by interaction of the central galaxy with the other components of the system. These possibilities are discussed in details in the next Section~\ref{sec:outflow}.
Enhanced line width (W80 $>300$~\kms) is also seen at the boundary between the central and north galaxies, possibly due to the superposition along the line-of-sight of the emission coming from two or more components with different velocities, or by an enhanced turbulence produced by interactions. 

\subsubsection{High-velocity dispersion regions: Outflow or gravitational interaction}
\label{sec:outflow}

The integrated spectrum of the main target (see Fig.~\ref{fig:int_spectrum}) shows two kinematic components: a narrow and a broad component clearly detected both in the Balmer lines \Ha\ and \Hb, and also in \OIII\ and \NII.
The broad component, which has a W80$\sim560$~\kms (velocity dispersion $\sigma \sim 220$~\kms), could be indicative of an outflow. 
Additionally, the W80 map of the total profile obtained from the spaxel-by-spaxel analysis shows two off-nuclear regions with large line widths (see Fig.~\ref{fig:Ha_map}), again compatible with an outflow.

\begin{figure*}[h!]
\centering 
\includegraphics[width=0.37\textwidth]{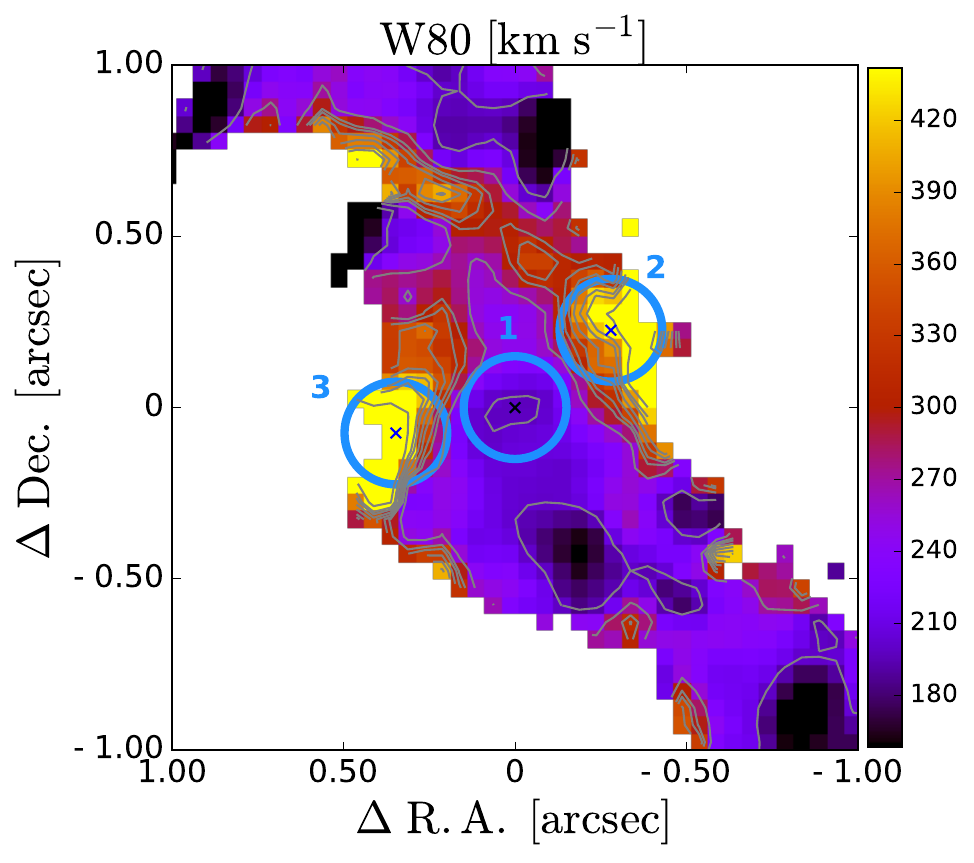}
\includegraphics[width=0.25\textwidth]{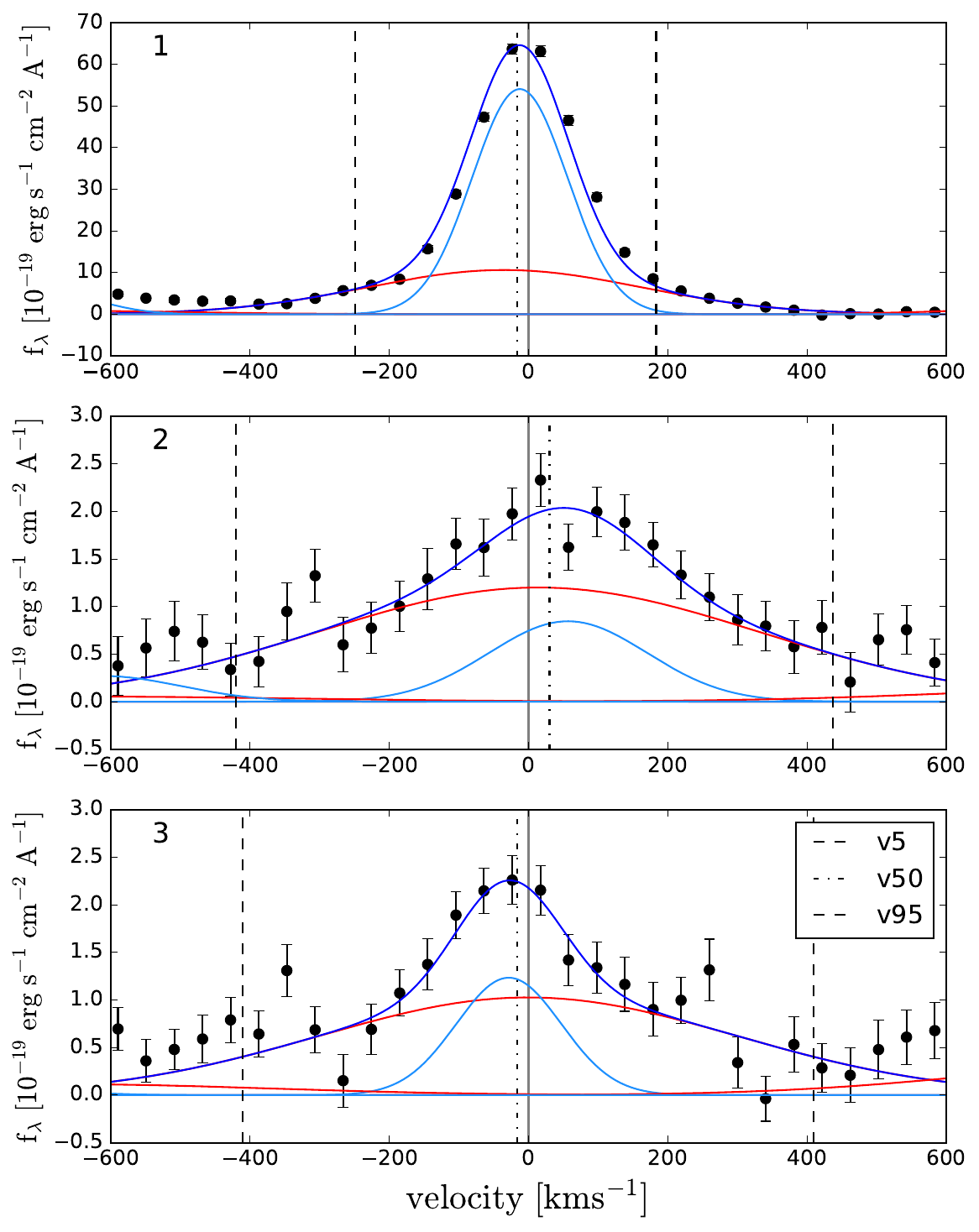}
\includegraphics[width=0.37\textwidth]{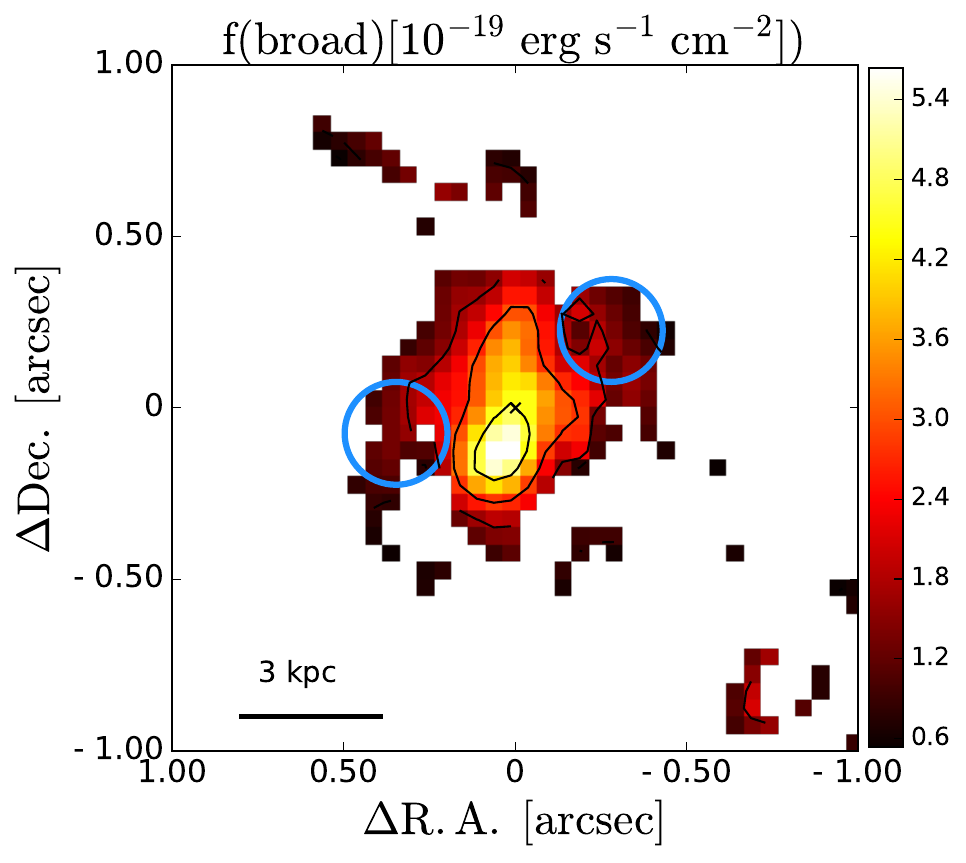}

\caption{Spectra and distribution of the \Ha\ broad component emission. \textit{Left panel}: Map of the line width (W80) of \Ha\ derived from the best-fit model. The three circles indicate the extraction regions of the spectra shown on the middle panel. 
\textit{Middle panel}: Comparison of the spectra extracted from the central region (upper row), and from the two regions with a high velocity dispersion (middle and bottom row). The extraction regions are shown in the left panel and have a radius of 0.15\arcsec.
\textit{Right panel}: Map of the \Ha\ broad component flux. This map is created by adding the flux of all Gaussian component with $\sigma > 140$~\kms. This map has not been corrected for obscuration. Light blue circles mark the positions of the regions 2 and 3.}
\label{fig:Ha_outflow}
\end{figure*}

We further analyse the \Ha\ line in order to characterise this broad component, since it is the line with the highest S/N. 
In Figure~\ref{fig:Ha_outflow} we show the \Ha\ spectra from the two regions with enhanced velocity dispersion ($\sigma> 160$~\kms, or W80 $>350$~\kms) in the E and W parts of the main galaxy,  together with the spectrum of the central region. The three spectra are extracted within circular apertures of radius 0.15$\arcsec$.
We fit these spectra using a two-component model. In the off-nuclear regions, the broad component is very prominent compared to the flux of the narrow component, while in the central region the narrow component dominates,  giving a much smaller W80 of the total profile ($\sim250$~\kms). The broad components have W80 $\sim500-770$~\kms (i.e. a velocity dispersion of $\sigma \sim200-300$~\kms). 
The broad component in these two external regions has a very small velocity shift with respect to the systemic velocity, making the shape of the total profile fairly symmetric.

All these morphological and kinematic characteristics are compatible with a bi-conical outflow. 
The selected off-nuclear regions with highest W80 (Fig.~\ref{fig:Ha_map}) are in the direction of the photometric minor axis, as was expected from star-formation-driven outflows, which typically expand perpendicular to the disc \citep[e.g.][]{Bellocchi2013}.

Another possibility is that the large line widths are caused by turbulence due to the interactions of the main galaxy with the north companion. The high W80 in the two regions at the east and west of the main galaxy could be due to galaxy-galaxy interaction. However, in this scenario we would expect to observe the highest W80 values in the region between the main and north galaxy. 
Moreover, we observe a prominent broad component also in the centre of the main galaxy (see central panel of Fig.~\ref{fig:Ha_outflow}). The right panel of Fig.~\ref{fig:Ha_outflow} shows a map of the flux of the broad \Ha\ emission\footnote{The map of the velocity ($v50$) and line width (W80) associated with the broad \Ha\ emission are reported in Figure~\ref{fig:Ha_maps_2comp} in the appendix.} 
(i.e. all the Gaussian components with $\sigma>140$~\kms, equivalent to W80 $\gtrsim 360$~\kms), illustrating that the broad component flux peaks a bit south of the centre of the main galaxy (defined as the centroid of the continuum emission in the HST/WFC3 F125W filter). This broad component in the centre is  difficult to explain with interactions, and instead could be explained by an outflow originating from a nuclear starburst. Indeed, the peak of the SFR density is located in a similar position as the peak of the broad component flux (see Sec.~\ref{sec:SFR} and Fig.~\ref{fig:SFR_map}, left panel). Thus, we favour the outflow scenario to explain the enhanced line widths in the off-nuclear regions of the main galaxy, even though we acknowledge the possibility that galaxy interactions may also contribute to the broadening of the lines. 
 We discuss the properties and possible origin of the outflow in Sec.~\ref{sec:outflow_discussion}.

\subsection{Properties of the interstellar medium}
\label{sec:ISM_prop}
\subsubsection{Dust attenuation}
\label{sec:obscuration}
We measure the dust attenuation in each spaxel using the Balmer decrement \Ha/\Hb. We use the results of the fit with only one component across the whole data cube, as we are interested in the obscuration of the dominant gas component (i.e. not the outflow).
We assume a theoretical value of \Ha/\Hb = 2.86, estimated for Case B recombination assuming a temperature of $10^4$~K and an electron density {$n_e=10^2-10^3$~cm$^{-3}$ \citep{Osterbrock2006}. 
We adopt the \citet{Cardelli1989} attenuation law.
 We derive the attenuation for the global emission inferred from the one-component fits, as the S/N of the broad component  is not high enough to derive reliable line ratios separately for the broad and narrow components in individual spaxels, nor binning $3\times3$ adjacent spaxels.  We shall discuss the attenuation of the outflow component detected in the integrated spectrum in Sec.~\ref{sec:outflow_discussion}.

Figure~\ref{fig:AV_map} shows the map of the visual attenuation $A_V$.  
We consider only spaxels with S/N(\Ha)$>3$ and S/N(\Hb)$>2$. There are some spaxels where \Ha\ is detected while \Hb\ is below the detection threshold (S/N(\Hb)$< 2$). 
Those spaxels are illustrated with a grey contour in Fig.~\ref{fig:AV_map}. In those regions, the attenuation could be even higher than the maximum value measured in the spaxels with \Hb\ detection  ($A_V=2.7$), but the low S/N in \Hb\ prevents use from measuring $A_V$.
The central galaxy has $A_V$ in the range $0.5-2.7$. The attenuation is enhanced in a region $\sim0.3-0.5''$ to the NW and in a region $\sim0.2-0.4"$ to the SW from the centre.  A region with lower attenuation is found close to the centre and extending towards the SE.
This  low-attenuation region corresponds to an area with high ionisation parameter (traced by  \SIII/\SII, see section \ref{sec:ionization}).

The high attenuation in the region between the main component and the north component corresponds to a dust lane that is visible in the composite NIRCam image shown in Fig.~\ref{fig:HST-NIRSpec_maps}. In Figure~\ref{fig:AV_map}, we show with  contours the location where the ratio of the F444W/F090W NIRCam maps is higher, to highlight the region with redder colour, which we interpret as a dust lane. 

The northern component (particularly the n2 and n3 clumps) and the south clump s3 show the highest attenuation values ($A_V> 2.4$, see Table~\ref{tab:integrated_prop}).  These components were not detected in the HST UV map (see Fig.~\ref{fig:HST-NIRSpec_maps}): the low UV fluxes could indeed be due to the high attenuation.
The clump s1 instead shows the lowest $A_V \sim 0.2-0.7$.

In summary, we observe a rather heterogeneous distribution  of dust over the system with typical sub-kiloparsec variations of  $A_V\sim1$~mag.   The least obscured regions ($A_V <0.5$~mag.) are found at the south, while those  most attenuated  ($A_V > 2$~mag) are in the north component and associated with a dust lane identified  by NIRCam broad band imaging (see Fig.~\ref{fig:AV_map}).
We also find a relatively low attenuated region in the main target, to the south-east of the centre, corresponding to an area with high ionisation parameter (see Sec.~\ref{sec:ionization}).
We use this $A_V$ map to correct for attenuation the fluxes of the measured emission lines.

\begin{figure}[!]
\centering 
\includegraphics[width=0.45\textwidth]{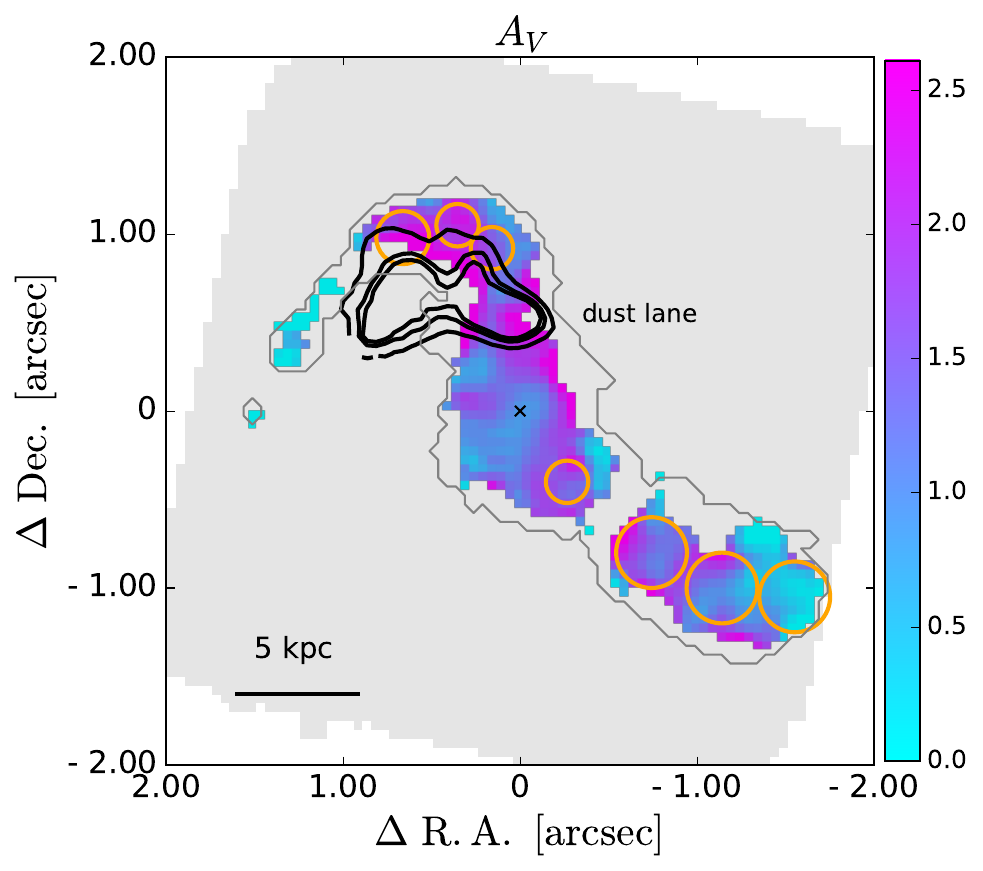}
\caption{Visual attenuation map inferred from the \Ha/\Hb\ line ratio, using the \citet{Cardelli1989} reddening law. We show spaxels with S/N(\Ha)$>3$ and S/N(\Hb)$>2$. Black contours indicates the dust lane identified in the map ratio of the NIRCam F444W/F090W filters. Orange circles indicate the position of the components identified in Fig.~\ref{fig:HST-NIRSpec_maps}. The grey contour indicate the region where \Ha\ is detected with S/N>3.}
\label{fig:AV_map}
\end{figure}

\subsubsection{Emission line diagnostics of gas excitation mechanisms}
\label{sec:BPT}
In this subsection, we investigate the source of ionisation in a spatially resolved way.
In Figure~\ref{fig:BPT_map}, we present the maps of the line ratios \NII/\Ha\ and \OIII/\Hb, together with the emission line diagnostic diagram \citep[BPT;][]{Baldwin1981}.
We show here all the spaxels with S/N>3 in the \Ha, \Hb, \NII\ and \OIII\ lines.

The line ratios in our target are all below the separation line between star formation and AGN found by \citet{Kewley2001} for local sources, thus, they are consistent with ionisation from HII regions. The line ratios are higher than the typical line ratios of star-forming galaxies in the local Universe (see SDSS contours), but they follow quite well the locus of star-forming galaxies at $z=2-3$ from the KBSS-MOSFIRE sample presented by \citet{Strom2017}. 
Recently, some high-redshift AGN discovered by JWST  have been reported to  be on the same location of the KBSS-MOSFIRE sample \citep{Uebler2023, Maiolino2023arXiv, Scholtz2023arXiv}. However, our target is at a lower redshift and more metal rich (see Sec.~\ref{sec:metallicity}) than the AGN explored in those JWST studies, thus the BPT classification should still be reliable.
The scatter in our points is partly due to uncertainties (the average uncertainties are of the order of 0.1 dex).
The  south and north companions, shown with light blue and red dots, respectively, have lower  \NII/\Ha\ and higher \OIII/\Hb\ values compared with the main target (orange dots). This is consistent with these targets having lower metallicity (see Sec.~\ref{sec:metallicity}).

The north part of the main target shows the highest values of \NII/\Ha\ and lowest \OIII/\Hb\ values.
We investigate two possible mechanisms that could explain these line ratios.
One possibility is that the \NII/\Ha\ ratio is enhanced by shocks. 
If the line ratios are driven by shocks, they are expected to correlate with the kinematics of the gas \citep[e.g][]{Monreal-Ibero2006, Arribas2014, Ho2014, McElroy2015, Perna2017b,Perna2020,  Mingozzi2019, Johnston2023}. The velocity dispersion in that region is indeed higher than in the south part of the main target (see W80 map in Fig.~\ref{fig:Ha_map}).
In Fig.~\ref{fig:shock_map} we show the line ratio \NII/\Ha\ versus the velocity dispersion. 
 The points in Fig.~\ref{fig:shock_map}  are colour-coded according to a combination of the two quantities on the x and y axes, using the expression from Eq.~1 in \citet{DAgostino2019a}:
\begin{equation}
Q = \frac{X-min(X)}{max(X)-min(X)}\times \frac{Y-min(Y)}{max(Y)-min(Y)},
\label{eq:ratio_sigma}
\end{equation}
where $X= \log$ \NII/\Ha\ and $Y$ is the velocity dispersion $\sigma$.
The upper panel of Fig.~\ref{fig:shock_map} shows the map of the system colour-coded by $Q$,
 illustrating that the highest values of velocity dispersion and \NII/\Ha\ are found in the northern part of the main galaxy.
There is an overall correlation between the two quantities (Spearman's rank correlation coefficient $r=0.71$,  $p$-value $<0.01$). The slope of the correlation is flatter at low values of $\log$ \NII/\Ha < 0.6, and then it becomes steeper and more scattered at higher \NII/\Ha\ values. This could indicate that the shocks are affecting the line ratios in regions with a high velocity dispersion ($\sigma > 100$~\kms), while at lower $\sigma$ photo-ionisation is probably the main mechanism driving the changes in the line ratio.

In Fig.~\ref{fig:shock_map}, we overlaid a set of shock models calculated with the code MAPPING V \citep{Sutherland2017, Sutherland2018}. We use the 3MdBs database \citep{Alarie2019} to download the grid of models presented in \citet{Allen2008}. We consider models with solar metallicity, magnetic fields values in the range 1-10~$\mu$G, and pre-shock densities in the range of $1-10$~cm$^{-3}$ (which correspond to \SII-based post-shock electron densities in the range $50-4000$ cm$^{-3}$). The models assume shock velocities >100~\kms. In the velocity range covered by both the observations and the models (100-160~\kms), the shock models match the observed \NII/\Ha\ line ratios.

Another possibility is that the enhanced \NII/\Ha\ and diminished \OIII/\Hb\ line ratios is due to higher metallicity in that region. We shall discuss this possibility further in the next section.
 In summary, the spatially resolved \OIII/\Hb\ and \NII/\Ha\ line ratios are consistent with emission due to star formation, with no sign of AGN ionisation.
We find a region with enhanced \NII/\Ha\ and low \OIII/\Hb, located in the northern part of the main target. This peculiar line ratios could be due to shocks, or to a higher metallicity. 

\begin{figure*}[h!]
\centering 
\includegraphics[width=0.999\textwidth]{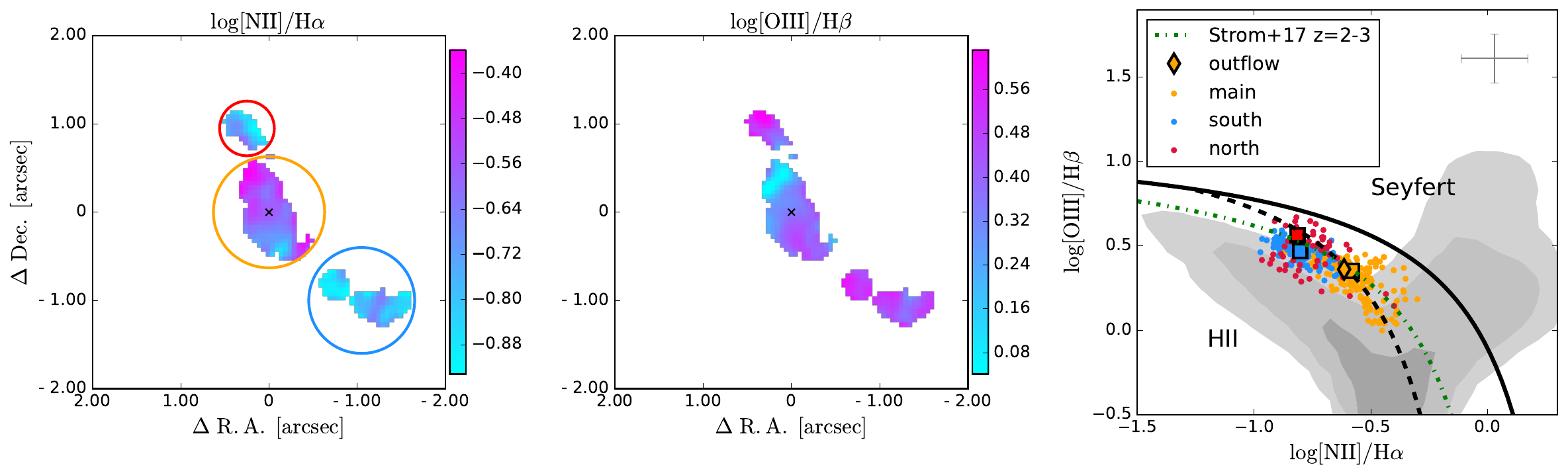}

\caption{Spatially resolved maps of the emission line ratios \NII/\Ha\ and \OIII/\Hb. The circles show the colour-coding for the spaxels in the BPT diagram in the right panel. \textit{Right:} Emission line diagnostic diagram for the single spaxels.  Spaxels belonging to the north, main, and south regions are indicated in red, orange, and light blue, respectively.   A representative (median) error bar is shown in the top right. The line ratios derived from the integrated spectra of the three regions are shown with squares. The orange diamond shows the line ratios of the outflow, derived from the broad component of the integrated spectrum of the main target. In black are shown the demarcation lines from \citet{Kewley2001} (solid) and \citet{Kauffmann2003} (dashed). The dashed green line show the locus of star-forming galaxies at $z=2-3$ from \citet{Strom2017}. The location of galaxies from the SDSS sample \citep{Abazajian2009} is illustrated with the grey contours showing the 50th, 90th, and  98th percentiles  of the sample.}
\label{fig:BPT_map}
\end{figure*}


\begin{figure}[h!]
\centering 
\includegraphics[width=0.49\textwidth]{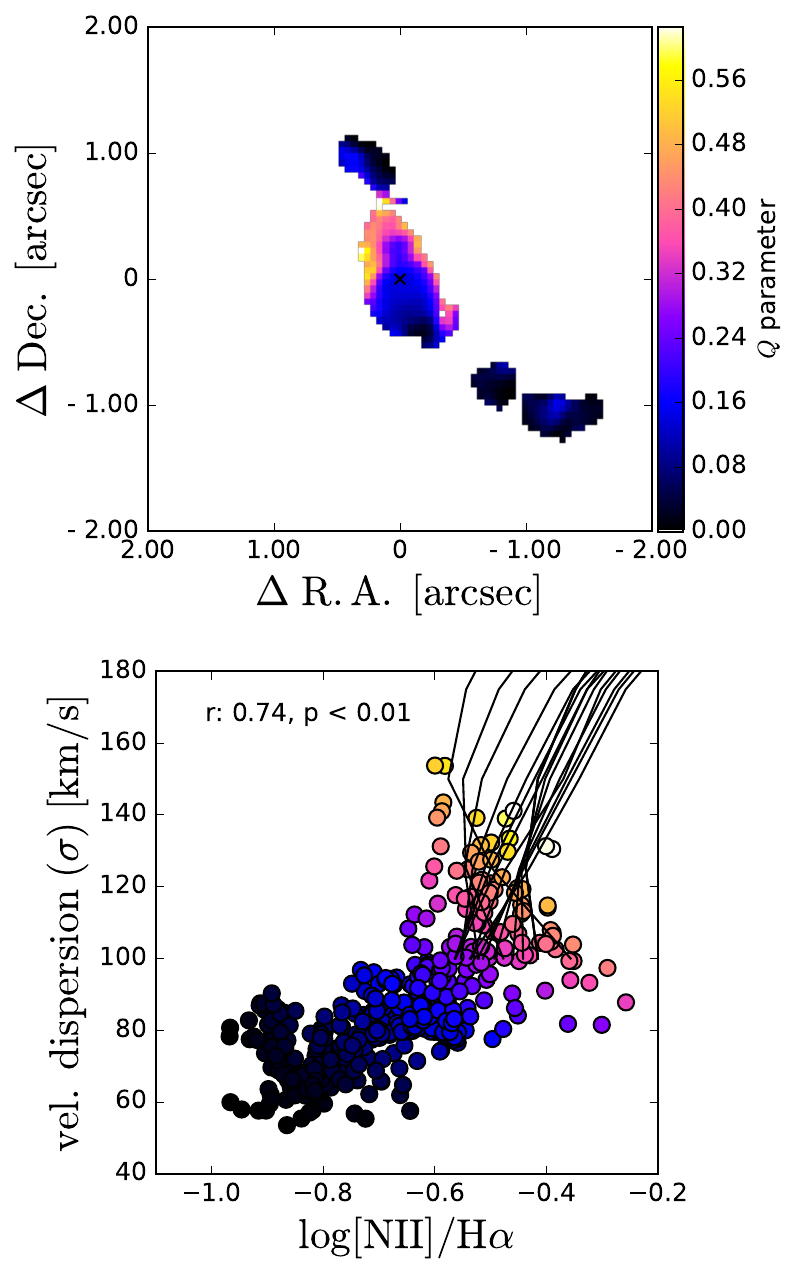}
\caption{Comparison of mission line ratio \NII/\Ha\ and velocity dispersion. \textit{Lower panel:} Emission line ratio \NII/\Ha\  versus the  velocity dispersion for individual spaxels.  Points are colour-coded based on the function $Q$ described in Equation~\ref{eq:ratio_sigma}, which combines \NII/\Ha\ and velocity dispersion. Black curves show shock models derived with MAPPINGS V (see text). \textit{Upper panel:} Map colour-coded as in the lower panel,  showing the region with high  \NII/\Ha\  and high velocity dispersion.}
\label{fig:shock_map}
\end{figure}

\subsubsection{Gas-phase metallicity}
\label{sec:metallicity}
In this section, we study the gas-phase metallicity in our system. 
We calculate the gas-phase metallicity following the prescriptions from \citet{Curti2020, Curti2024}. We prefer to use these prescriptions even if they are calibrated for local galaxies, because prescriptions for higher redshift galaxies do not cover well the high metallicity regime  \citep[12+log(O/H)> 8.4, e.g.][]{Sanders2024}. 
We consider the metallicity indicators based on the line ratios  R3=\OIII$\lambda$5007/\Hb,  O3O2=\OIII$\lambda$5007/\OII$\lambda$3727,3729,  and R23= (\OII$\lambda$3727,3729+ \OIII$\lambda$4959,5007)/\Hb. We decide to use this set of line ratios, because they are appropriate for the metallicity regime of our target. 
We did not include the N2 = \NII/\Ha\ line ratio because, as we have seen in Sec.~\ref{sec:BPT}, it may be affected by shocks. 
We fit simultaneously the R3, O3O2 and R23 parameters to reduce the uncertainties, following the procedure described in \citet{Curti2024}.

From the integrated spectrum of the main target we measure  $\text{12+log(O/H)}=8.45\pm0.04$. For the stellar mass of GS5001 - that is, $\log (M_{\star}/M_{\odot}) =10.0-10.6$ (see Sec.~\ref{sec:intro}) - 
the mass-metallicity relation derived by \citet{Sanders2021} for galaxies at $z=3.3$ predicts a  metallicity 12+log(O/H)$ = 8.42-8.49$\footnote{We re-scale the values by --0.1~dex, to account for the fact that \citet{Sanders2021} use a different metallicity calibration from \citet{Bian2018} \citep[see also][]{Curti2024}.}, consistent with our measurement.
The north and south companions show slightly lower metallicities (12+log(O/H)$\sim8.34-8.44$) compared to the central target. This is expected considering that the companions have a lower stellar mass than the central galaxy.
We do not have estimates of the stellar masses for the individual companions; however, for the south source (which includes s1, s2, and s3) there is an estimate from the literature ($\log (M_{\star}/M_{\odot}) = 9.7-10.4$, see Sec.~\ref{sec:intro}). 
For this stellar mass, the mass-metallicity relation by \citet{Sanders2021} would predict 12+log(O/H)$\sim8.4-8.6$, in agreement within the uncertainties with the measured values.

Figure~\ref{fig:metallicity_map} shows a resolved map of the metallicity in the system.
The main galaxy shows a metallicity gradient, with lower metallicity in the south (12+log(O/H)=8.4) and a higher metallicity in the NE region, reaching values of 12+log(O/H)=8.58.  The lower metallicity in the south could be due to an inflow of lower metallicity gas. 
We note that  in this position we also identify a small clump (\clump) that is merging with the main galaxy. This clump could be part of an inflowing stream of gas, or be a lower metallicity companion.  We note that the region with low-metallicity observed in the south part of the main galaxy is more extended than the clump `\clump'.
Similar metallicity gradients have been recently reported in galaxies at $z=3.6-7.9$ \citep{Arribas2024, RodriguezDelPino2024, Venturi2024arxiv}, and were explained by possible accretion of low metallicity gas and merger event with a lower-metallicity satellite.

Another possibility is that the north part has been enriched in metals,
 due to a past episode of enhanced SF activity in the interacting region between the `main' galaxy and the northern component.  We note however that the possible presence of shocks in this region (see Fig.~\ref{fig:shock_map}) may also increase the line ratios mimicking a further rise in metallicity.  Then, the derived high metallicities may be a consequence of both enhanced SF and the presence of shocks. 

We note that the interacting region shows also a higher level of obscuration with respect to the rest of the galaxy, and that it may be coincident with a dust lane (see Fig.~\ref{fig:AV_map}).
Previous works have found a correlation between the obscuration in the UV (traced by the UV slope) and the  gas-phase metallicity \citep{Heckman1998, Reddy2010}. 
Their interpretation is that the extinction of the UV emission increases as the dust-to-gas ratio increases with gas-phase metallicity. Similarly, the higher $A_V$ and higher metallicity in the interacting region of our system could be related to a higher dust-to-gas ratio.  

In summary, we find that the main galaxy has an average metallicity of 12+log(O/H)$=8.45$ and the companions have slightly lower metallicities ($8.34-8.44$), consistent with the mass-metallicity relation at $z\sim3$. The main target shows a metallicity gradient with lower values in the SW region and higher in the NE, the region close to the north target.
This metallicity gradient could be explained by an inflow of low-metallicity gas or accretion of a lower metallicity companion  from the south, or by an increase in the metallicity in the north due to a past star formation episode triggered by the interaction with the north component.

\begin{figure}
\centering 
\includegraphics[width=0.48\textwidth]{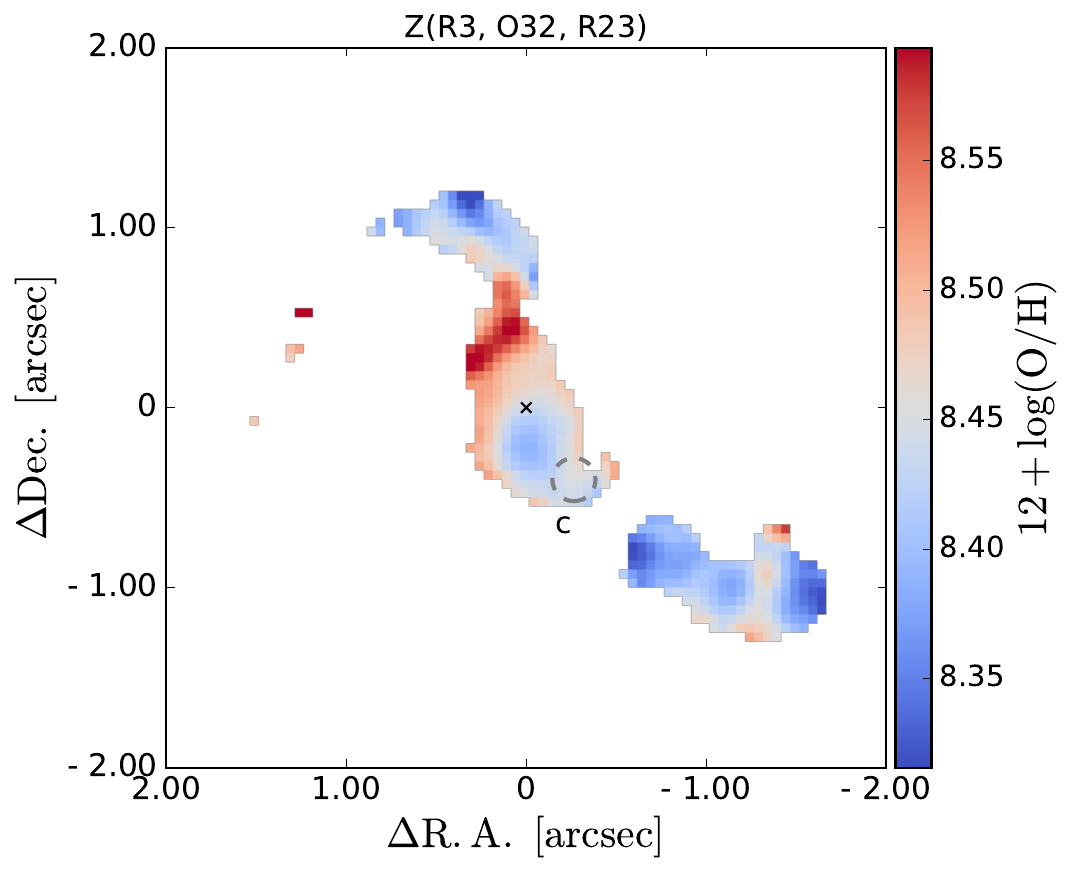}
\caption{Gas-phase metallicity map calculated using  three metallicity calibrations R3, O32, and R23, following the prescriptions from \citet{Curti2020}. The dashed circle shows the position of the clump `\clump'.}
\label{fig:metallicity_map}
\end{figure}

\subsubsection{Star formation rates}
\label{sec:SFR}
In Section~\ref{sec:BPT}, we showed that the ionisation in our system is dominated by star formation, thus, we can use the \Ha\ luminosity to trace the SFR.
We first correct the \Ha\ fluxes for obscuration using the Balmer decrement (see Sec.~\ref{sec:obscuration}).
Then, we convert the \Ha\ luminosity into SFR using the relation from \citet{Kennicutt2012} \citep[see also][]{Murphy2011, Hao2011}, assuming a \citet{Chabrier2003} IMF.

In Table~\ref{tab:integrated_prop} we report the SFRs derived from the integrated spectra of the different components.
For the main target, the SFR derived from the narrow component flux is 100~\Msunyr.
Considering the star-formation main-sequence definition at $z=3.5$ from \citet{Schreiber2015} and a stellar mass $\log M_{\star}/M_{\odot}=10-10.6$ (see Sec.~\ref{sec:intro}), GS5001 would be 0.1-0.2~dex above the main-sequence.
The south clumps have SFRs between 12-30~\Msunyr, for a combined SFR of $\sim76$~\Msunyr, while the north component has a total SFR of 106~\Msun.
Thus, the north and south companions have a total SFR comparable to the main target, and the total SFR of the system is $\sim300$~\Msunyr.  
Previous estimates of the SFR, derived from SED fitting,  were in the range 150-240~\Msunyr\ for the central galaxy and in the range 60-110~\Msunyr\ for the south component, consistent with our measurements within a factor of two (see Sec.~\ref{sec:intro}).

Figure~\ref{fig:SFR_map} shows the maps of the SFR surface density ($\Sigma_{SFR}$) derived from \Ha, together with the F090W NIRCam image, tracing the UV flux at $~\sim2000$~\AA\  rest-frame.
In the F090W image, we can identify three peaks in the central target, aligned in the N-S direction. 
The region with the highest $\Sigma_{SFR}$ is also aligned in the same direction.
Overall, the $\Sigma_{SFR}$ is very high in the central target, where it reaches values of $\sim10$~\Msunyr kpc$^{-2}$.
Also in the companions the values are relatively high, in the range $0.3-6$~\Msunyr kpc$^{-2}$.

SFR densities for galaxies at redshift $z>3$ have been reported in the literature, mainly from integrated measurements \citep[e.g.][]{Reddy2023b, Reddy2023c, Morishita2024}. 
Thanks to NIRSpec IFS, it is now possible to derive spatially resolved maps. 
\citet{RodriguezDelPino2024} reported similar high values ($\Sigma_{SFR}\sim8$~\Msunyr kpc$^{-2}$) in the central region of GS4891, a star-forming galaxy at z=3.7, also  part of the GA-NIFS sample. 
 \citet{Morishita2024} finds similarly high values ($\Sigma_{SFR}>10$~\Msunyr kpc$^{-2}$) in a sample of compact sources at $z>5$.

\begin{figure*}[h!]
\centering 
\includegraphics[width=0.9\textwidth]{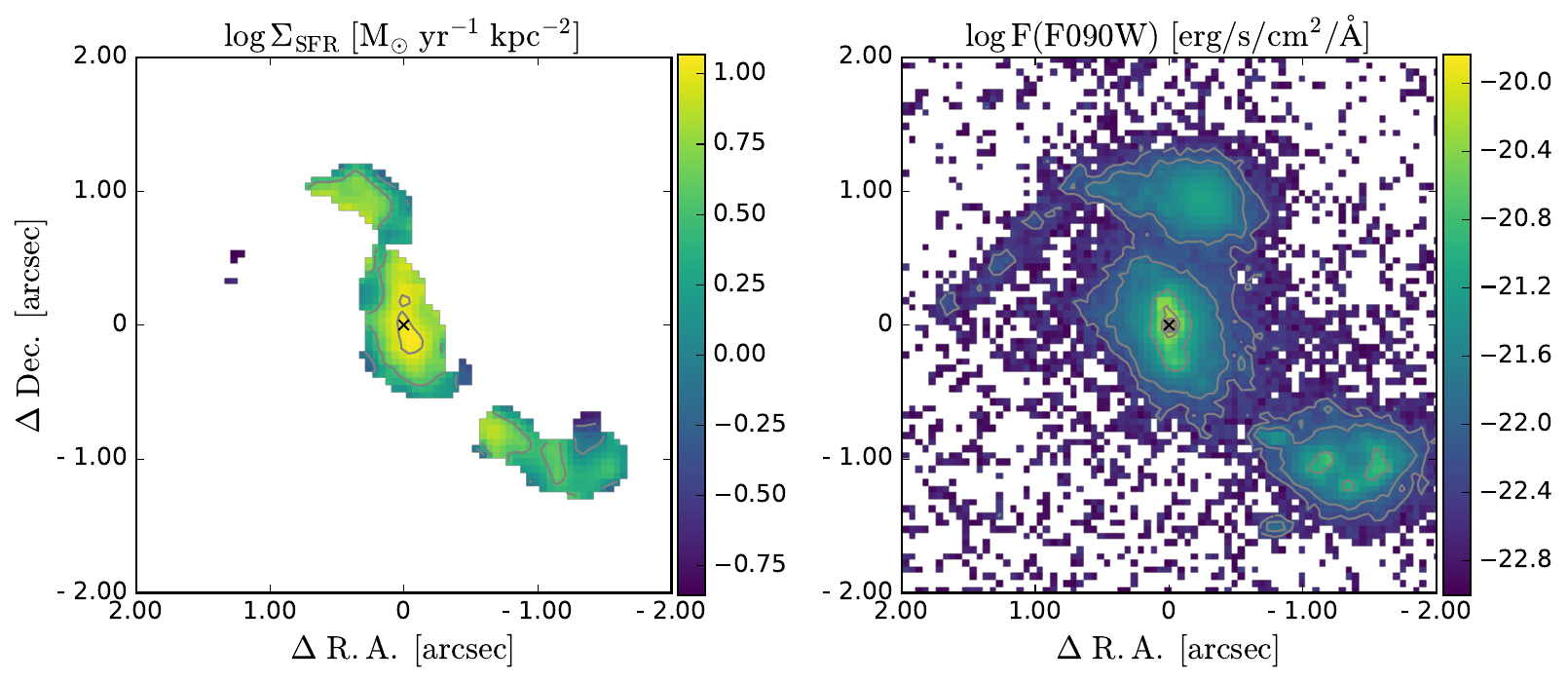}
\caption{Maps of star formation rate tracers. \textit{Left:} Map of the SFR surface density, inferred from the dust-corrected \Ha\ luminosity. \textit{Right:} Flux in the NIRCam filter F090W.}
\label{fig:SFR_map}
\end{figure*}

\subsubsection{Ionisation parameter}
\label{sec:ionization}
We calculated the ionisation parameter, U - in other words, the ratio of (hydrogen-)ionising photons over the gas density - using the prescriptions from  \citet{Diaz1991, Diaz1999}. 
 We consider the calibrations involving the flux line ratio S2S3=(\SII$\lambda$6717+\SII$\lambda$6731)/(\SIII$\lambda$9069+\SIII$\lambda$9531),  measured from the R100 data cube.
 The S2S3 line ratio is less dependent on metallicity than O3O2, thus it is considered to be a more accurate estimator of the ionisation parameter \citep{Mathis1982, Dopita1986, Kewley2002, Kewley2019, Mingozzi2020}. 
 
 The values of the ionisation parameter derived from the integrated spectra of the different components of the system are reported in Table~\ref{tab:integrated_prop}.
 The values of the ionisation parameter span a range from log U = [-3.1, -2.7].
 These values are within the range measured by \citet{Reddy2023b} using the O3O2 ratio in a sample of galaxies at $z=1.9-3.7$ (log U = [-3.25, -2.05]). 

Figure~\ref{fig:ionization_map} shows the map of the ionisation parameter obtained using the S2S3 calibrations. We show only spaxels with S/N>3 in the sum of the \SIII\ and \SII\ lines (i.e. S/N(\SIII$\lambda$9069)+S/N(\SIII$\lambda9531)>3$ and  S/N(\SII$\lambda$6717)+S/N(\SII$\lambda6731)> 3$) and with S/N(\Hb$)>2$, to ensure that we can correct the fluxes for obscuration using the Balmer decrement.
The values of the ionisation parameter span a range from log U = --3.8 to --2.3.
In the main target, the ionisation parameter is enhanced (by $\sim0.4$~dex) with respect to the rest of the galaxy in a region south of the continuum peak pointing toward the south-east. We note that this region has lower obscuration ($A_V$) than the rest of the main galaxy (see Fig.~\ref{fig:AV_map}).
The clump s3 also shows a similar enhancement of the ionisation parameter.
We note that these regions corresponds to the regions with brighter \Ha\ flux, and therefore higher SFR (see Fig.~\ref{fig:SFR_map}). 
The north companion and the clumps s1 and s2 have low ionisation parameter (log U $<-3$) similar to the values in the north part of the main target.

We calculated the ionisation parameter maps also using the O3O2, S2\Hb\ and O2\Hb\ calibrations from \citet{Diaz2000}. These maps show similar spatial variations, even though the normalisation is different. 

In our target, we observe that the regions with high ionisation parameter (region south of the continuum peak, and clump s3) correspond to regions with elevated SFR surface density (see Fig.~\ref{fig:SFR_map}).
 A correlation between ionisation parameter and SFR surface density has been observed in previous spatially resolved  studies of nearby galaxies \citep[][]{Dopita2014, Kaplan2016}, but also within a galaxy sample at $z=1.9-3.7$  \citep{Reddy2023b}. This correlation suggests that SFR plays an important role in regulating the ionisation parameter \citep{Reddy2023b}.

\begin{figure}[t!]
\centering 
\includegraphics[width=0.48\textwidth]{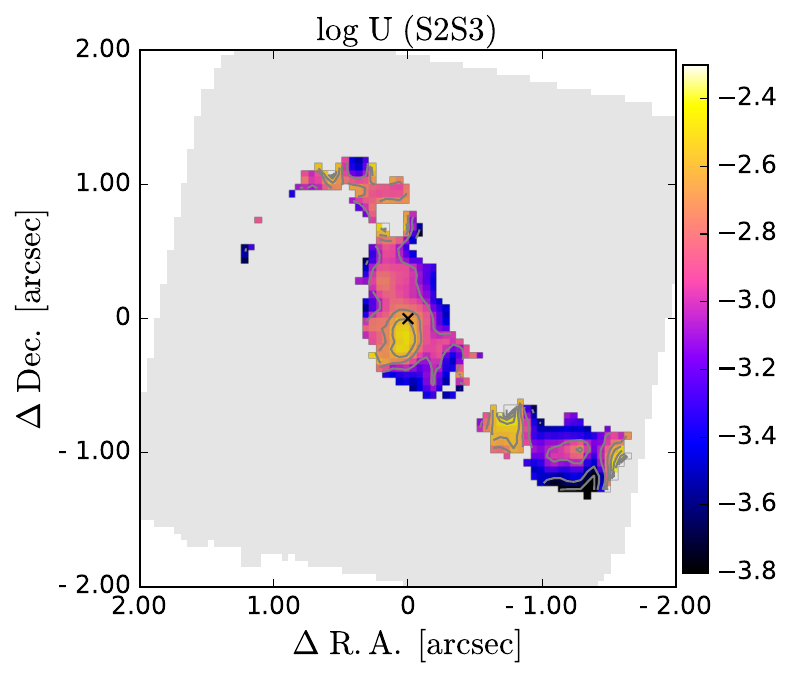}
\caption{Maps of the ionisation parameter derived from the S2S3 line ratio. We show only spaxels with S/N>2 in \Hb, and S/N>3 in the sum of the \SIII\ and \SII\ lines.}
\label{fig:ionization_map}
\end{figure}

\subsubsection{Electron density}
\label{sec:electron_density}

The electron density can be derived using the ratios of the \SII $\lambda \lambda$6716,31  or the \OII$\lambda\lambda$3726,29 doublets \citep{Osterbrock2006}.
In our case, the S/N is not enough to study these line ratios spaxel-by-spaxel, thus, we focus on the integrated spectra.
We note that the two \OII\ lines are not fully spectroscopically resolved in our data. On the other hand, the \SII\ lines are well resolved, but have lower S/N. 
In order to better constrain the electron density, we fit simultaneously the \SII\ and \OII\ lines, forcing the line ratios to agree on the same $n_e$ parameter, following the same approach used by \citet{RodriguezDelPino2024}.
We use the formulas presented in \citet{Sanders2016a} that relate the \SII\ and \OII\ line ratios  to the electron density. We do not attempt to model the emission lines with two components (broad and narrow), because even in the integrated spectrum of the main galaxy the S/N of the broad component is not high enough to derive meaningful constraints on the electron density (see Fig.~\ref{fig:SII_OII_2comp_fit} in the appendix). 
The results of the fit of the \OII\ and \SII\ doublets of the main target are shown in Fig.~\ref{fig:SII_OII_fit}.

For the main target, we find $n_e =540_{-100}^{+120}$~cm$^{-3}$.
The independent modelling of the \SII\ line ratio would give $n_e$(\SII)$= 73_{-50}^{+70}$ cm$^{-3}$, while the \OII\ line ratio gives  $n_e$(\OII)$ = 1480 _{-320}^{+400}$ cm$^{-3}$. These estimates have large uncertainties due to the issues mentioned above.
 We note that the electron densities derived from \OII\ are systematically higher than the ones derived from \SII\ for all the sources in our system. 
\citet{Kewley2019} highlight that the \SII\ and \OII\ lines are produced in different regions of ionised nebulae: the \SII\ lines are produced in the extended partially ionised region of the nebula, while the \OII\ lines are produced closer to the ionising source. If the density is higher close to the ionising source than further out, this could explain the different $n_e$ estimates from \SII\ and \OII.

We note that the fit of the main galaxy shows significant residuals ($>3\sigma$) on the blue side of the \OII\ doublet (see left panel of Fig.~\ref{fig:SII_OII_fit}). This is probably due to the outflow component that we are not considering in the fit. 
 An attempt to fit simultaneously the \SII\ and \OII\ lines with two components (narrow+broad) gives similar results for the electron density of the narrow component ($n_e=540_{-280}^{+270}$~cm$^{-3}$), while the electron density of the broad component is unconstrained ($n_e=340_{-280}^{+1400}$~cm$^{-3}$, see Fig.~F.2. 

For the companions, we find electron densities in the range 200-600~cm$^{-3}$ from the simultaneous fit of \OII\ and \SII. The results of the fit of the \OII\ and \SII\ doublets of the companions are shown in Fig.~\ref{fig:SII_OII_fit_appendix} 
in the appendix, while the values of the electron densities are reported in Table~\ref{tab:integrated_prop}.

Our results are consistent with the electron densities of star-forming galaxies at redshifts of $z\sim1-2.5$ derived from the \OII\ or \SII\ ratios from the literature \citep[e.g.][]{Masters2014, Steidel2014,Shimakawa2015, Sanders2016a, Kaasinen2017, Kashino2017}.
Recently, \citet{RodriguezDelPino2024} measured an electron density of $n_e = 776\pm307$~cm$^{-3}$ in a star-forming galaxy at a similar redshift of $z=3.7$ (GS4891), using NIRSpec data, in agreement with our measurement.

\citet{Reddy2023c}  study a sample of $\sim50$ galaxies at $z=2.7-6.3$, and report average electron densities of $\sim100-500$~cm$^{-3}$. 
Interestingly, they find higher electron densities ($n_e\sim500$~cm$^{-3}$) in galaxies with higher SFR surface density \citep[$\Sigma_{SFR}> 1$~\Msunyr kpc$^{-2}$, see also][]{Shimakawa2015, Reddy2023b}, similar to the $\Sigma_{SFR}$ values observed in GS5001 (see Fig.~\ref{fig:SFR_map}). 
Recently, \citet{Isobe2023} measure $n_e \gtrsim 300$~cm$^{-3}$ in a sample of 14 galaxies at $z=4-9$ \citep[see also][]{Marconcini2024}. They find that the $n_e$ are higher than those of lower-redshift galaxies with similar values of stellar mass, SFR or specific SFR.
GS5001 has a SFR comparable to that of local ULIRGs (SFR > 150 \Msunyr), which have an average electron density $\sim300$~cm$^{-3}$ \citep{Arribas2014}, a bit lower than our measurement. This  suggests that, apart from an increase in SFR with redshift, the gas conditions at high redshift also lead to higher electron densities (but see also \citealt{Kaasinen2017}).

In summary, we take advantage of the large wavelength range covered by NIRSpec to measured the electron densities by modelling simultaneously the \SII\ and \OII\ doublets. 
We measure an electron density of $540^{+120}_{-100}$~cm$^{-3}$ in the main targets, and $n_e=200-600$~cm$^{-3}$ in the companions. We note that using only the \SII\ or \OII\ lines  would provide different results, with larger uncertainties.

\begin{figure}[t!]
\centering 
\includegraphics[width=0.47\textwidth]{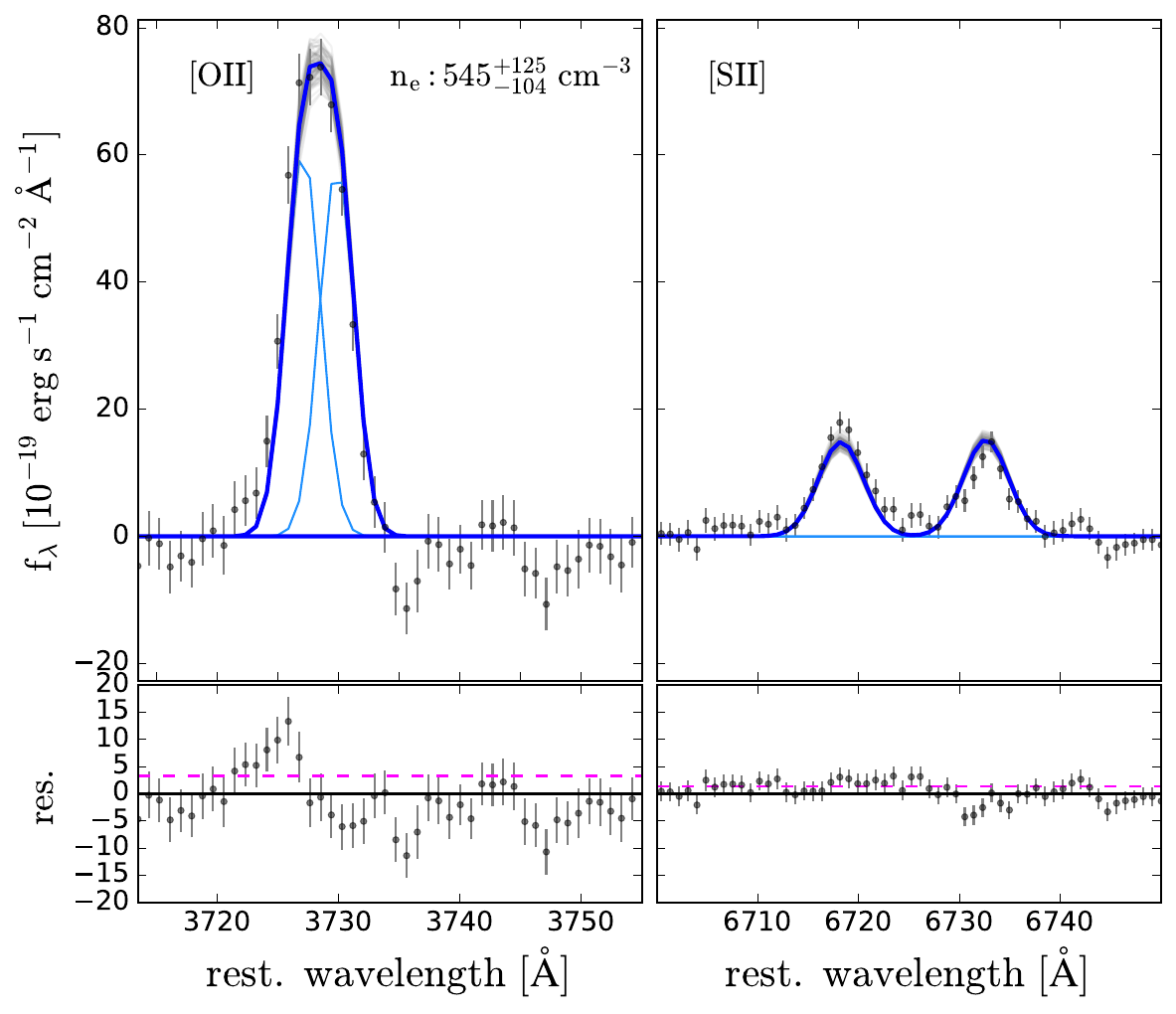}
\caption{Simultaneous spectral fit of the \OII$\lambda\lambda3726,29$ and \SII$\lambda \lambda6716,31$ doublets  to derive the electron density from the integrated spectrum of the main target. The blue curve shows the total best-fit model, the light blue curves show the best-fit Gaussians for the individual emission lines, and the grey curves the uncertainties of the MCMC fit. The bottom panels show the residuals, the dashed magenta line indicates the average 1$\sigma$ noise level.}
\label{fig:SII_OII_fit}
\end{figure}

\section{Discussion}
\label{sec:discussion}

\subsection{Dynamical and physical status of the system: \\ A pre-coalescence merger}
\label{disc:dyn}
Our NIRSpec observations confirm the interesting nature of this system and allow  the spatially resolved study of the central galaxy and its close companions ($< 15$~kpc).
From the ionised gas (Fig.~\ref{fig:Ha_map}) and continuum emission maps (Fig.~\ref{fig:cont_maps}), we identified two additional companions other than the central galaxy. The south component is blueshifted by $\sim130$~\kms\ with respect to the central galaxy, and shows three sub-structures identified as peaks in the \OIII\ flux and continuum emission maps. 
 The north companion is redshifted by $\sim 200$~\kms\ with respect to the main galaxy, and its ionised gas emission shows three peaks. We also detect continuum emission associated with this source. The presence of continuum emission in both the south and north companions indicates that these are not just gas clumps but are likely galaxies with a significant stellar component. Indeed, the estimated stellar mass for the south companion from the literature is in the range $M_{\star}= 0.5-2.3\times 10^{10}$~\Msun\ \citep[][]{Maiolino2008, Perez-Gonzalez2005, Guo2013}. 

We observe signs of interactions between the northern companion and the central galaxy: in this region, the ionised gas  is more turbulent and the ISM shows peculiar properties (higher \NII/\Ha, higher metallicity, and higher obscuration than elsewhere). Additionally, the north companion has a tidal tail extending for 10~kpc, which can be interpreted as a sign of interactions.
Given the low relative velocities of the companions ($< 200$~\kms) and projected distance ($< 10$~kpc), we can expect that they will coalesce with the central galaxy in the future. We note however that we can only measure projected velocities, thus, the `true' relative velocities are uncertain.

Moreover, we identify one clump (`\clump'), which is probably in the process of merging with the main galaxy.
This clump could be situated between the main galaxy and us. Since its emission is redshifted with respect to the main target by $\sim70$~\kms, in this scenario the clump would be moving towards the main galaxy and may be in the process of merging.
Given that its \Ha\ luminosity is only $3\%$ of the one of the main galaxy, we can consider it a minor merger event.
From the gas-phase metallicity map shown in Fig.~\ref{fig:metallicity_map}, we see that at the position of this clump the metallicity is lower compared to the values measured in the north-east part of the main galaxy. Thus,  this clump could be part of a low-metallicity inflow, or be a low-metallicity clump falling onto the main target.
As an alternative scenario, this clump could be located behind the main galaxy, moving away from us. In this case, it could be interpreted as a low-velocity clump associated with ejected (outflowing) gas, or, more in general, a debris formed by galaxy interactions.

The north and south companions could be part of a large scale gas filament that is feeding the central galaxy. \citet{Ginolfi2017} identify several CO emitters oriented along the NE-SW within a radius of $\sim250$~kpc. They interpret them as tracers of a cold gas stream feeding the central galaxy. The north and south companions,  which are broadly oriented along the same direction (see Fig.~\ref{fig:ALMA_map}), could have formed in this gas filament and be in the process of moving towards the central galaxy and merging with it. 
This scenario would explain the peculiar ISM properties in the region between the main and north component, and the presence of the extended tidal tail towards the north-east and the line emitter `\clump'. 
Hence, this system may be comparable to the Spiderweb Galaxy at $z\sim2.2$,  where a cloud of molecular gas extending for tens of kiloparsecs seems to be feeding the innermost (merging) galaxies in the proto-cluster \citep{Emonts2016}.

Recently, \citet{Jin2023} identified an overdensity consisting of a group of six galaxies within a FoV of $10\times20$~kpc$^2$ at a redshift of $z\sim5.2$, CGG-z5,  using NIRCam observations.
Five companions with $\log (M_\star/M_{\odot}) \sim8.4-9.2$ are aligned along two directions around a central more massive galaxy ($\log (M_\star/M_{\odot}) \sim9.8$).
The geometry looks remarkably similar to our system. \citet{Jin2023} look for similar compact structures  in the  EAGLE cosmological simulations \citep{Crain2015, Schaye2015}, and follow their evolution with cosmic time. They find that the identified structures will merge into a single galaxy by $z\sim3$.
Therefore, GS5001 could be a similar system, in which all the currently identified structures will merge at a later cosmic epoch.

\begin{figure}[!]
\centering 
\includegraphics[width=0.48\textwidth]{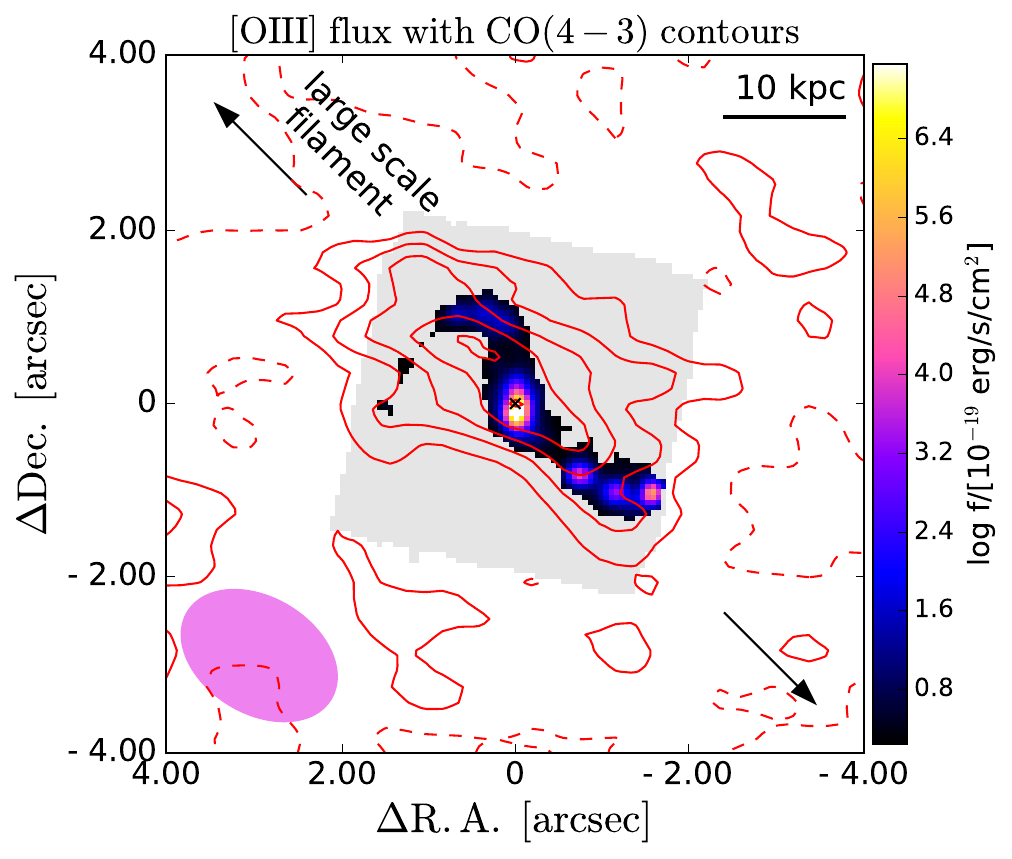}
\caption{Map of the  \OIII\ integrated flux  with overlaid contours of the  CO(4-3) emission from ALMA. Solid contours are at levels $[1,2,3,4,5]$ $\times$ rms,  and dashed contours show the $ -1\times$ rms level. The magenta oval shows the  $1.9''\times 1.3''$ ALMA beam. Arrows indicate the direction (NW-SE)  of the large scale filament traced by CO emitters discovered by \citet{Ginolfi2017}.}
\label{fig:ALMA_map}
\end{figure}

\subsection{The kiloparsec-scale outflow in GS5001: properties, origin and effects}
\label{sec:outflow_discussion}
In section~\ref{sec:outflow}, we present the regions with high velocity dispersion in the main galaxy and we discuss whether this enhancement is due to an outflow or galaxy-galaxy interaction.  Although both scenarios are plausible, we favour the outflow explanation based on geometric considerations, while acknowledging that we cannot entirely exclude the interaction scenario.
In this section, we describe the properties of gas in the ionised outflow, and we estimate the outflow energetics.
 As the limited signal-to-noise prevents us from studying the physical properties of the outflow spaxel-by-spaxel, we consider the integrated spectrum of the main target, and use the two-component fit to separate the flux due to the outflow (see Fig.~\ref{fig:int_spectrum}).
 
\subsubsection{Outflow properties}
\label{sec:outflow_prop}
The spatially integrated properties of the broad outflow component are reported in Table~\ref{tab:integrated_prop}.
The dust attenuation  measured in the broad component, tracing the outflow, is higher ($A_{V}=2.3\pm0.5$) compared to the rest of the galaxy ($A_{V}=1.4\pm0.1$).
Higher dust attenuation in the outflows than in the galaxy disc has been observed in several studies at lower redshift \citep[e.g.][]{Holt2011, VillarMartin2014, Perna2015a, Perna2019}.

In GS5001 the metallicity of the outflow ($12+\log \text{(O/H)} = 8.47\pm0.07$) is similar to the one derived from the narrow component ($12+\log \text{(O/H)} = 8.45\pm0.04$). 
The line ratios  \NII/\Ha\ and \OIII/\Hb\ in the broad component are also very similar to the ones observed in the narrow component ($\log$ \NII/\Ha $= -0.61\pm0.07$, $\log$ \OIII/\Hb $ = 0.36\pm0.07 $), thus, the ionisation source of the outflow is compatible with star formation (see Sec.~\ref{sec:BPT}).

We provide here an estimate of the outflow mass and mass outflow rate. 
We estimate a (projected) outflow radius of 0.4" (3~kpc), corresponding to the distance from the centre to the regions with high velocity dispersion (regions 2 and 3 in Fig.~\ref{fig:Ha_outflow}).
To estimate the maximum outflow velocity, we use the average of the absolute values of $v05$ and $v95$ (5th and 95th percentiles) velocities estimated from the integrated spectra of the high-velocity dispersion regions (see Fig.~\ref{fig:Ha_outflow}),  following  \citet{Cresci2015a} and \citet{Harrison2016}. 
The assumption behind this method is  that the outflow broadens the line profile due to having components in multiple directions, and the maximum velocity in the wings is produced by those components directed closer to line of sight, and is hence less affected by projection effects. Thus, the maximum velocity is a good approximation of the intrinsic outflow velocity \citep{Cresci2023}.
As in our case the profiles are symmetric and roughly centred around the zero velocity, $v05$ and $v95$ have a similar absolute value of $v_{max} = 400$~\kms.

Another common definition of the outflow velocity used in the literature is $v_\mathrm{out}=v_{broad}+2\sigma_{broad}$ \citep[e.g.][]{Fiore2017}. This definition would give $v_\mathrm{out} = 470$~\kms\ using the velocity parameters derived from the broad component of the integrated spectrum of the `main' galaxy, which is slightly higher than the value estimated from $v05$ and $v95$.
We cannot estimate the electron density in the outflow component as the broad component of \SII\ and \OII\ are too faint, so we use the electron density estimated for the total line profile ($n_e=540\pm110$~cm$^{-3}$, see Sec.~\ref{sec:electron_density}). 
We calculate the outflow mass following \citet{Cresci2017}: $M_\mathrm{out}=3.2\cdot 10^5 \cdot(L_\mathrm{broad}(\mathrm{H}\alpha)/10^{40}\text{ erg/s})\cdot (100 \text{ cm}^{-3}/n_e) \text{ M}_{\odot} =(1.7\pm0.4) \times 10^8$ \Msun. 
The corresponding mass outflow rate is  $\dot{M}_\mathrm{out}= M_\mathrm{out}\cdot v_\mathrm{out}/R_\mathrm{out}=23\pm5$~\Msunyr.  

We estimate the escape velocity following \citet{Arribas2014}. We assume the dynamical mass is twice the stellar mass  \citep{Erb2006}; that is, $M_{dyn}=2-8\times 10^{10}$~\Msun. We calculate the escape velocity at a radius of 3~kpc (size of the outflow), assuming a truncation radius of 30~kpc. We obtain escape velocities in the range $\sim330-660$~\kms.  On the basis of this calculation, we estimate that less than 15\% of the outflowing gas has velocities large enough to escape from the galaxy potential well.

\subsubsection{Origin of the outflow}

As was shown in Section~\ref{sec:outflow}, the outflow reaches maximum projected velocities  (as traced by $v05$ and $v95$) of $\sim\pm400$~\kms. 
These velocities are comparable to the velocities measured in ionised outflows driven by star formation \citep[e.g.][]{Arribas2014, ForsterSchreiber2019, Swinbank2019}.  
The SFR surface density, estimated from the \Ha\ flux, in the central region of GS5001 reaches 10~\Msunyr\ kpc$^{-2}$ (see Fig.~\ref{fig:SFR_map}). 
This high $\Sigma_{SFR}$ can explain the observed outflow, since it is known that the  prevalence of outflows in star-forming galaxies increases with $\Sigma_{SFR}$ \citep{ForsterSchreiber2019}.
Additionally, the global geometry with the outflow direction roughly aligned with the minor axis is what is expected for a SF-driven outflow. 

The mass loading factor, assuming the SFR from the narrow \Ha\ component SFR$\sim100$~\Msunyr, is $\eta=\dot{M}_{out}/\text{SFR}= 0.23$.
The moderate mass loading factor (< 1) means that the outflow is not having a significant impact on the total star formation of GS5001. We note however that we are only tracing the ionised outflow, while the molecular and atomic phases of the outflow could also contribute significantly to the total mass outflow rate \citep[e.g.][]{Fiore2017, Herrera-Camus2019, Ginolfi2020, Fluetsch2021, Belli2023arXiv, DEugenio2023arXiv}.

The  mass loading factor in GS5001 is in agreement with other studies of outflows in star-forming galaxies at lower redshift ($z=0.5-3$). For instance, \citet{ForsterSchreiber2019} find average $\eta=0.1-0.25$ in a sample at $z=0.6-2.7$, while \citet{Swinbank2019} report  $\eta=0.2-0.4$ in star-forming galaxies at $z\sim1$. 
\citet{RodriguezDelPino2024} detected a resolved ionised outflow in the star-forming galaxy GS4891 at $z=3.7$, which has a SFR of  $\sim45$~ \Msunyr.  This outflow has a similar outflow velocity of 400~\kms, but lower mass outflow rate $\dot{M}_\mathrm{out}=2$~\Msunyr\ and   $\eta=0.02$  compared to GS5001, which could be related to the lower SFR of GS4891 compared to GS5001.

\subsection{No evidence of an active galactic nucleus in GS5001}
As was discussed in the Introduction, from previous observations it is not clear whether an AGN in present in GS5001.
 We do not find evidence of AGN activity in the NIRSpec data of GS5001.
The emission line diagnostic diagram (BPT, see Fig.~\ref{fig:BPT_map}) shows line ratios similar to other star-forming galaxies at the same redshift, with no spaxels above the AGN separation line. 
We note that the line ratios for GS5001 do not fall in the region of the BPT diagram where low-metallicity AGN and star-forming galaxies tend to overlap at high redshifts \citep{Maiolino2023arXiv, Scholtz2023arXiv}. Our target is at lower redshift and have higher metallicity than the AGN presented in those works; therefore, the classical BPT diagram can still be effectively used to distinguish between AGN and star-forming galaxies. 

GS5001 has been detected in the X-ray, with an X-ray luminosity of $L_{0.5-7keV}=(9.5\pm 2.8) \times 10^{42}$ erg s$^{-1}$ \citep{Fiore2012,Luo2017}. 
 In the Chandra Deep Field-South 7Ms catalogue \citep{Luo2017}, it has been classified as an AGN, based on an intrinsic X-ray luminosity $L_{0.5-7keV}> 3\times 10^{42}$ erg/s. However, for this target no additional criteria could be applied.
Therefore, \citet{Luo2017} caution that X-ray emission in this sources may come from intense star formation. A SFR$\sim 520\pm 180$ \Msunyr\ would be required to explain the observed X-ray luminosity of this target by assuming the \citet{Ranalli2003} relation adapted for a \citet{Chabrier2003} IMF \citep{Kennicutt2012}.  
Thus, the majority of this X-ray luminosity could be explained by the observed SFR of this system (total SFR of the main, south and north components: $\sim 300$~\Msunyr).
\citet{Lyu2022} classify this galaxy as an AGN based on the X-ray-to-radio luminosity ratio $L_\textrm{X-ray} \text{ [erg s$^{-1}$]}/L_\textrm{3\,GHz} \text{ [W Hz$^{-1}$]} > 8\times 10^{18}$. However, we note that GS5001 is only $\sim0.25$ dex above this threshold and that the X-ray luminosity has been inferred from an extremely small number of counts.

A recent SED analysis  including JWST/NIRcam and MIRI photometry up to restframe wavelengths $< 6$~$\mu$m classify this galaxy as an AGN,  because the AGN template dominates in the range $3-6$~$\mu$m  \citep{Lyu2024}. However, given the limited coverage in the infra-red, a fit with a galaxy dust template and a weaker AGN component would still be acceptable. 
From an SED fitting analysis with the Code Investigating GALaxy Emission \citep[\cigale;][]{Burgarella2005, Noll2009, Boquien2019} from the UV to the FIR , we find that a strong AGN component in the MIR is not required to produce a good fit (C.~Circosta et al., in prep.).
 In summary, we do not find evidence for an AGN in GS5001, even though we cannot rule out that a weak and possibly obscured AGN is present.

\section{Summary and conclusions}
\label{sec:conclusions}

In this work, we present JWST/NIRSpec IFS observations of the galaxy GS5001 and its companions at redshift $z=3.47$ within a FoV of $4"\times4"$ ($30\times30$~kpc$^2$).
We analysed the properties of the emission lines using the high-resolution (R2700) data, complemented by the low-resolution (R100) data, which allowed us to cover the optical emission lines from \OII$\lambda \lambda3726,29$  to \SIII$\lambda9531$.
We fit the data cube to derive the maps of the emission line fluxes as well as kinematic maps of the ionised gas.

The main results of this study are:
\begin{itemize}
\item We identify several companions close to GS5001 (main target). In particular, we identify a companion in the south, with three sub-structures (s1, s2, s3), and a companion in the north, with three sub-structures (n1, n2, n3), showing also an extended tail.
The south components show velocities blueshifted by [-160, -153, -80]~\kms\ with respect to the main target, while the north companion is redshifted by $190$~\kms\ (see Sec.~\ref{sec:kinematics}).
\item The spatially resolved emission line ratios are in the star-forming region of the BPT diagram, with no sign of AGN excitation (see Sec.~\ref{sec:BPT}).
\item We estimate the gas-phase metallicity using the R3, R23, and O32 line ratios. We find that the main galaxy has a metallicity  of 12+log(O/H)$= 8.45 \pm0.04$, and the companions show slightly lower metallicities 12+log(O/H)$= 8.32-8.42$, consistent with the mass-metallicity relation at $z\sim3$. The main galaxy shows higher metallicity in the north-east region, and lower in the south-east (see Sec.~\ref{sec:metallicity}).
\item From the dust-corrected \Ha\ luminosity, we infer a total SFR of 100~\Msunyr in the main target. The south companions have a combined SFR of $76\pm2$~\Msunyr\ and the north companion a SFR of $106\pm1$~\Msunyr. The SFR surface density reaches values of $\Sigma_{SFR}=10$~\Msunyr kpc$^{-2}$ in the central region of the main galaxy (see Sec.~\ref{sec:SFR}).
\item We find that the region between the north companion and the main target has high \NII/\Ha\ and low \OIII/\Hb\ line ratios, compared with the rest of the main galaxy. These line ratios could be due to the higher metallicity or to shocks. The gas in this region also shows enhanced velocity dispersion, probably due to the interaction between the main and north components  (see Sec.~\ref{sec:BPT}).
\item 
We identify an outflow traced by a broad symmetric component clearly visible in \Ha\ and \OIII\ (with velocity dispersion $\sim220$~\kms, or FWHM of $\sim520$~\kms).
The outflow originates in the nucleus of the main galaxy, and extends up to $\sim3$~kpc, where interaction with the north companion might also contribute to the broadening of the lines.
The outflow has maximum velocity of $\sim400$~\kms, outflow mass of $(1.7\pm0.4)\times 10^8$~\Msun, mass outflow rate $23\pm5$ \Msunyr, and a mass loading factor of 0.23  (see Sec.~\ref{sec:outflow_discussion}). This indicates that probably the outflow is not going to impact significantly the star formation in the host galaxy. We note however that we are not tracing the neutral and molecular phases of the outflow.

\end{itemize}

 JWST NIRSpec IFS data allowed us to obtain  unprecedented insights into the interplay between star formation, galactic outflows and interactions in the core of a $z\sim 3.5$  candidate protocluster. This valuable information could be used to inform hydro-dynamical simulations and therefore follow the evolutionary path of GS5001 and other similar systems, like the Spiderweb Galaxy \citep{Emonts2016} and CGG-z5 \citep{Jin2023}.

\begin{acknowledgements}


We thank the anonymous referee for a constructive report that helped to improve the quality of this manuscript.
This paper makes use of the following ALMA data: 
ADS/JAO.ALMA$\#$2012.1.00423.S. 
ALMA is a partnership of ESO (representing its member states), NSF (USA) and NINS (Japan), together with NRC (Canada), MOST and ASIAA (Taiwan), and KASI (Republic of Korea), in cooperation with the Republic of Chile. The Joint ALMA Observatory is operated by ESO, AUI/NRAO and NAOJ.
This work has made use of data from the European Space Agency (ESA) mission
{\it Gaia} (\url{https://www.cosmos.esa.int/gaia}), processed by the {\it Gaia} Data Processing and Analysis Consortium (DPAC,
\url{https://www.cosmos.esa.int/web/gaia/dpac/consortium}). Funding for the DPAC has been provided by national institutions, in particular the institutions participating in the {\it Gaia} Multilateral Agreement.
This work has made use of the Rainbow Cosmological Surveys Database, which is operated by the Centro de Astrobiología (CAB), CSIC-INTA, partnered with the University of California Observatories at Santa Cruz (UCO/Lick,UCSC). The project leading to this publication has received support from ORP, that is funded by the European Union’s Horizon 2020 research and innovation programme under grant agreement No 101004719 [ORP].
This research made use of Astropy,  a community-developed core Python package for Astronomy \citep{astropy},  {\tt Matplotlib} \citep{Hunter2007}, {\tt NumPy} \citep{VanDerWalt2011}, {\tt corner} \citep{corner}. 
This research has made use of "Aladin sky atlas" developed at CDS, Strasbourg Observatory, France \citep{Bonnarel2000}.
IL acknowledges support from PID2022-140483NB-C22 funded by AEI 10.13039/501100011033, from BDC 20221289 funded by MCIN by the Recovery, Transformation and Resilience Plan from the Spanish State, and by NextGenerationEU from the European Union through the Recovery and Resilience Facility, and from PRIN-MUR project "PROMETEUS" (202223XPZM).
 SA, MP and BRdP acknowledges grant PID2021-127718NB-I00 funded by the Spanish Ministry of Science and Innovation/State Agency of Research (MICIN/AEI/ 10.13039/501100011033).
  PGP-G acknowledges support  from  Spanish  Ministerio  de  Ciencia e Innovaci\'on MCIN/AEI/10.13039/501100011033 through grant PGC2018-093499-B-I00.
 AJB, JC and GCJ acknowledges funding from the “FirstGalaxies” Advanced Grant from the European Research Council (ERC) under the European Union’s Horizon 2020 research and innovation programme (Grant agreement No. 789056).
 SCa and GV acknowledges support from the European Union (ERC, WINGS,101040227). 
RM, FDE and JS acknowledge support by the Science and Technology Facilities Council (STFC), by the ERC through Advanced Grant 695671 ``QUENCH'', and by the UKRI Frontier Research grant RISEandFALL.
H{\"U} gratefully acknowledges support by the Isaac Newton Trust and by the Kavli Foundation through a Newton-Kavli Junior Fellowship.
 EB and GC acknowledges the support of the INAF Large Grant 2022 "The metal circle: a new sharp view of the baryon cycle up to Cosmic Dawn with the latest generation IFU facilities".
\end{acknowledgements}

%
%

\bibliographystyle{aa} 
\bibliography{main.bib}


\begin{appendix} 

\section{Tables of the emission line fluxes}
In Tables~\ref{tab:integrated_fluxes} and ~\ref{tab:integrated_fluxes_R100}  we report the emission line fluxes measured from the spectra integrated over the different regions shown in Fig.~1, from the R2700 and R100 data cubes, respectively.
\begin{table*}
\centering
\caption{Emission line fluxes of the different components (and sub-components) of the system, in units of [$\times 10^{-18}$ erg s$^{-1}$ cm$^{-2}$].}
\setlength{\tabcolsep}{3pt}
\begin{tabular}{lccccccccccc}
\hline
region & \OII $\lambda3726$  & \OII $\lambda3729$ & \Hg\ & \Hb & \OIII $\lambda5007$ & \OI $\lambda6300$ &  \Ha & \NII $\lambda6584$ &  \SII $\lambda6716$  & \SII $\lambda6731$ &\\ 
  \hline \hline

main & $180.7\pm 1.9$  & $127.3\pm 0.8$  & $45.2\pm 0.4$  & $106.8\pm 0.4$  & $227.6\pm 0.4$  & $7.6\pm 0.5$  & $305.3\pm 0.8$  & $78.6\pm 0.5$  & $28.1\pm 0.7$  & $20.9\pm 0.3$  & \\
main n & $77.9\pm 1.7$  & $60.9\pm 0.9$  & $28.4\pm 0.4$  & $59.6\pm 0.6$  & $125.8\pm 0.6$  & $5.1\pm 0.4$  & $170.3\pm 1.2$  & $43.1\pm 0.7$  & $14.6\pm 0.6$  & $11.0\pm 0.4$  & \\
main b & $316.1\pm 4.0$  & $134.0\pm 1.4$  & $11.2\pm 0.6$  & $89.7\pm 1.1$  & $197.5\pm 1.8$  & $3.1\pm 0.5$  & $256.4\pm 4.0$  & $63.8\pm 1.7$  & $26.7\pm 1.1$  & $21.1\pm 0.8$  & \\
south & $81.9\pm 1.6$  & $60.1\pm 0.7$  & $21.3\pm 0.3$  & $45.0\pm 0.3$  & $125.0\pm 0.4$  & $4.5\pm 0.4$  & $128.6\pm 0.6$  & $18.8\pm 0.4$  & $13.2\pm 0.7$  & $10.7\pm 0.3$  & \\
north & $189.8\pm 1.0$  & $210.9\pm 0.5$  & $42.1\pm 0.3$  & $63.2\pm 0.2$  & $202.5\pm 0.3$  & $2.5\pm 0.3$  & $180.7\pm 0.4$  & $29.2\pm 0.3$  & $20.9\pm 0.4$  & $15.7\pm 0.2$  & \\
\hline
s1 & $8.7\pm 1.0$  & $5.7\pm 0.4$  & $2.1\pm 0.2$  & $7.0\pm 0.2$  & $20.4\pm 0.3$  & $0.6\pm 0.2$  & $20.0\pm 0.4$  & $2.1\pm 0.3$  & $1.6\pm 0.3$  & $1.8\pm 0.2$  & \\
s2 & $23.4\pm 0.5$  & $19.6\pm 0.2$  & $5.0\pm 0.1$  & $11.5\pm 0.1$  & $31.2\pm 0.1$  & $1.3\pm 0.1$  & $32.9\pm 0.2$  & $6.0\pm 0.1$  & $3.4\pm 0.2$  & $2.8\pm 0.1$  & \\
s3 & $30.8\pm 0.4$  & $24.2\pm 0.2$  & $9.3\pm 0.1$  & $18.0\pm 0.1$  & $54.6\pm 0.1$  & $1.7\pm 0.1$  & $51.5\pm 0.2$  & $6.9\pm 0.1$  & $4.9\pm 0.1$  & $3.5\pm 0.1$  & \\
n1 & $12.0\pm 0.2$  & $10.2\pm 0.1$  & $2.0\pm 0.1$  & $7.2\pm 0.1$  & $17.6\pm 0.1$  & $0.3\pm 0.1$  & $20.5\pm 0.1$  & $2.6\pm 0.1$  & $1.8\pm 0.1$  & $1.4\pm 0.1$  & \\
n2 & $24.5\pm 0.2$  & $21.4\pm 0.1$  & $4.1\pm 0.1$  & $7.5\pm 0.1$  & $27.2\pm 0.1$  & $0.9\pm 0.1$  & $21.5\pm 0.1$  & $3.8\pm 0.1$  & $2.1\pm 0.1$  & $1.8\pm 0.1$  & \\
n3 & $99.2\pm 0.4$  & $79.1\pm 0.2$  & $31.3\pm 0.1$  & $16.2\pm 0.1$  & $78.8\pm 0.1$  & $1.6\pm 0.1$  & $46.2\pm 0.1$  & $4.3\pm 0.1$  & $3.7\pm 0.1$  & $3.3\pm 0.1$  & \\
\clump & $6.2\pm 0.2$  & $5.6\pm 0.1$  & $1.4\pm 0.1$  & $3.5\pm 0.1$  & $8.2\pm 0.1$  & $0.5\pm 0.1$  & $10.0\pm 0.1$  & $2.2\pm 0.1$  & $0.7\pm 0.1$  & $0.6\pm 0.1$  & \\

\hline
\end{tabular} 
\label{tab:integrated_fluxes}
\tablefoot{
Fluxes of the different components identified in Fig.~1. For the main galaxy, we report the fluxes obtained with one Gaussian component  model (main) and the fluxes obtained with a two-component model (narrow (main n) and broad (main b) components). The label `main n' and `main b' refer to the narrow and broad component of the two-component model, respectively. For the other regions, the fluxes have been obtained with the one-component model fit. Fluxes have been corrected for obscuration using a \citet{Cardelli1989} attenuation law as described in Sec.~5.2.1.
}
\end{table*}

\begin{table}
\centering
\caption{Fluxes of the different components of the system derived from the R100 data cube, in units of [$\times 10^{-18}$ erg s$^{-1}$ cm$^{-2}$].}
\setlength{\tabcolsep}{3pt}
\begin{tabular}{lccccccccccc}
\hline
region &  \SII $\lambda6716$  + $\lambda6731$ & \SIII $\lambda9069$  & \SIII $\lambda9531$ \\ %
  \hline \hline

main & $55.3\pm 6.8$  & $17.2\pm 3.3$  & $43.1\pm 4.1$   \\
south & $27.9\pm 3.6$  & $6.8\pm 1.8$  & $19.9\pm 2.3$ \\
north & $23.0\pm 3.1$  & $8.5\pm 1.2$  & $21.0\pm 1.4$  \\
\hline
s1 & $4.0\pm 0.8$  & $2.0\pm 0.6$  & $3.8\pm 0.8$  & \\
s2 & $8.2\pm 1.0$  & $1.5\pm 0.4$  & $4.7\pm 0.6$  & \\
s3 & $6.5\pm 1.4$  & $2.4\pm 0.7$  & $7.6\pm 0.8$  & \\
n1 & $3.0\pm 0.6$  & $1.1\pm 0.3$  & $2.7\pm 0.3$  & \\
n2 & $3.8\pm 0.6$  & $1.1\pm 0.3$  & $2.6\pm 0.3$  & \\
n3 & $6.7\pm 0.9$  & $1.1\pm 0.2$  & $4.4\pm 0.3$  & \\
\clump & $2.6\pm 0.4$  & $0.6\pm 0.2$  & $1.4\pm 0.3$  & \\

\hline
\end{tabular} 
\label{tab:integrated_fluxes_R100}
\tablefoot{
Fluxes of the different components identified in Fig.~1. Fluxes have been corrected for obscuration using a \citet{Cardelli1989} attenuation law as described in Sec.~5.2.1. We report the total flux of the  \SII $\lambda6716$  + \SII $\lambda6731$ doublet, since the two lines are not resolved in the R100 data.
}
\end{table}

\section{Example of spectral fit of one spaxel}
Figure~\ref{fig:spectrum_fit_example} shows the spectrum extracted from one spaxel close to the centre of the main galaxy ($x, y= 44,46$), together with the best-fit model obtained with the python routine `scipy.optimize.curve$\_$fit'.

\begin{figure*}[!]
\centering 
\includegraphics[width=0.99\textwidth]{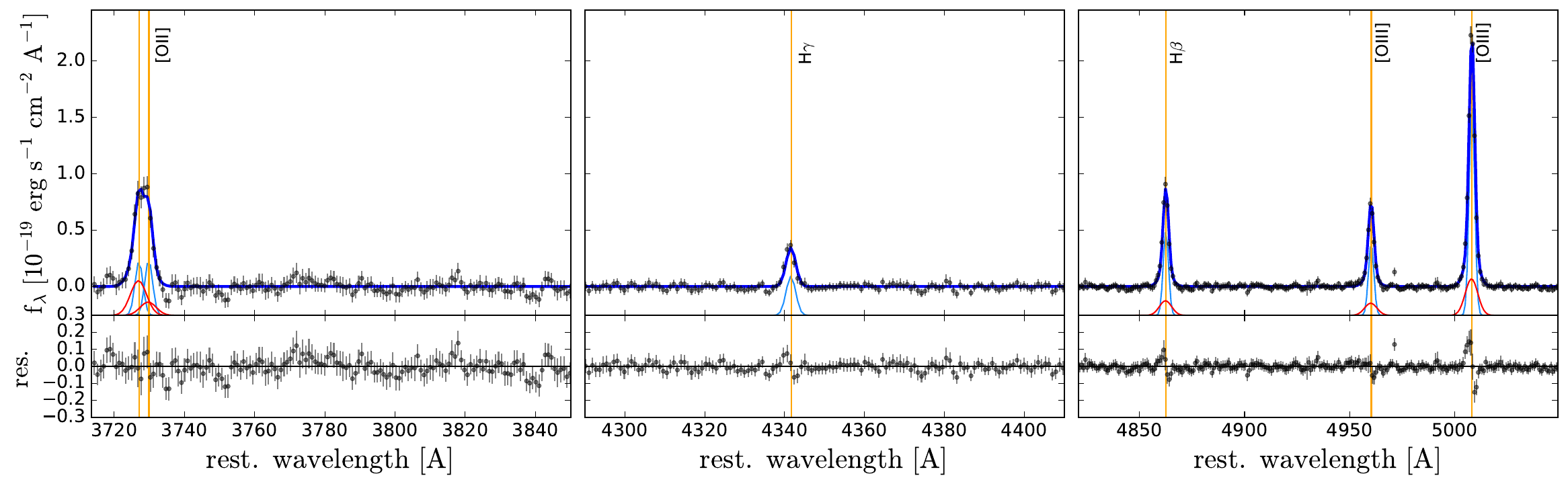}
\includegraphics[width=0.99\textwidth]{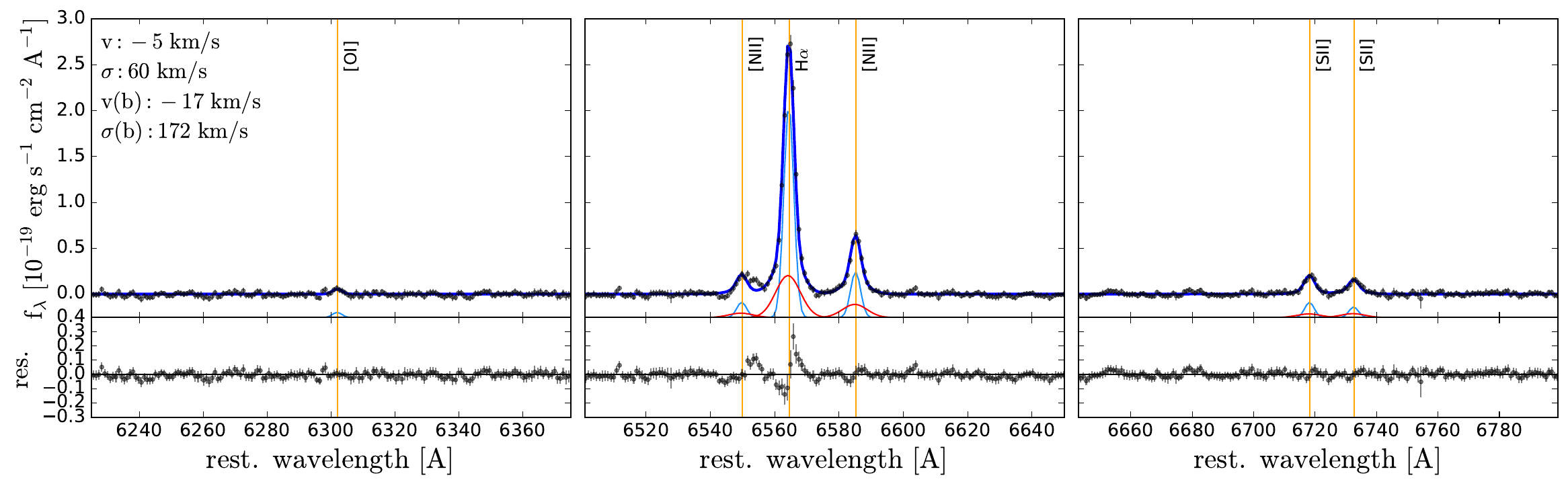}
\caption{Example of fit of the average spectrum of nine adjacent spaxels close to the centre of the `main' galaxy, used for the spatially-resolved fit. Data are shown in black, the total best fit in blue, the narrow and broad components are shown in lightblue and red, respectively, shifted vertically for visual purposes.  
The vertical lines mark the wavelength positions of the emission lines at the systemic redshift of the source ($z=3.4705)$. The fitting residuals are shown in the bottom panel.}
\label{fig:spectrum_fit_example}
\end{figure*}

\section{\texorpdfstring{H$\alpha$}{Ha} maps from the one-component Gaussian fit}
In this section, we show the maps of the \Ha\ flux, velocity (v50) and velocity dispersion, obtained from the fit with one Gaussian component (Figure~\ref{fig:Ha_map_1comp}). These maps are qualitatively similar to the maps obtained with a variable number of components (one or two) shown in Fig.~5 of the paper. 

\begin{figure*}[!]
\centering 
\includegraphics[width=0.9\textwidth]{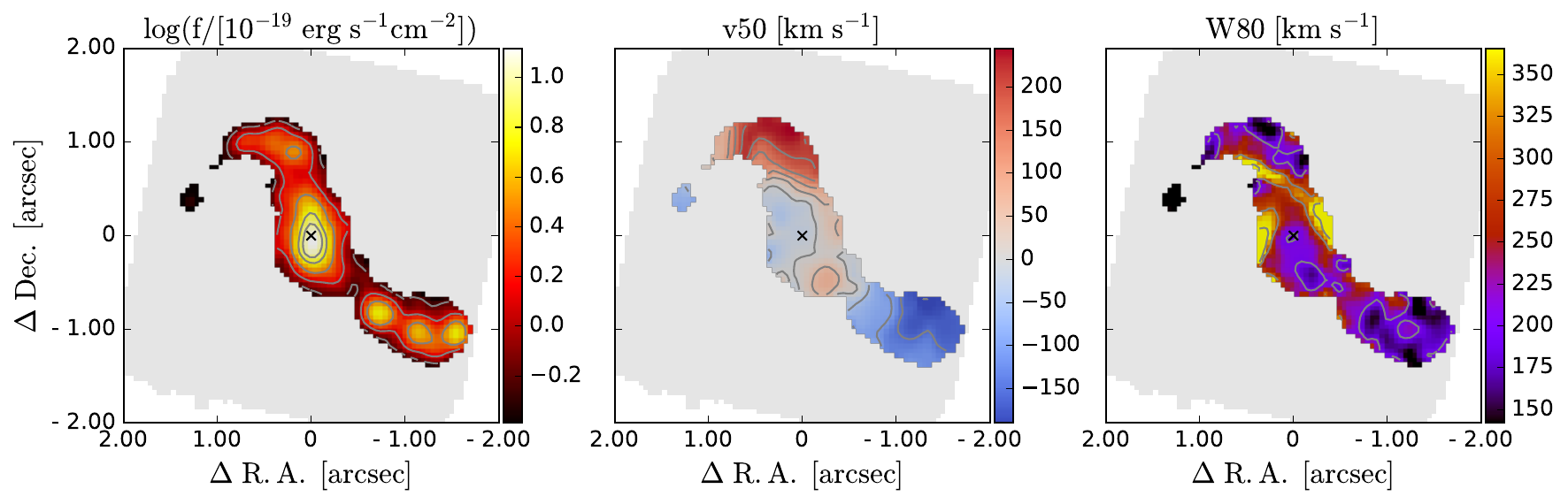}
\caption{Maps of the observed \Ha\ emission  obtained from the spatially resolved emission line fit with one Gaussian component.  
From left to right: integrated observed flux (not corrected for obscuration), 50th percentile velocity ($v50$), line width W80 (difference between 10th and 90th percentile velocity). 
Contours in the flux map show the [20,40, 60, 80, 90] percentiles, contours in the $v50$ map start at -150~\kms and increase every 50~\kms, contours in the W80 map are at [200, 300, 400]~\kms.}
\label{fig:Ha_map_1comp}
\end{figure*}


\section{Channel maps to highlight the position of clump `\clump'}

In this section we show channel maps that we use to identified a clump (`\clump') at the south-west of the main galaxy. Figure~\ref{fig:clump_SW} shows continuum-subtracted channel maps of \Ha. The channels have a width of 40~\kms and span the range from --80 to 240~\kms\ with respect to the systemic redshift of the main galaxy ($z=3.4705$). The emission at the position of the clump is visible in the channels from $-40$ to 160~\kms. In particular, the morphology in the channel 120-160~\kms shows that at the clump position there is a clear peak,  separated from the emission of the main galaxy.
\begin{figure*}[!]
\centering 
\includegraphics[width=0.24\textwidth]{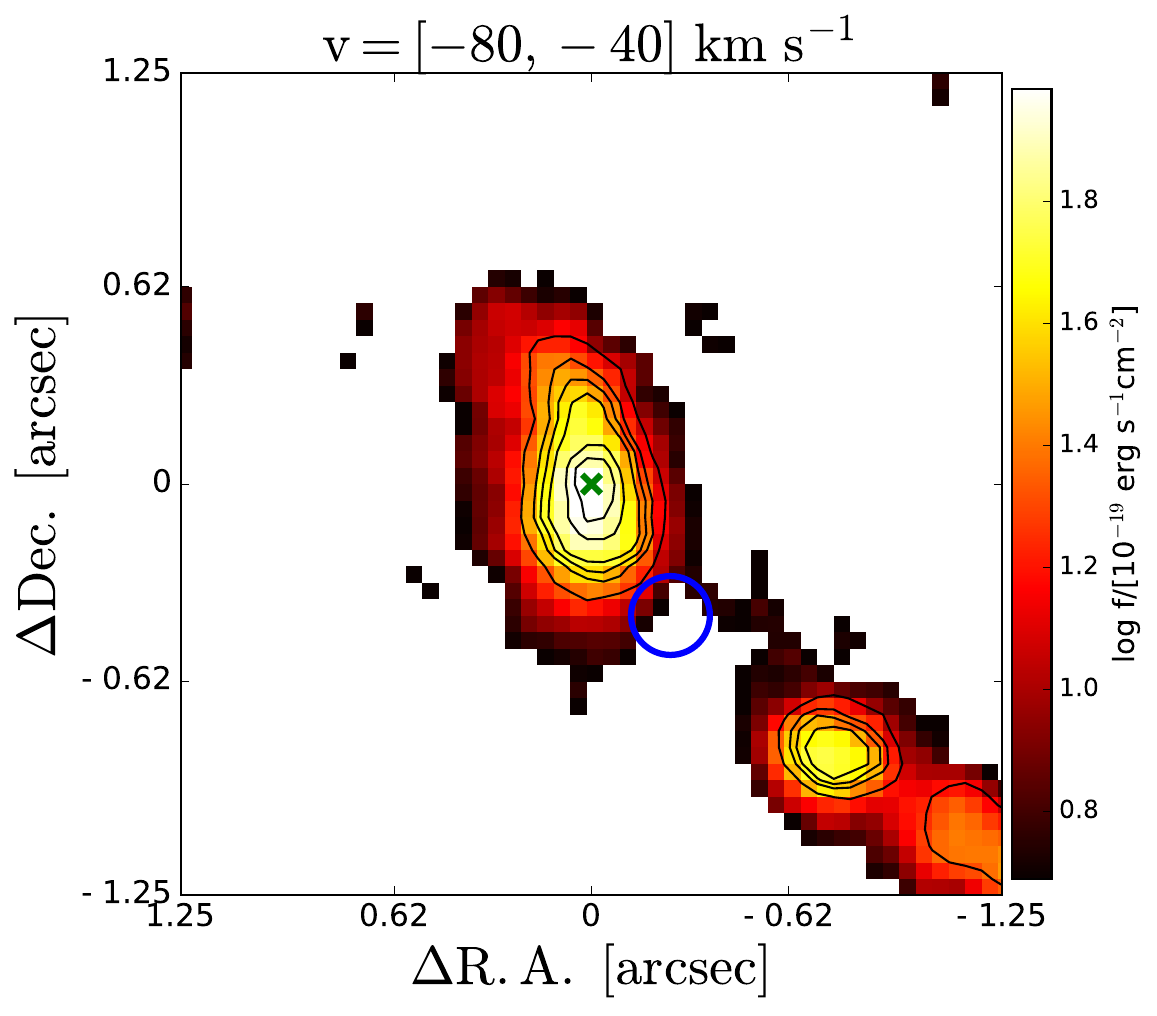}
\includegraphics[width=0.24\textwidth]{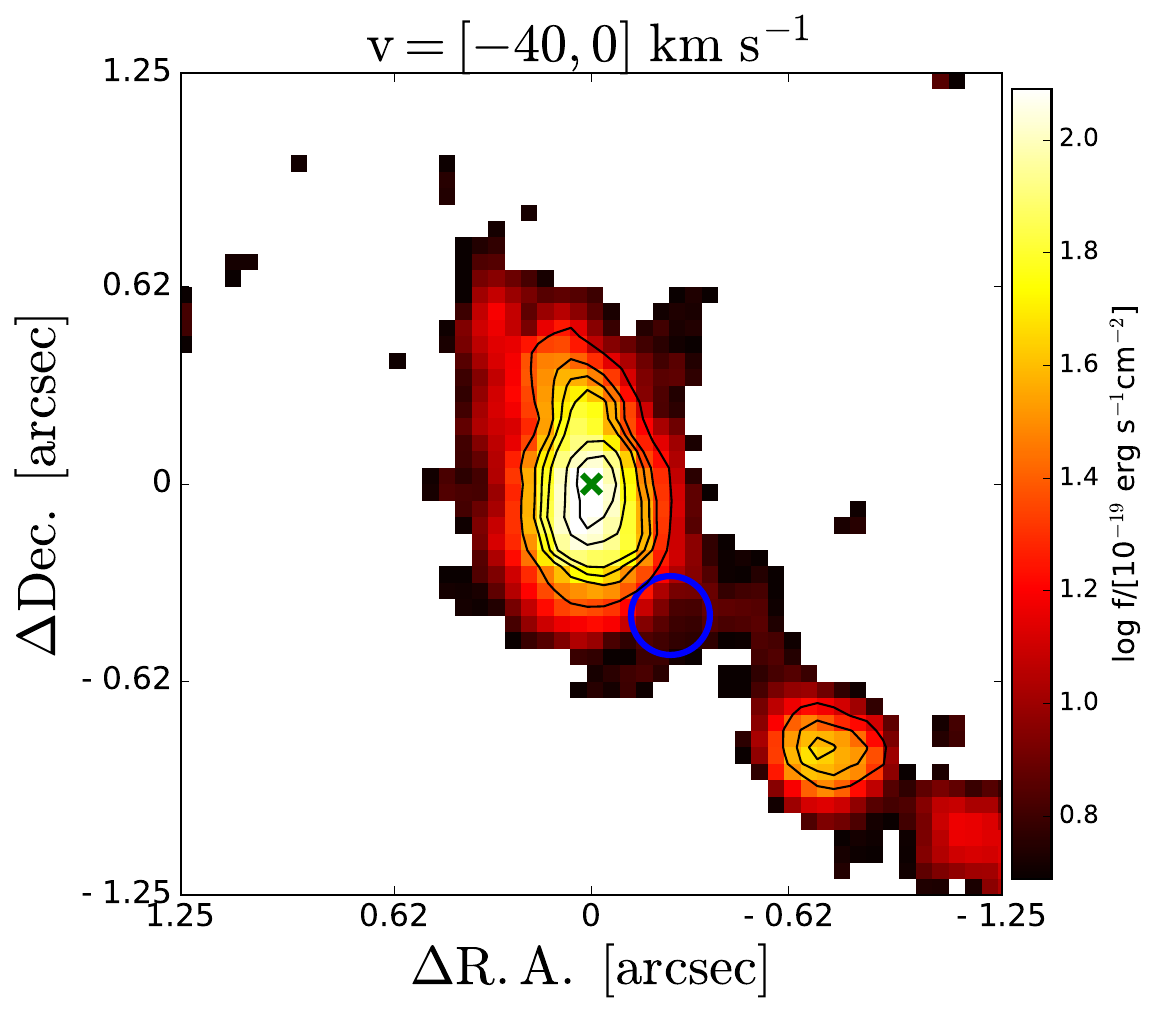}
\includegraphics[width=0.24\textwidth]{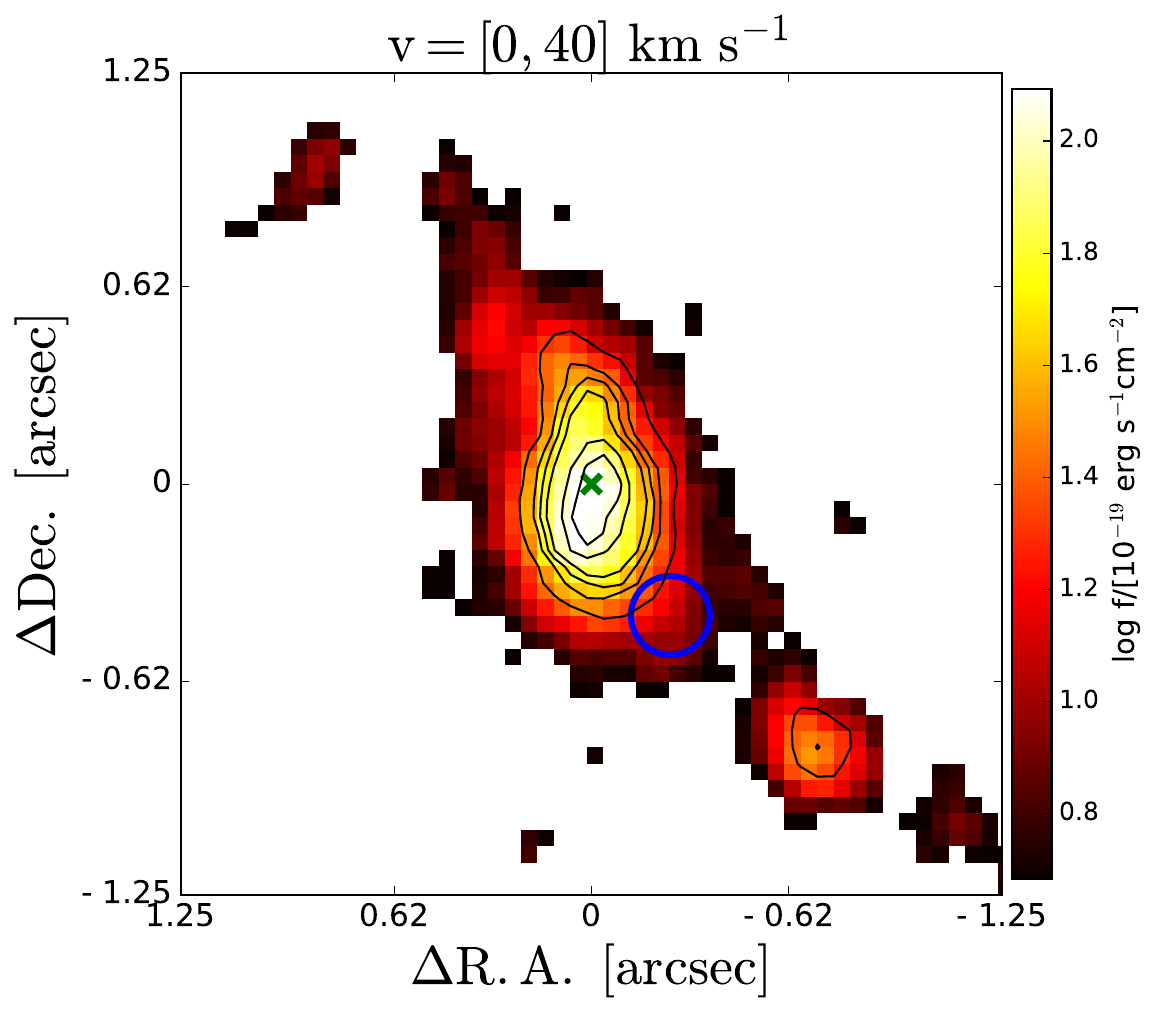}
\includegraphics[width=0.24\textwidth]{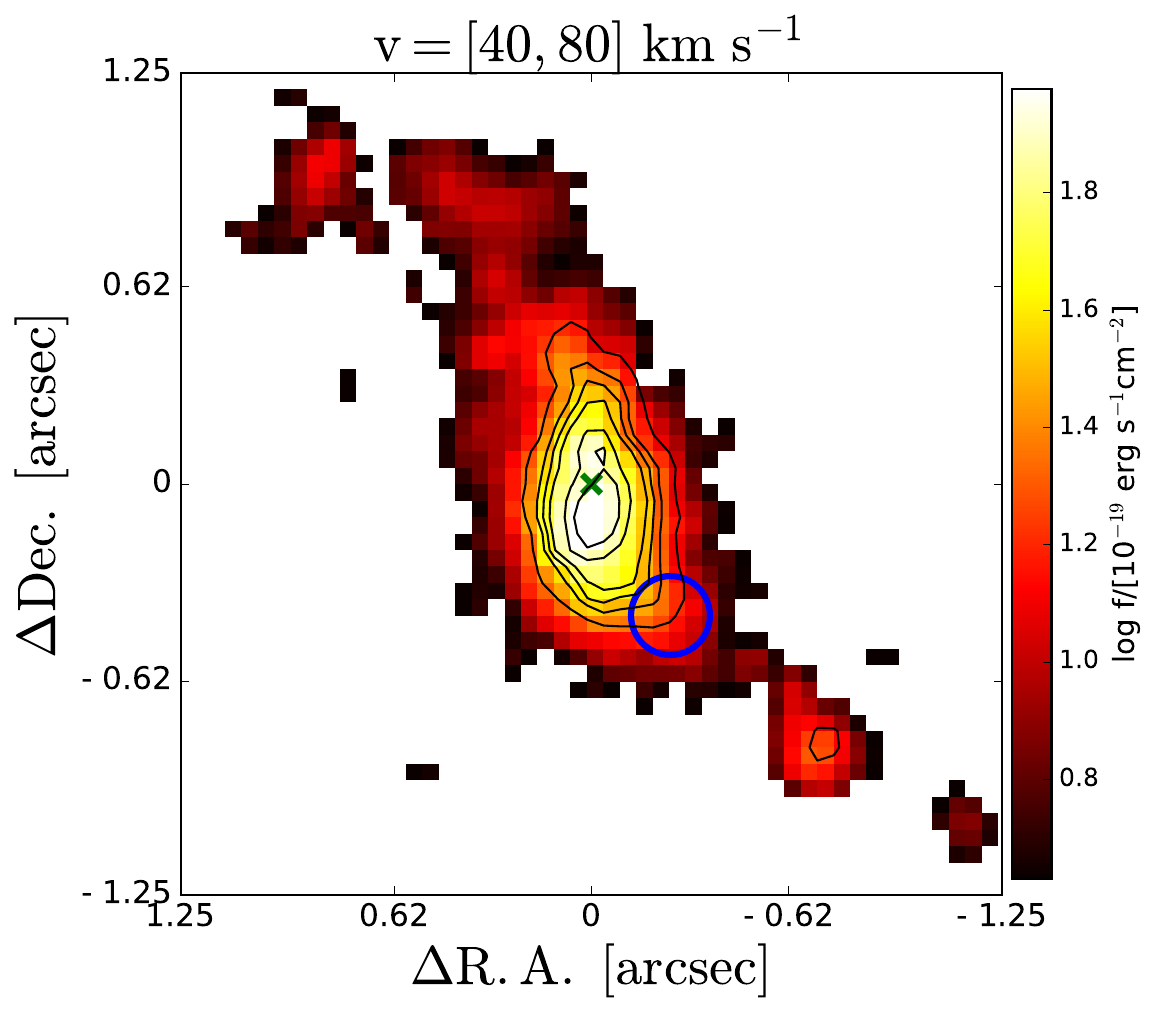}
\includegraphics[width=0.24\textwidth]{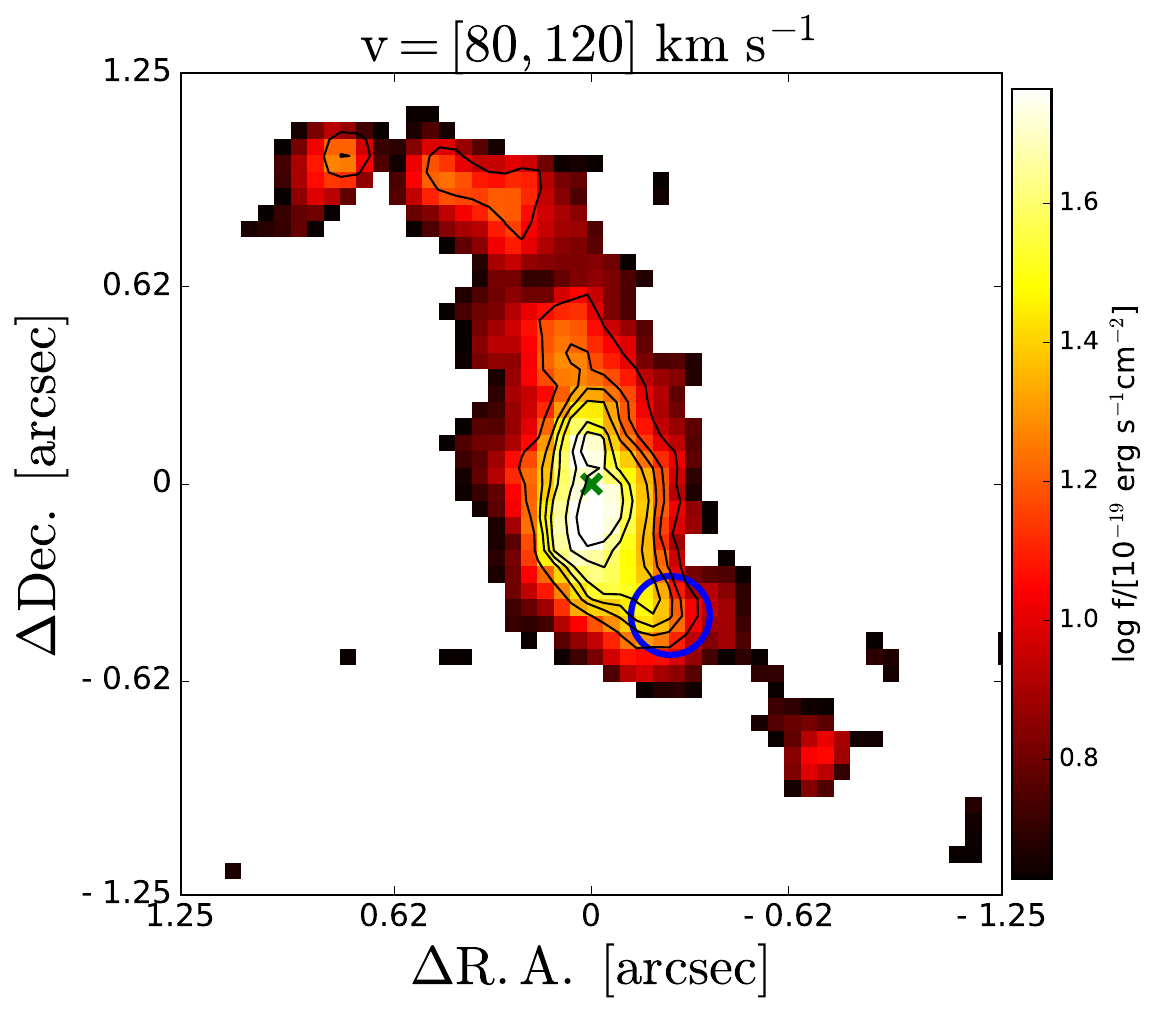}
\includegraphics[width=0.24\textwidth]{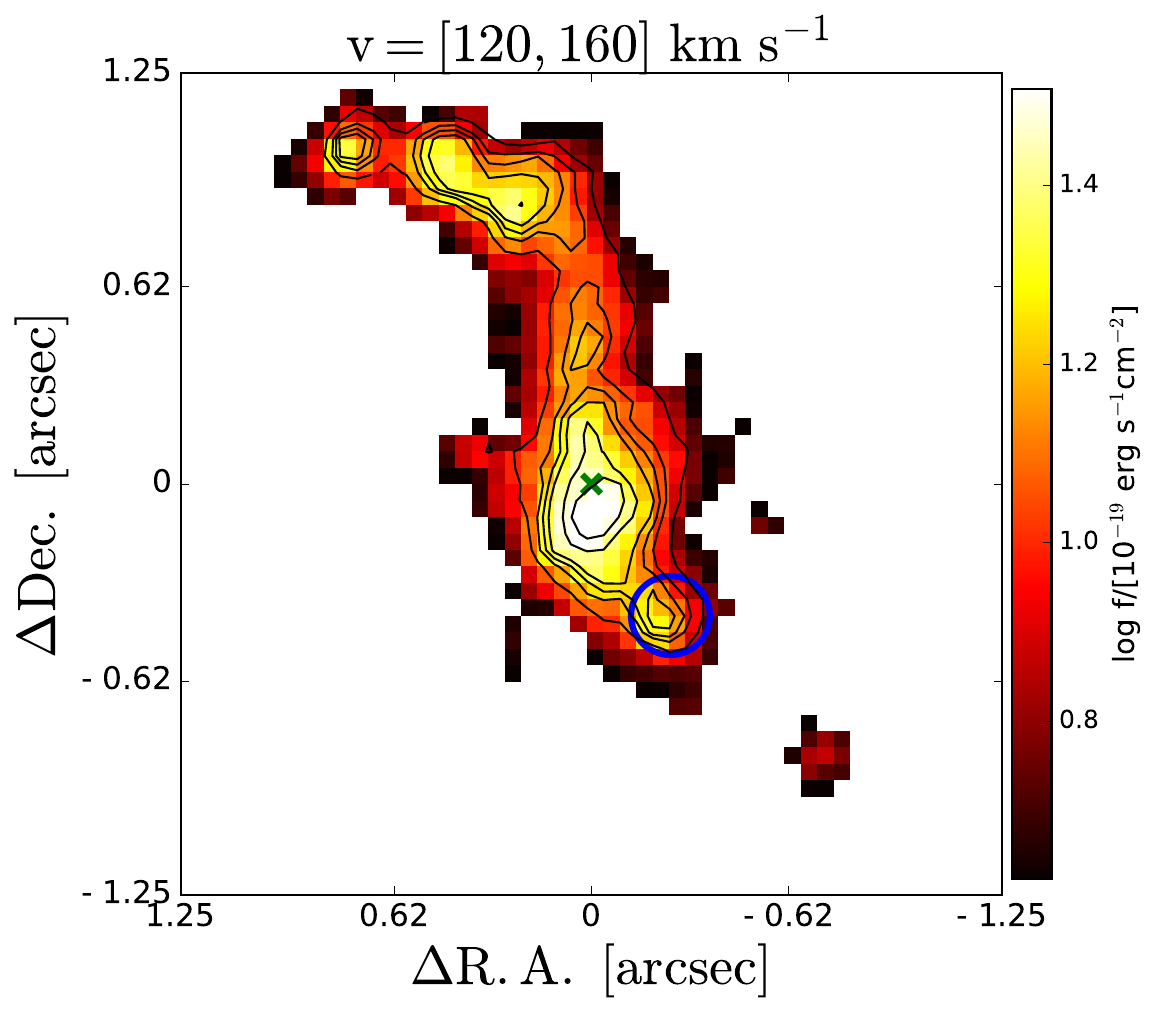}
\includegraphics[width=0.24\textwidth]{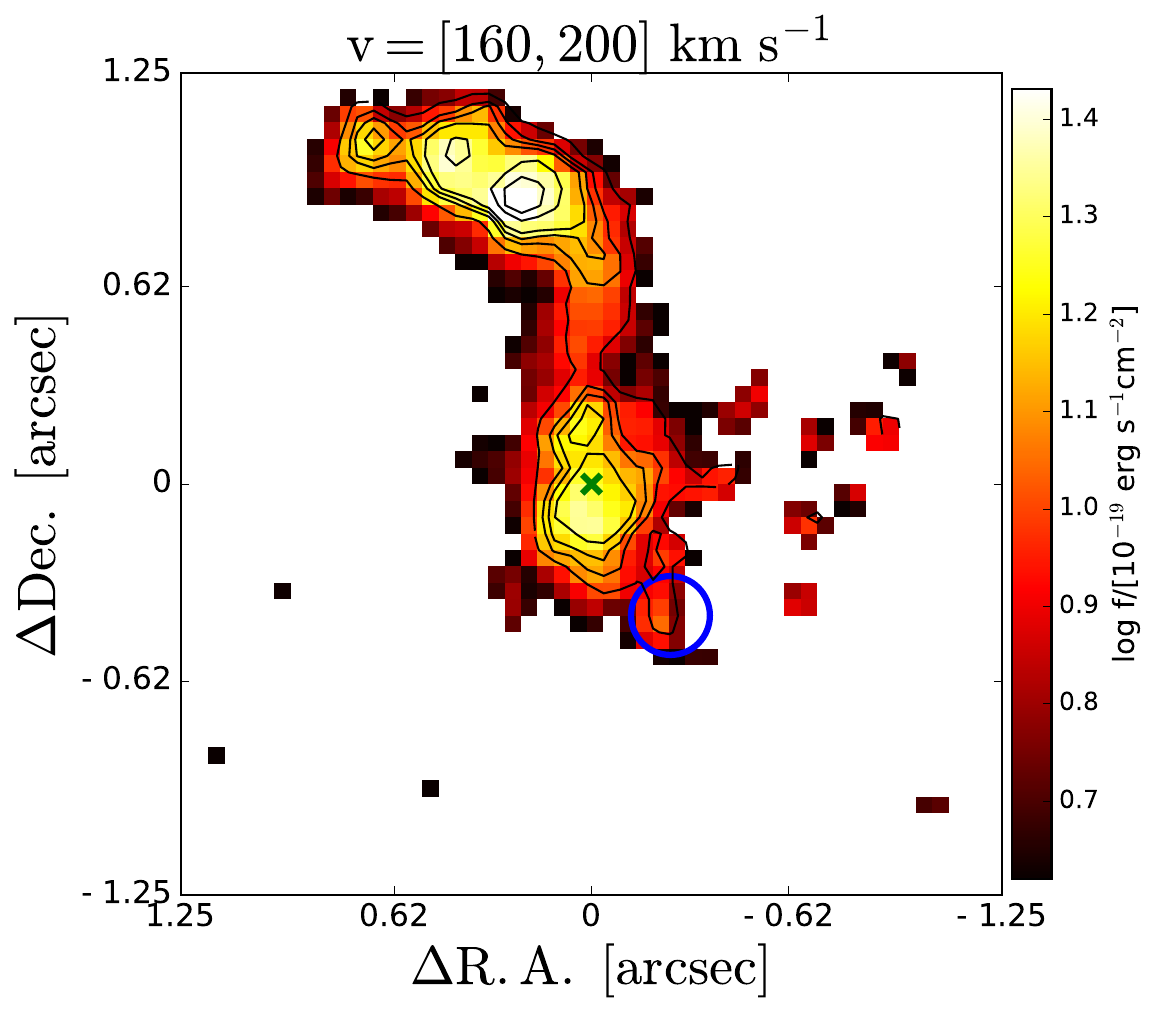}
\includegraphics[width=0.24\textwidth]{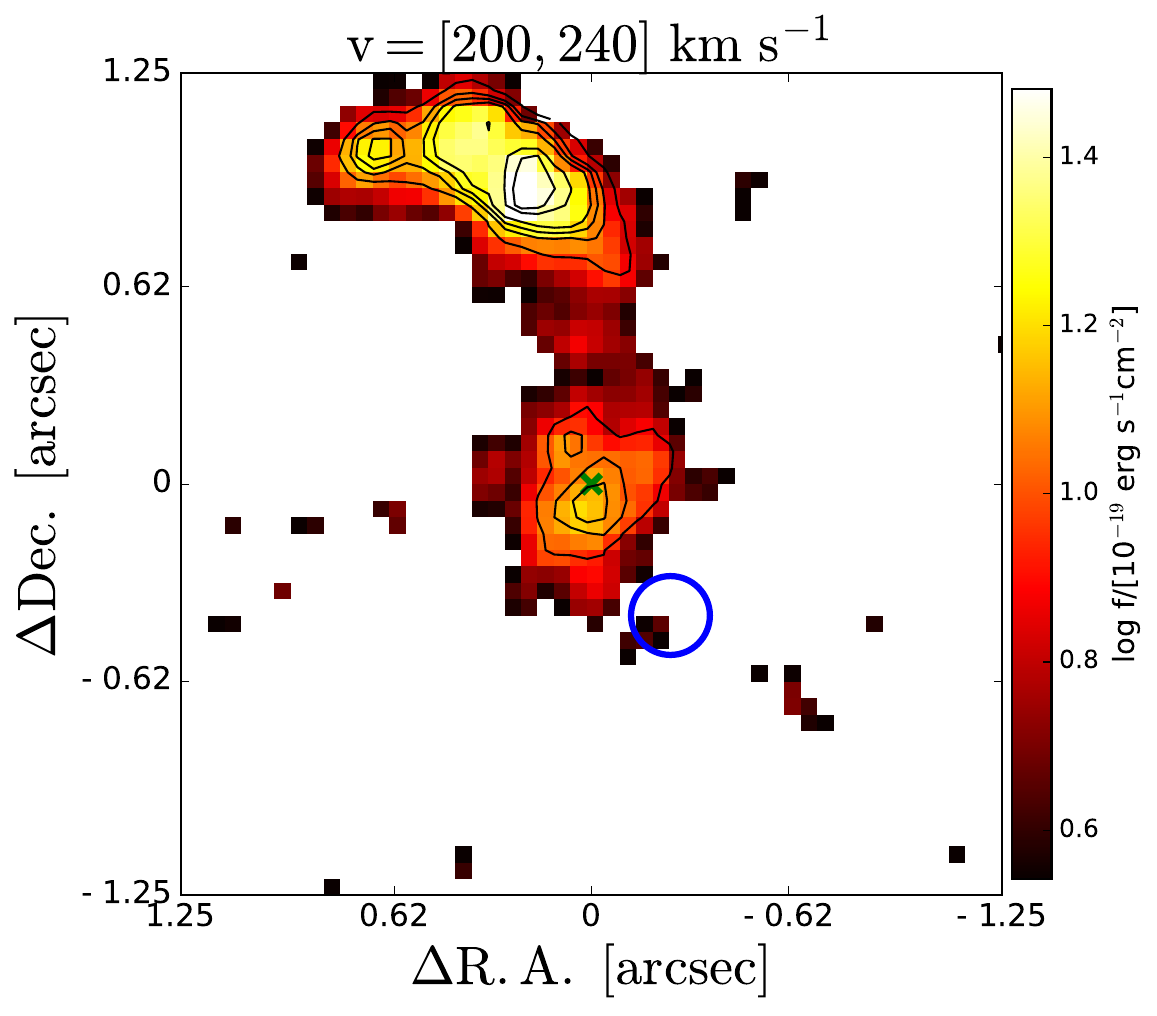}
\caption{Channel maps of the \Ha\ emission to illustrate the position of the clump `\clump', on the south-west of the central target. The channels have a width of 40~\kms\ and the velocities are given with respect to the systemic velocity.  The clump `\clump' is indicated with a blue circle and is visible in the velocity channels [0--200]~\kms.}
\label{fig:clump_SW}
\end{figure*}

\section{Two component fit: kinematic maps of the broad component of \texorpdfstring{\Ha}{Ha}}
\label{app:two_comp_model}
Figure~\ref{fig:Ha_maps_2comp} shows the maps of the broad emission of \Ha, including all Gaussian components with $\sigma > 140$~\kms. The three panels show the flux, central velocity ($v50$) and the line width (W80).

\begin{figure*}
    \centering
    \includegraphics[width=0.95\linewidth]{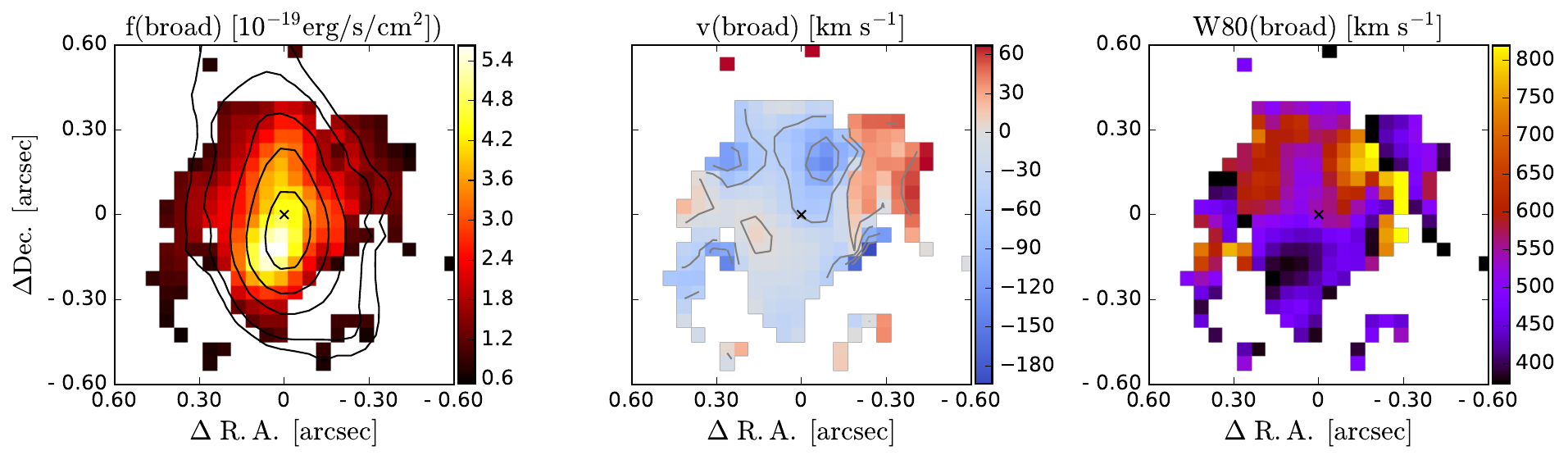}
    \caption{Maps of the broad emission of \Ha, including all Gaussian components with $\sigma > 140$~\kms. For the spaxels fitted with a two-Gaussian model, the broader component is shown. For the spaxels fitted with a one-component model, the component  is shown only if  $\sigma > 140$~\kms. From left to right:  observed  flux (not corrected for attenuation), centroid velocity, and velocity dispersion (W80). Only spaxels with S/N> 3 are shown. 
    }
    \label{fig:Ha_maps_2comp}
\end{figure*}

\section{Fit of \texorpdfstring{\SII}{S2} and \texorpdfstring{\OII}{O2} for deriving the electron densities}

In this section, we show  the fit of \SII\ and \OII\ for the four regions in which we could estimate the electron density: south, north, s2, s3, n1, and n2.

\begin{figure*}[h!]
\centering 
\includegraphics[width=0.43\textwidth]{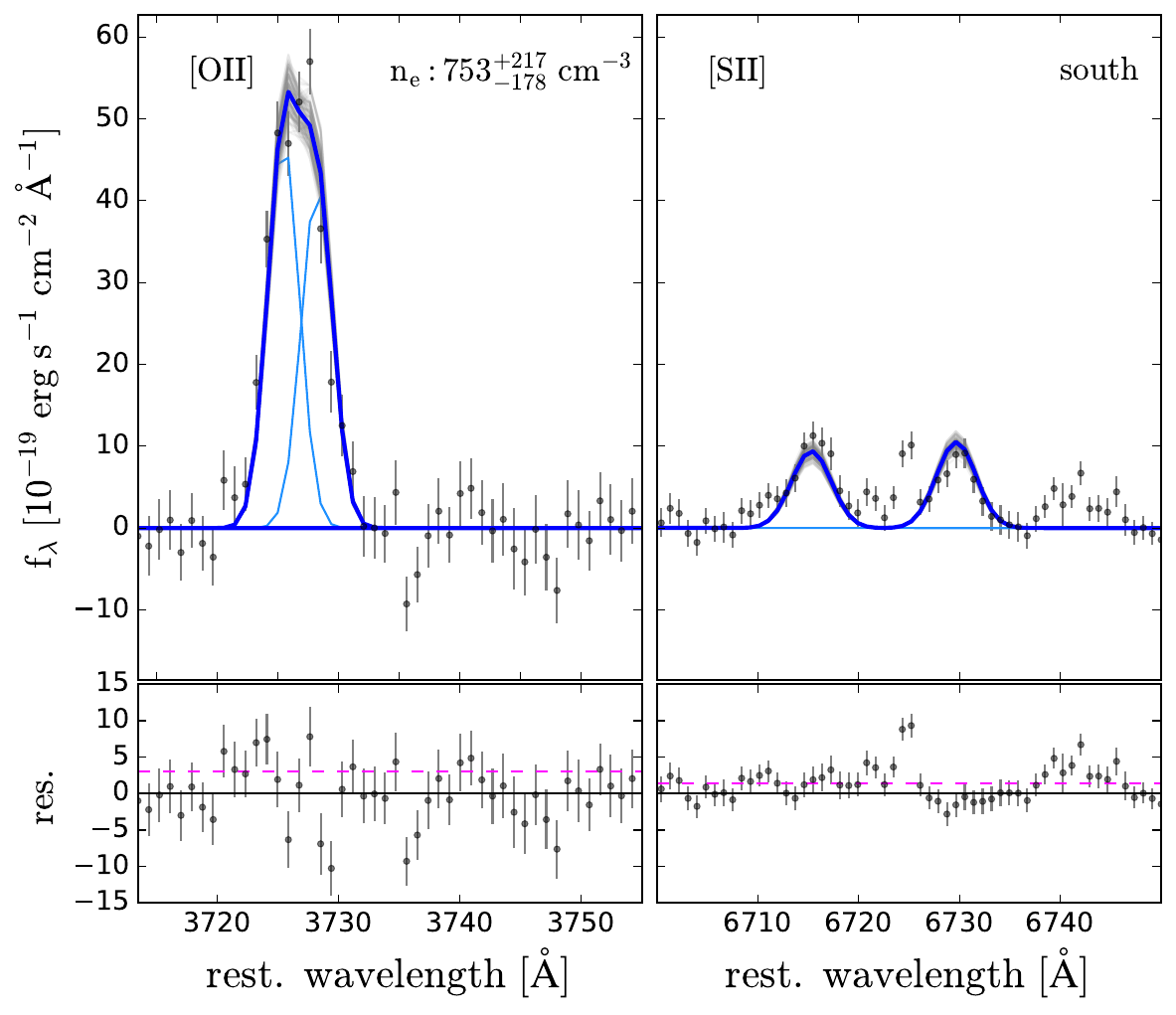}
\includegraphics[width=0.43\textwidth]{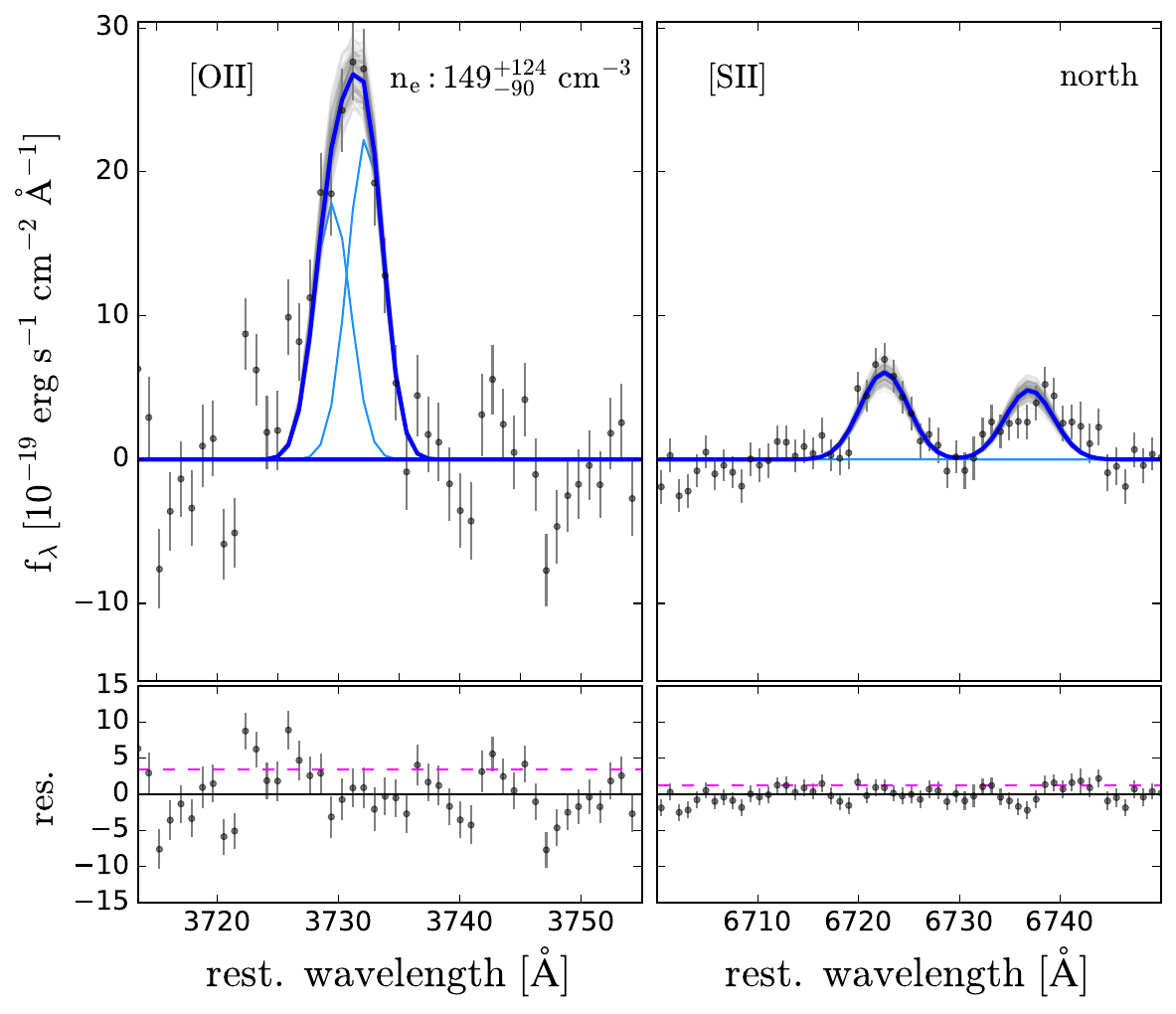}
\includegraphics[width=0.43\textwidth]{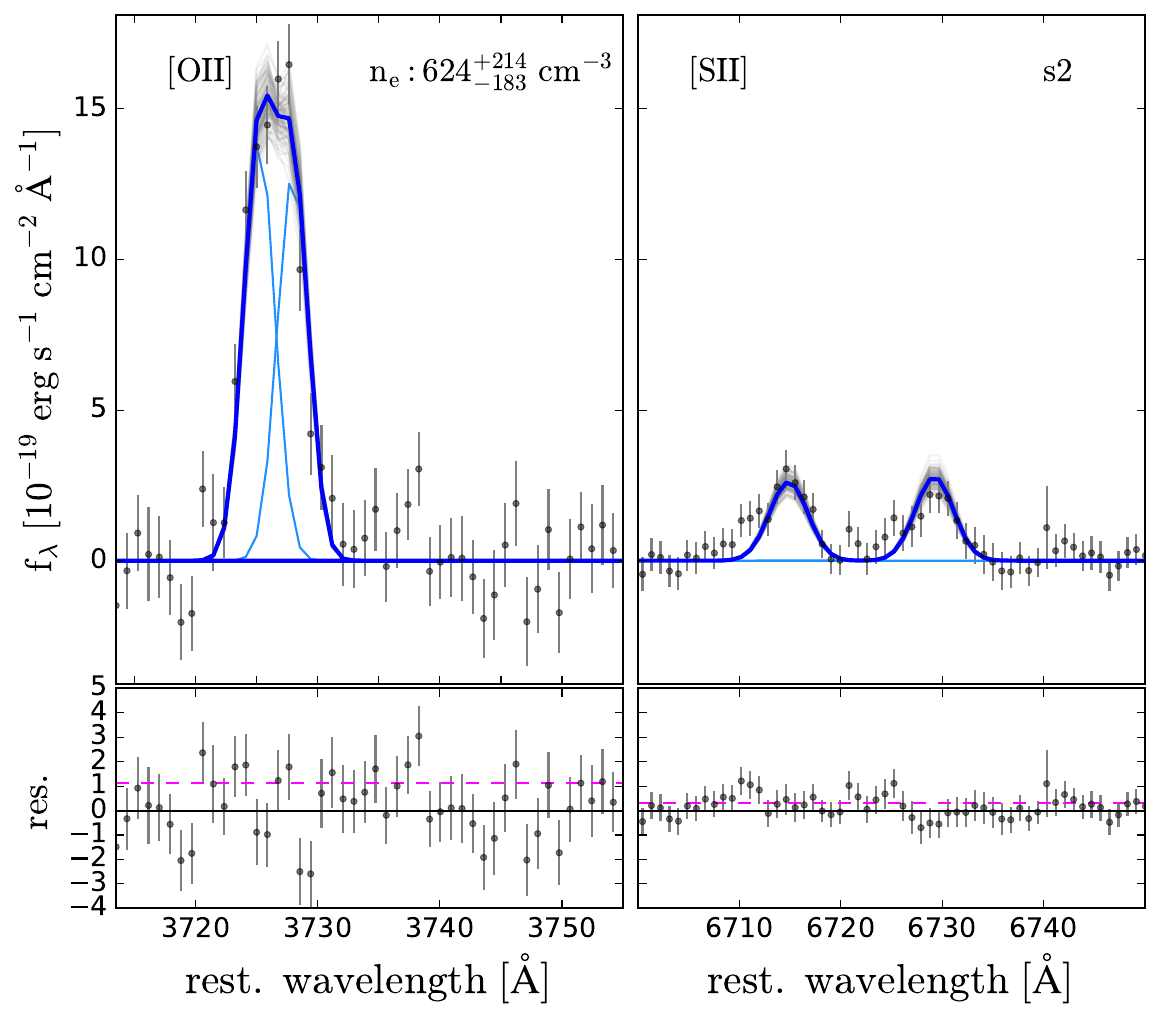}
\includegraphics[width=0.43\textwidth]{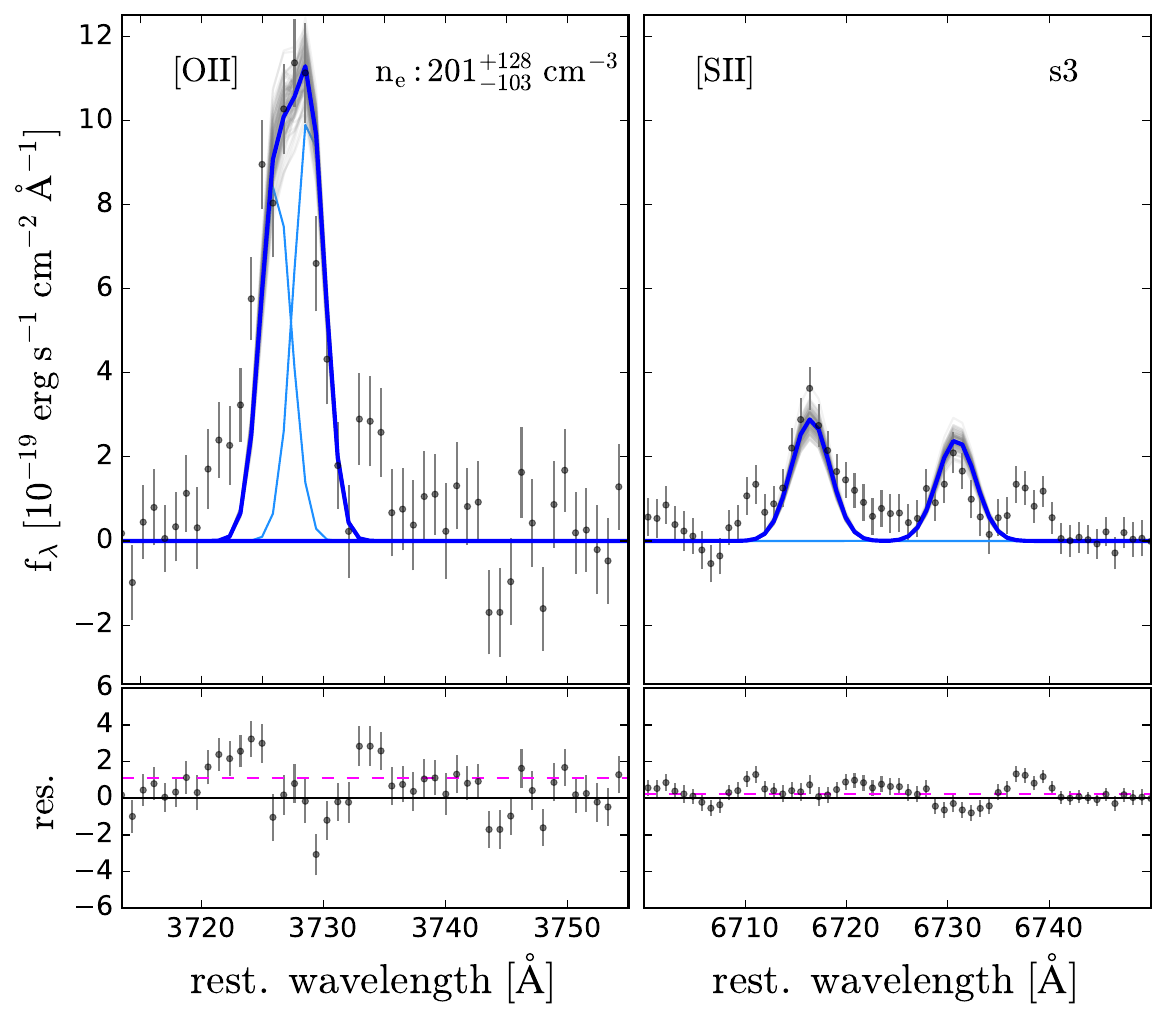}
\includegraphics[width=0.43\textwidth]{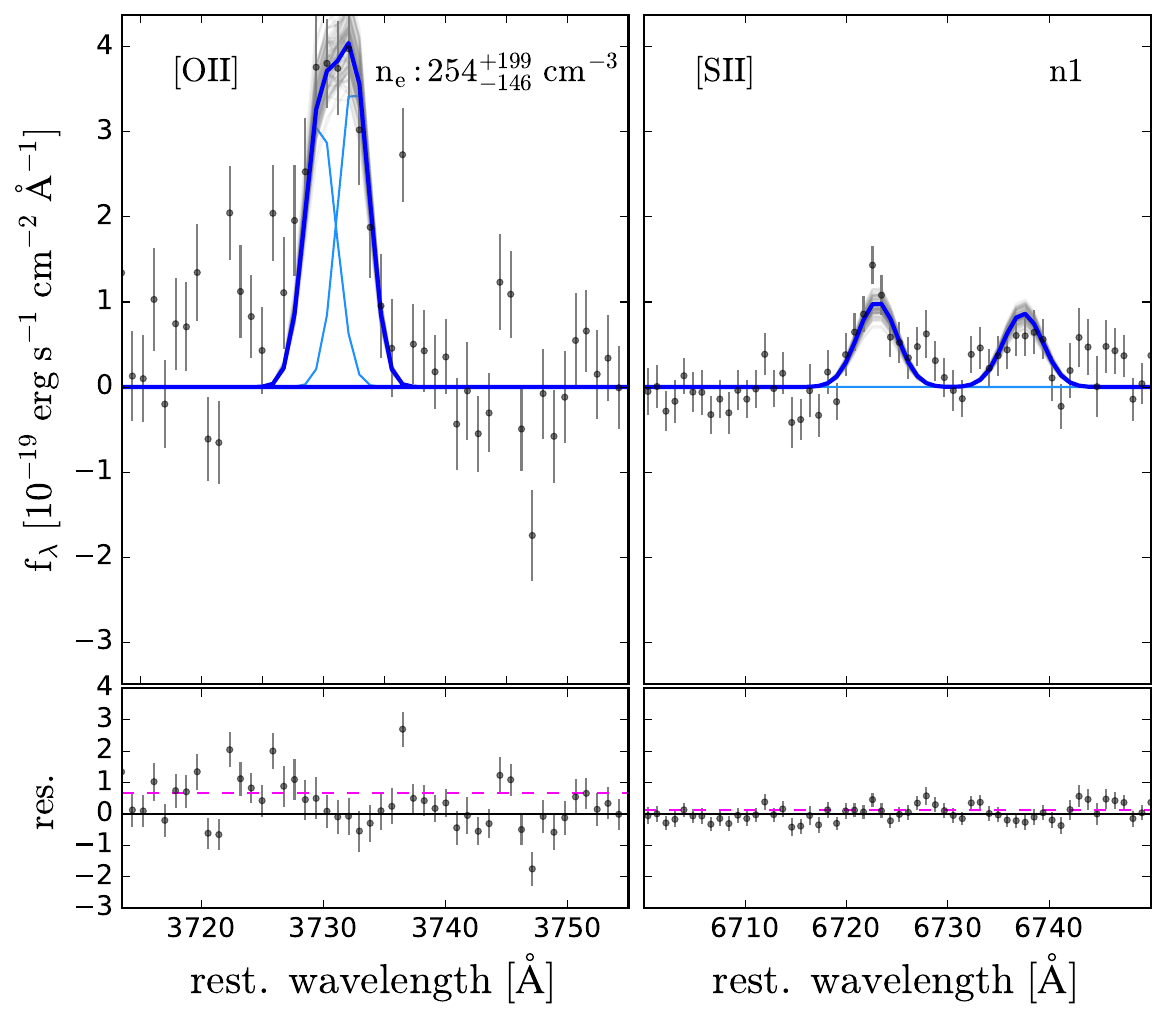}
\includegraphics[width=0.43\textwidth]{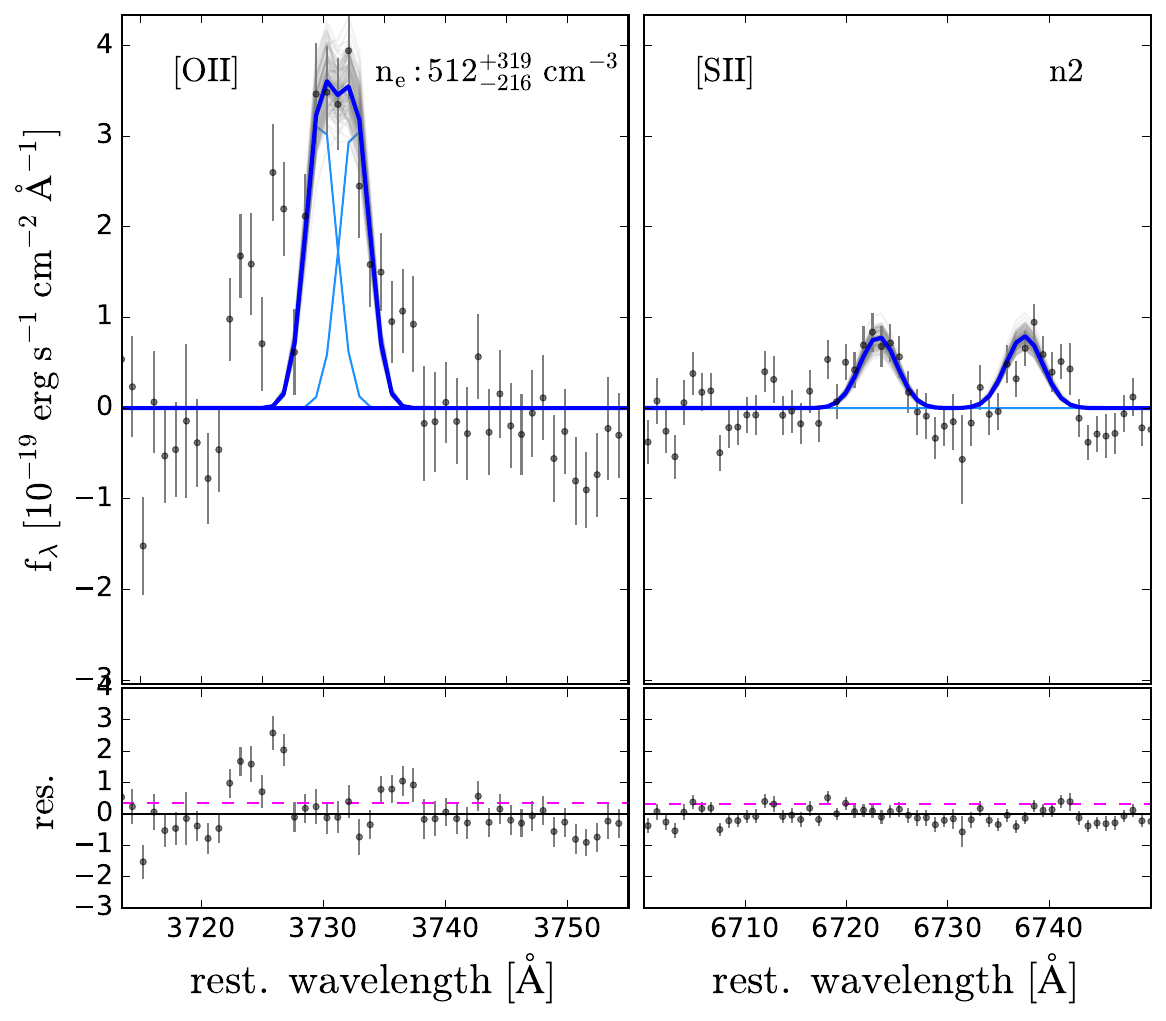}
\caption{Simultaneous spectral fit of the \OII$\lambda\lambda3726,29$ and \SII$\lambda \lambda6716,31$ doublets  to derive the electron density from the integrated spectrum of the targets south and north (upper raw), s2 and s3 (middle raw), n1 and n2 (lower raw). The blue curve shows the total best-fit model, the light-blue curves show the best-fit Gaussians for the individual emission lines and the gray curves show the uncertainties of the MCMC fit. The bottom panels show the residuals, the dashed line shows the one-sigma error level.}
\label{fig:SII_OII_fit_appendix}
\end{figure*}

We also show the fit of \SII\ and \OII\ lines of the main target using two components (narrow+broad). 
The electron density derived for the narrow component is similar to the value inferred from the one-component fit but with larger uncertainties ($n_e=540_{-280}^{+270}$~cm$^{-3}$). The electron density of the broad component is unconstrained ($n_e=340_{-280}^{+1400}$~cm$^{-3}$).

\begin{figure}[t!]
\centering 
\includegraphics[width=0.47\textwidth]{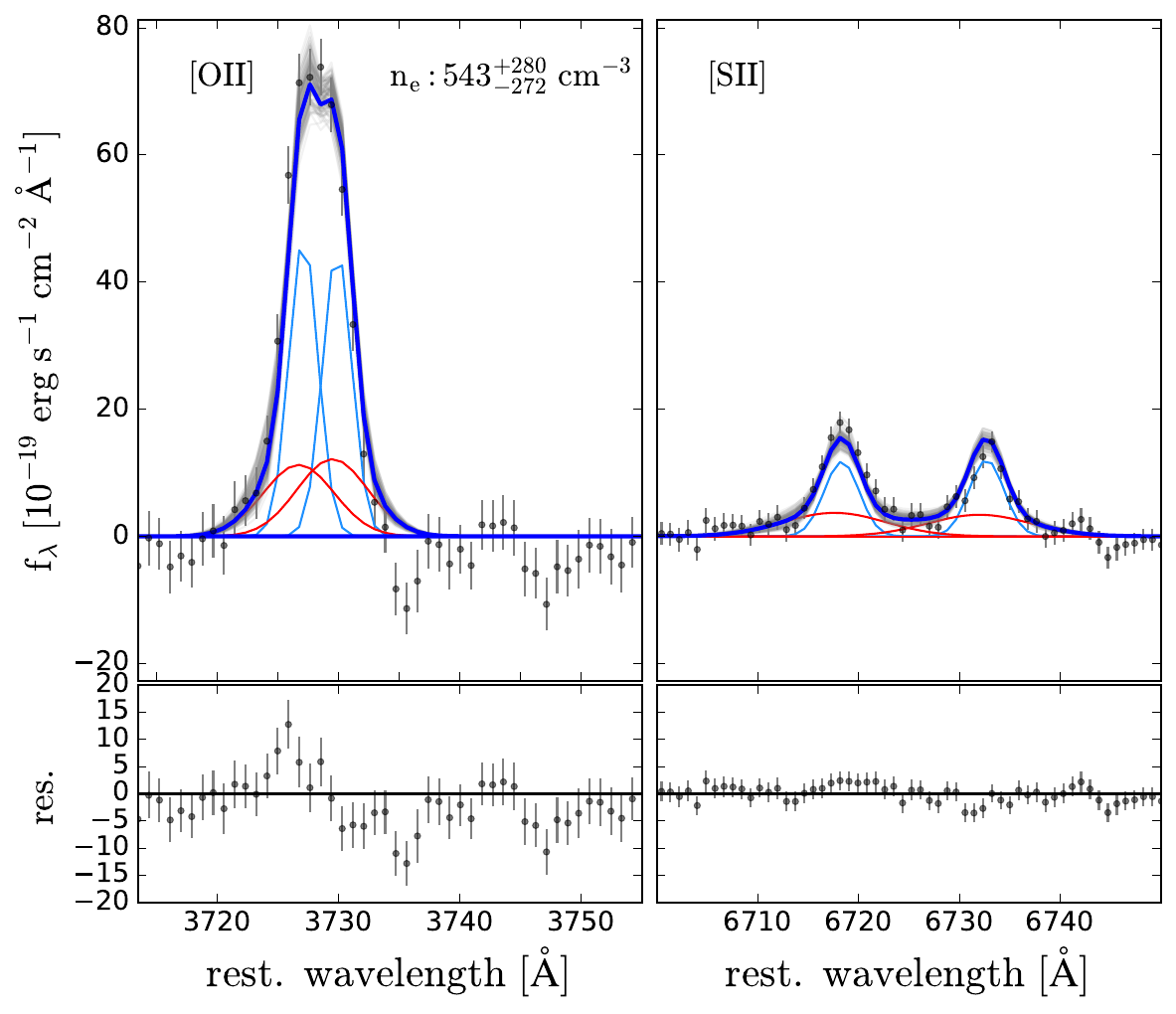}
\caption{Simultaneous spectral fit of the \OII$\lambda\lambda3726,29$ and \SII$\lambda \lambda6716,31$ doublets  to derive the electron density from the integrated spectrum of the main target using two components (broad and narrow). The blue curve shows the total best-fit model, the light-blue and red curves show the best-fit narrow and broad Gaussian components, respectively, for the individual emission lines. The grey curves the uncertainties of the MCMC fit. The bottom panels show the residuals.}
\label{fig:SII_OII_2comp_fit}
\end{figure}

\section{Integrated spectra}
In this section we show the integrated spectra of the regions defined in Fig.~1, together with the best-fit model. In particular, we show the spectra of the regions `south' and  `north', the sub-structures s1, s2, s3, n1, n2, n3, and the clump `\clump'.
\begin{figure*}
\centering 
\includegraphics[width=0.9\textwidth]{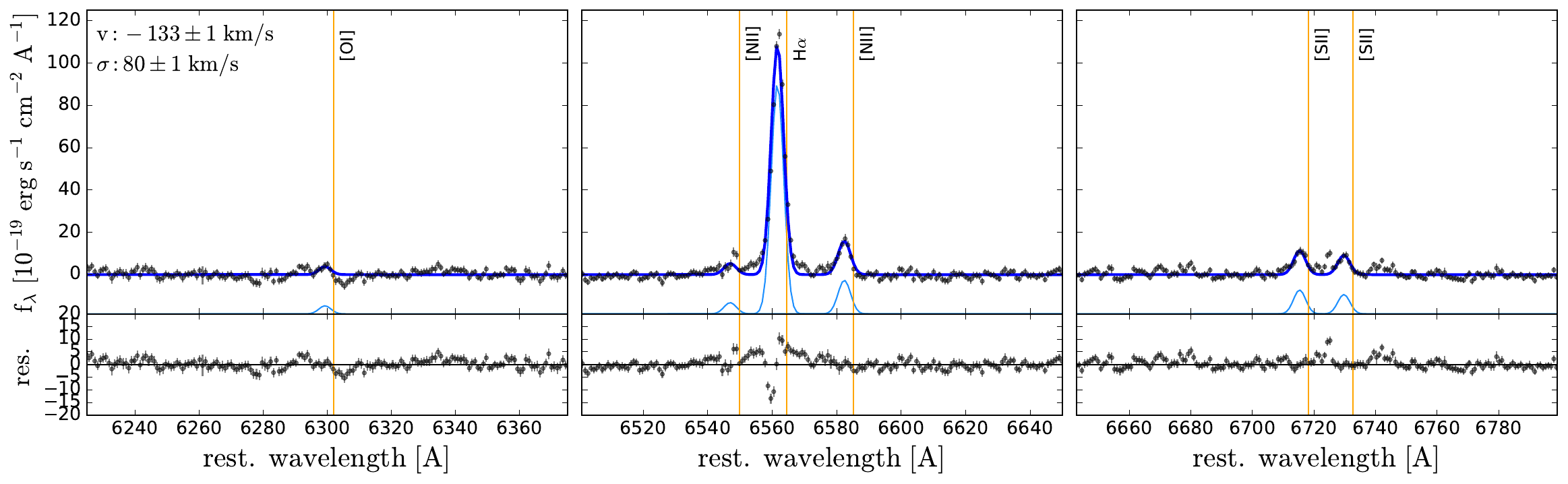}
\includegraphics[width=0.9\textwidth]{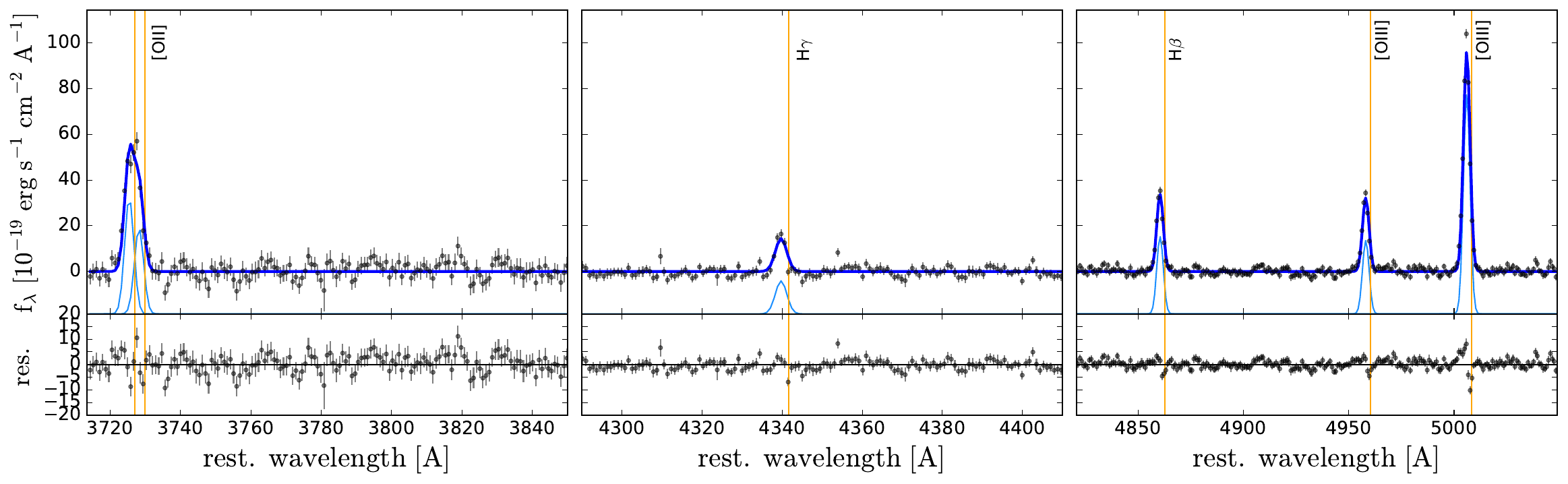}
\caption{Integrated spectrum of source `south'. The total best fit is shown in blue. The individual narrow components are shown in lightblue, shifted vertically for visual purposes. 
 The vertical lines mark the wavelength positions of the emission lines at the systemic redshift of the main source ($z=3.4705)$. The fitting residuals are shown in the bottom panel.}
\end{figure*}

\begin{figure*}\ContinuedFloat 
\centering 
\includegraphics[width=0.9\textwidth]{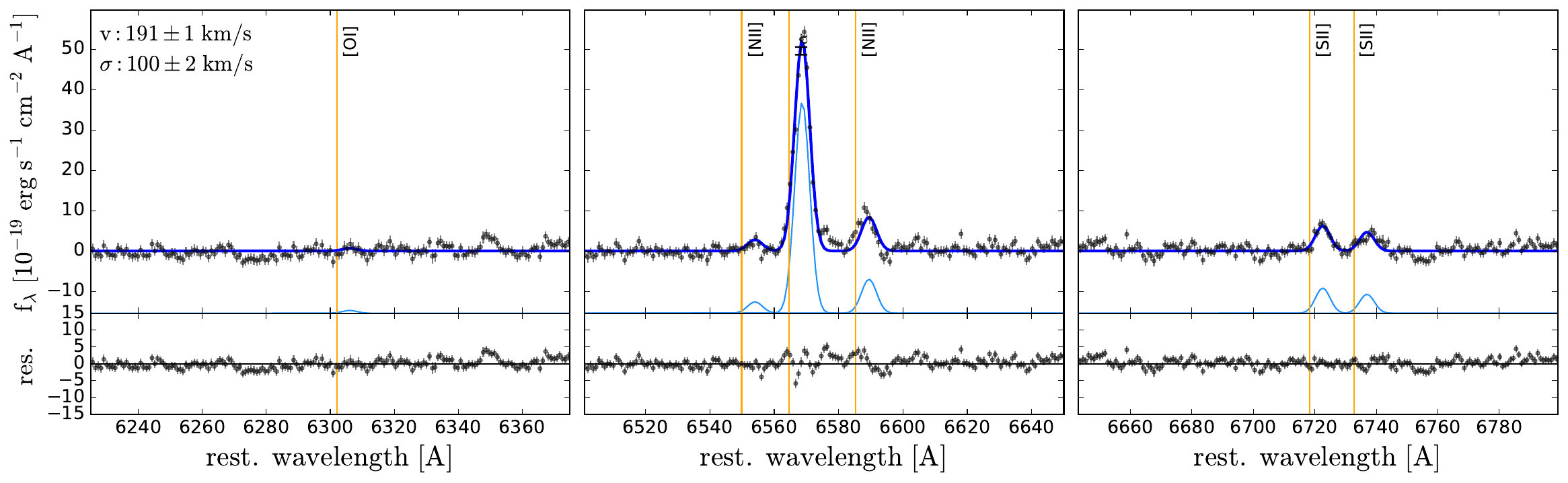}
\includegraphics[width=0.9\textwidth]{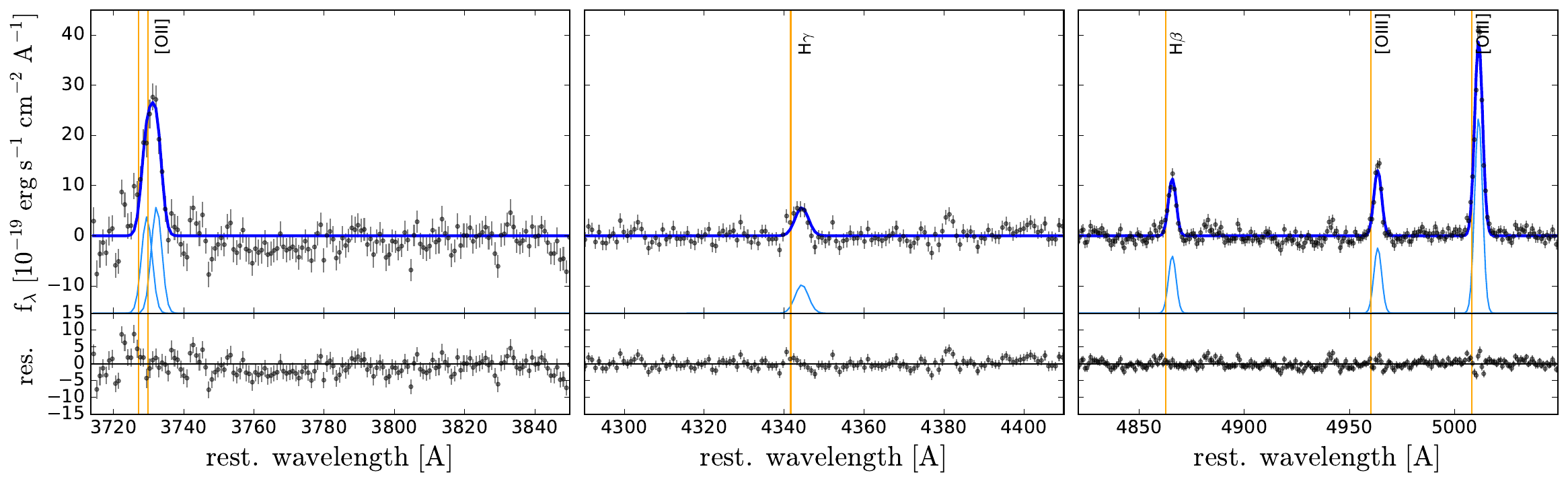}
\caption{Integrated spectrum of source `north'.}
\end{figure*}

\begin{figure*}\ContinuedFloat 
\centering 
\includegraphics[width=0.9\textwidth]{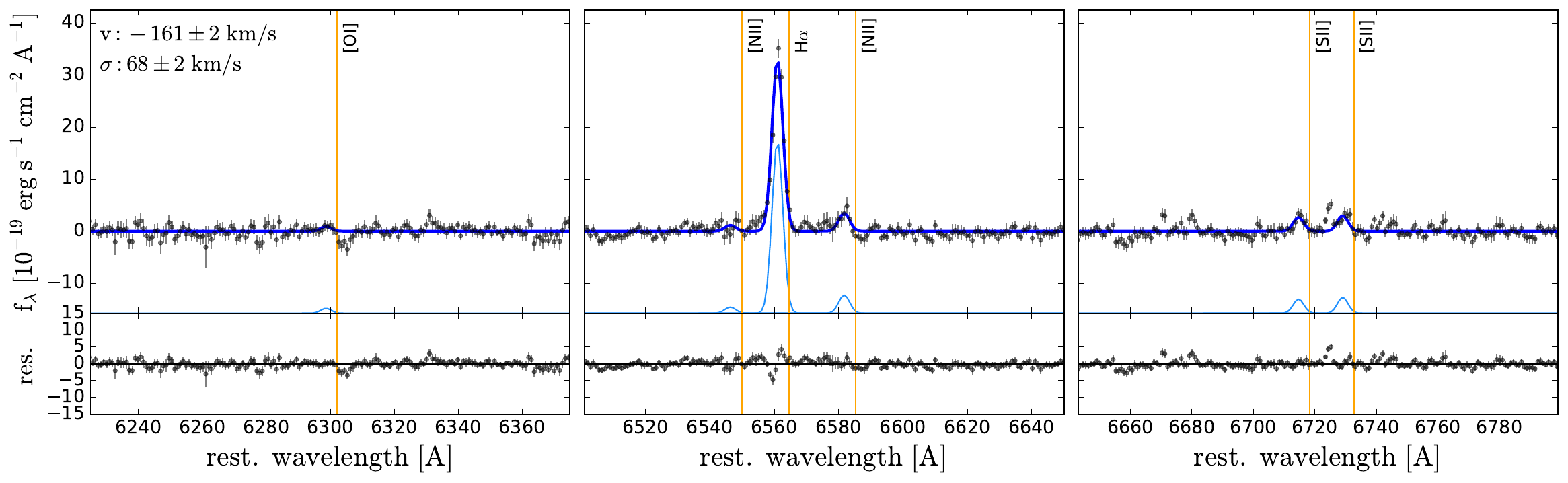}
\includegraphics[width=0.9\textwidth]{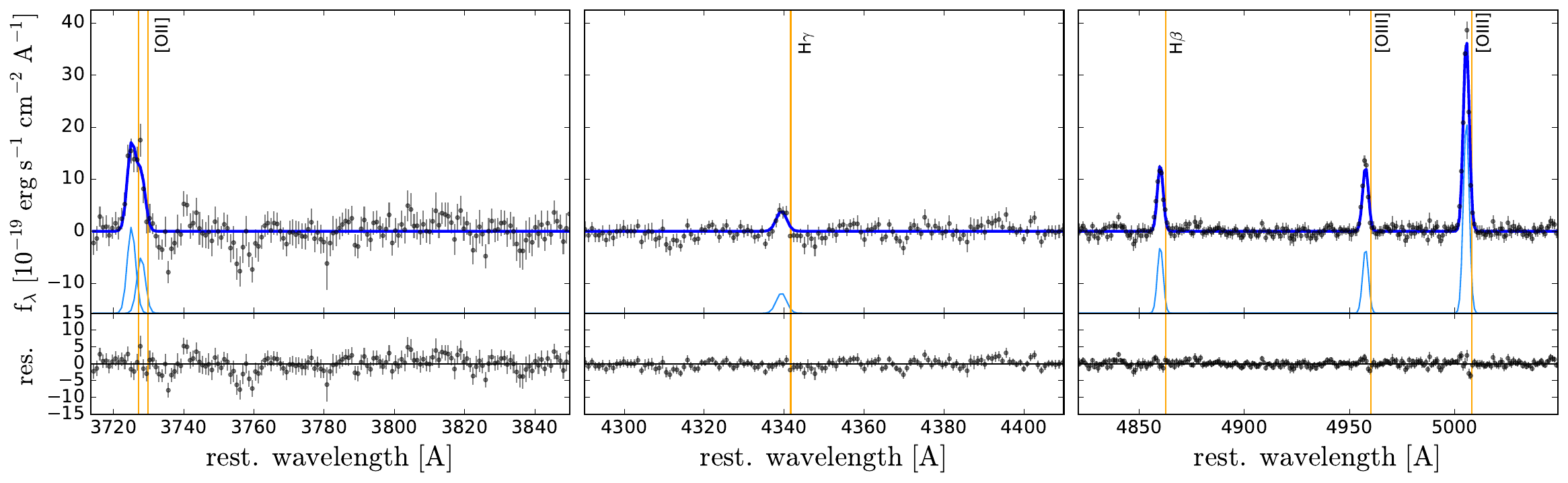}
\caption{Integrated spectrum of source `s1'. }
\label{fig:int_spectra}
\end{figure*}

\begin{figure*}\ContinuedFloat 
\centering 
\includegraphics[width=0.9\textwidth]{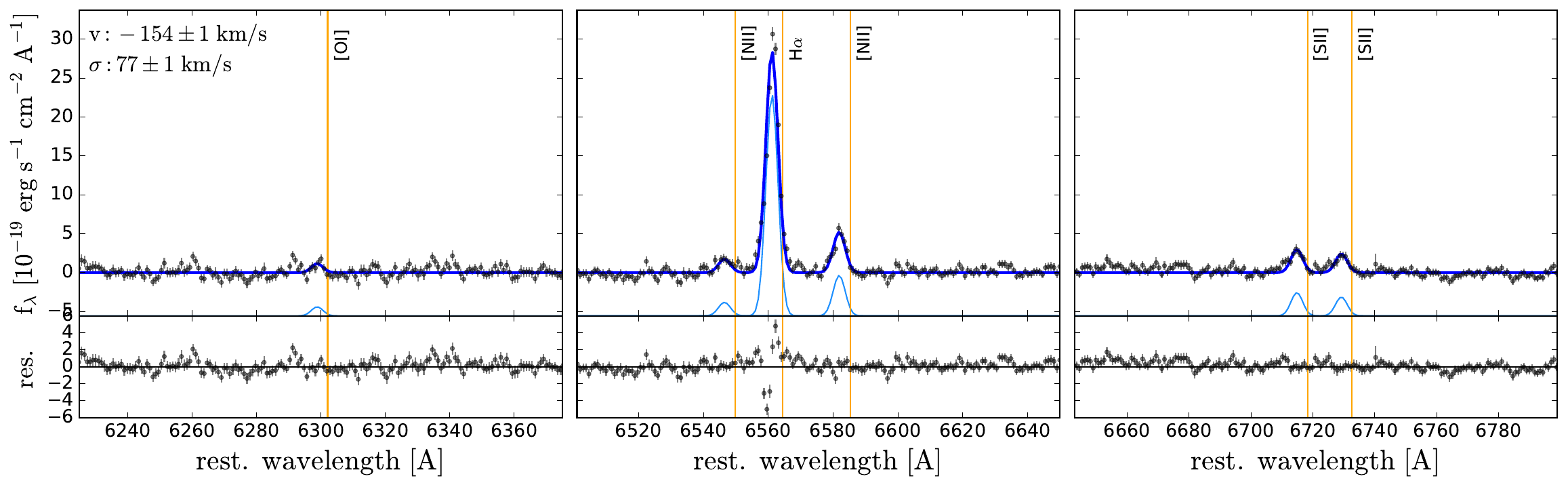}
\includegraphics[width=0.9\textwidth]{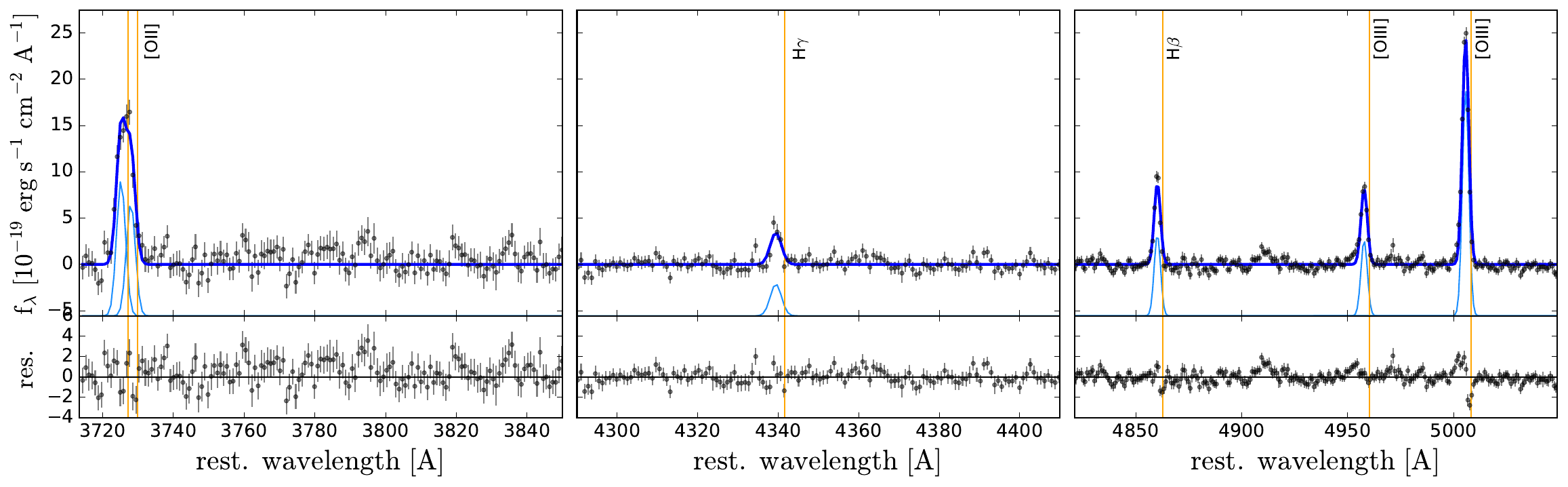}
\caption{Integrated spectrum of source `s2'.}
\end{figure*}

\begin{figure*}\ContinuedFloat 
\centering 
\includegraphics[width=0.9\textwidth]{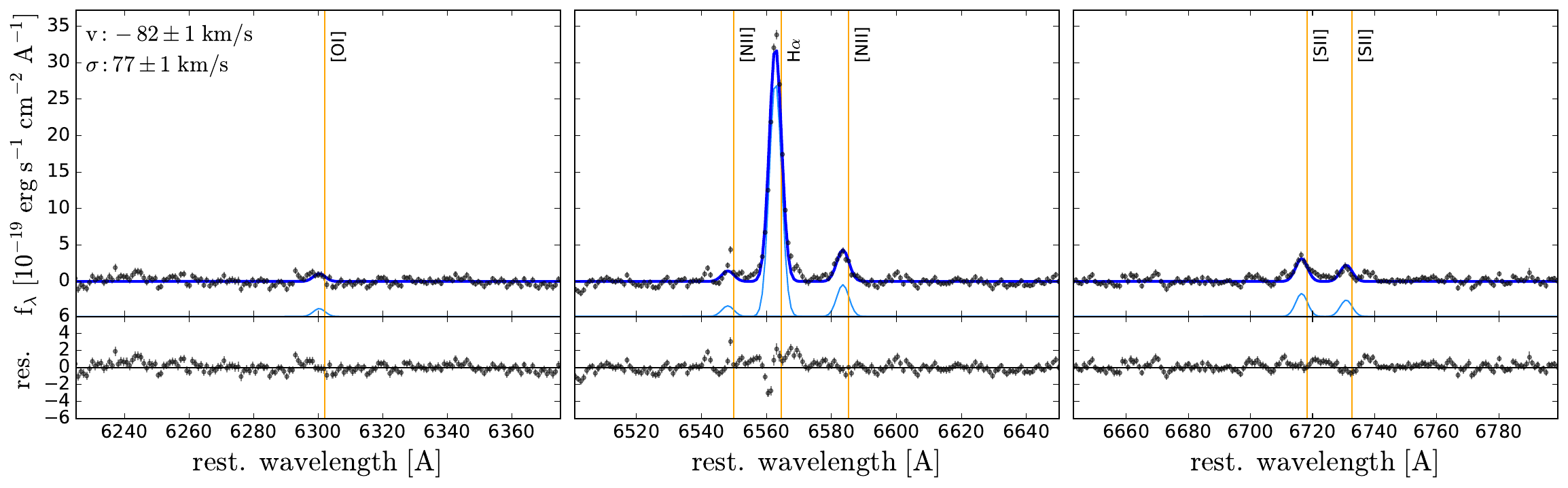}
\includegraphics[width=0.9\textwidth]{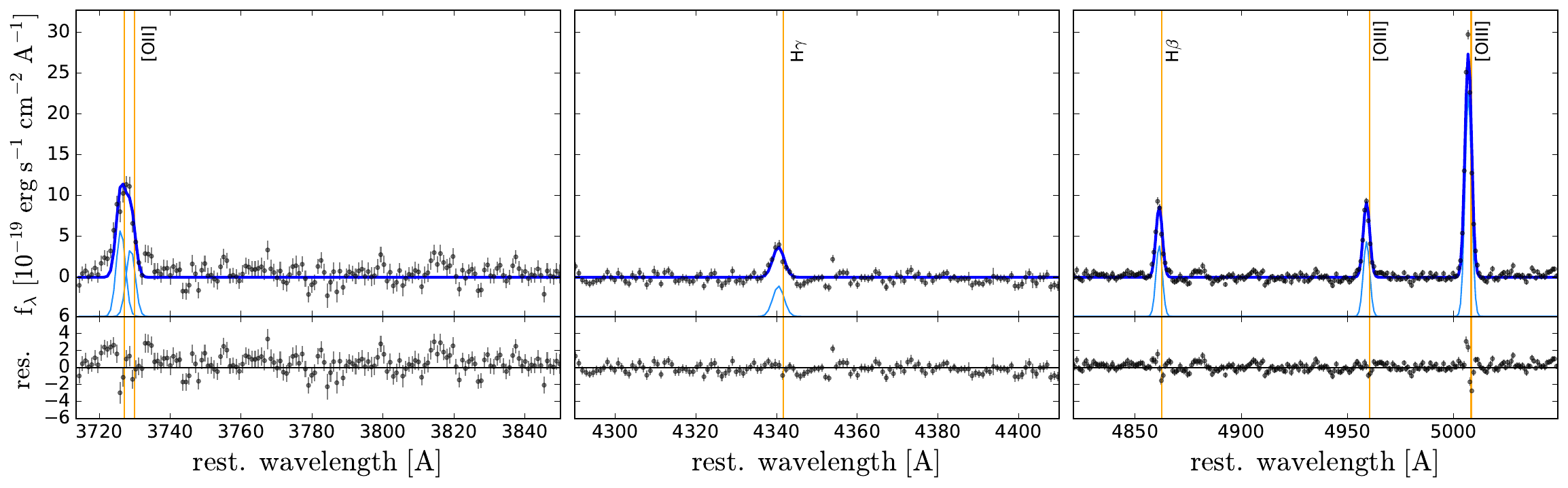}
\caption{Integrated spectrum of source `s3'.}
\end{figure*}

\begin{figure*}\ContinuedFloat 
\centering 
\includegraphics[width=0.9\textwidth]{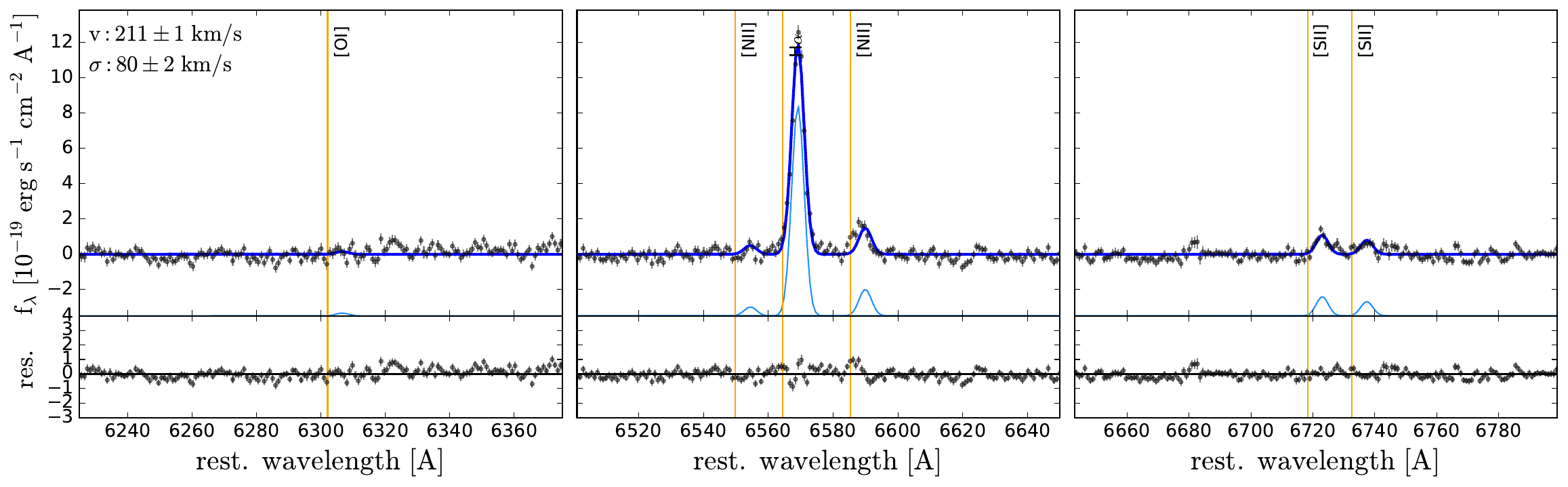}
\includegraphics[width=0.9\textwidth]{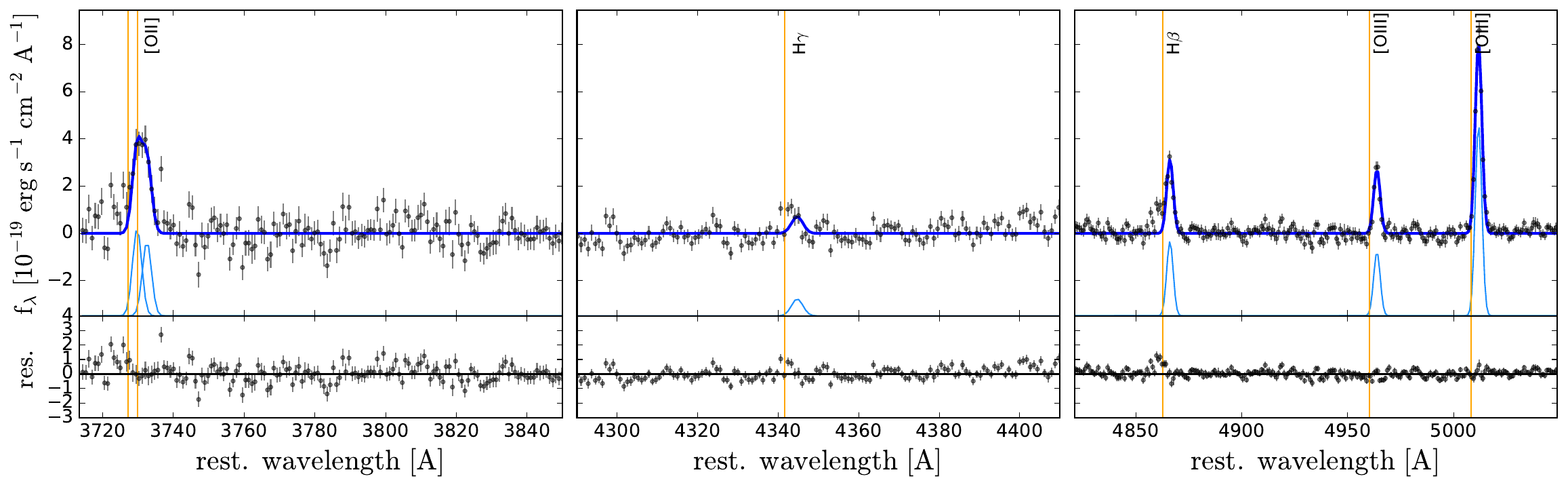}
\caption{Integrated spectrum of source `n1'.}
\end{figure*}

\begin{figure*}[!]\ContinuedFloat 
\centering 
\includegraphics[width=0.9\textwidth]{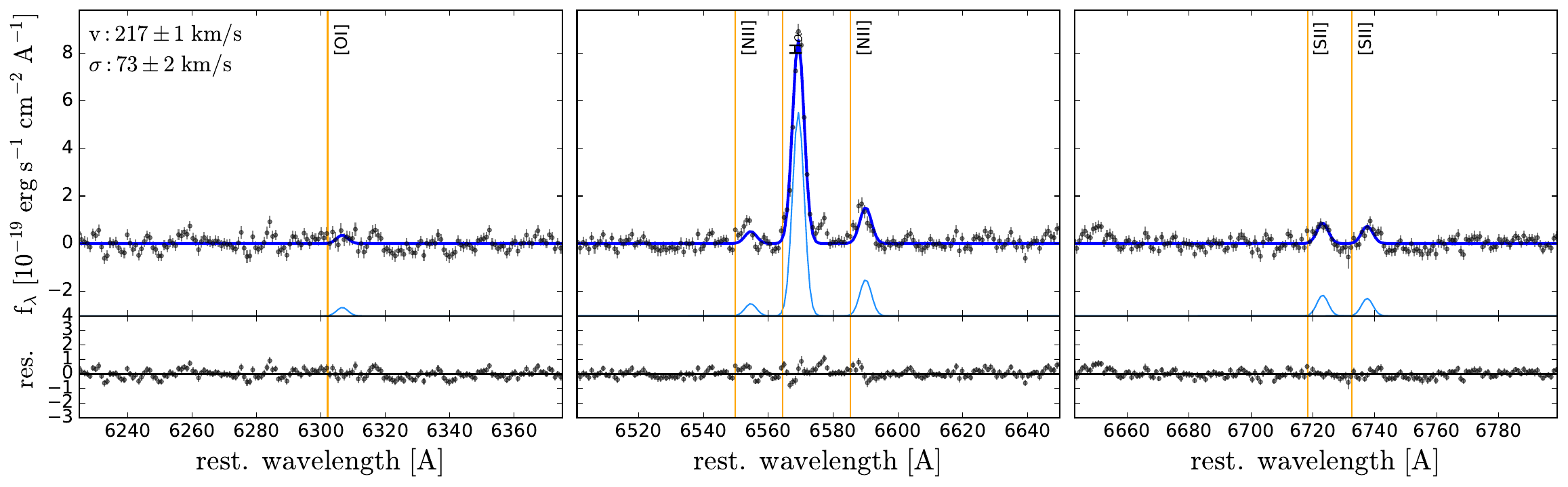}
\includegraphics[width=0.9\textwidth]{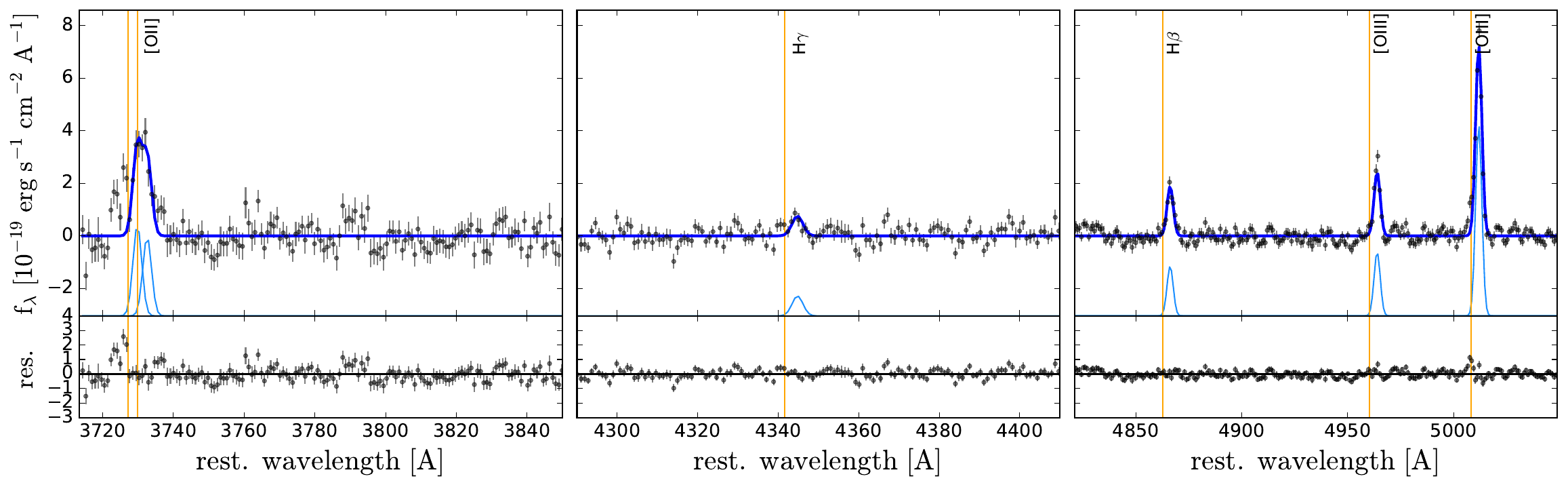}
\caption{Integrated spectrum of source `n2'.}
\end{figure*}

\begin{figure*}[!]\ContinuedFloat 
\centering 
\includegraphics[width=0.9\textwidth]{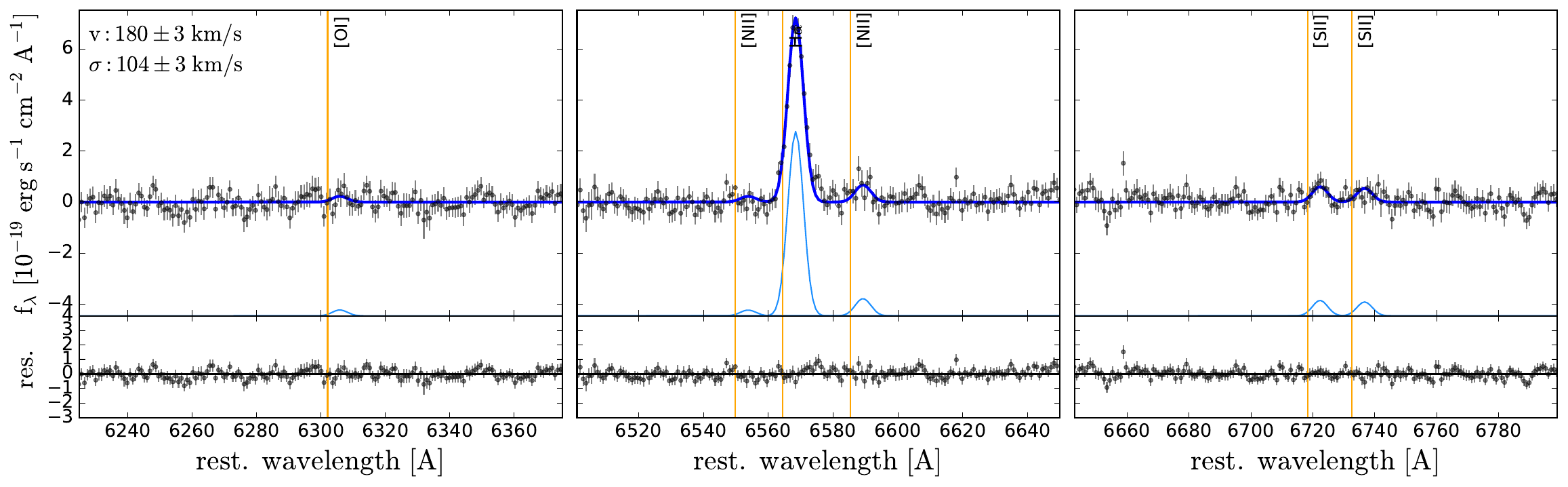}
\includegraphics[width=0.9\textwidth]{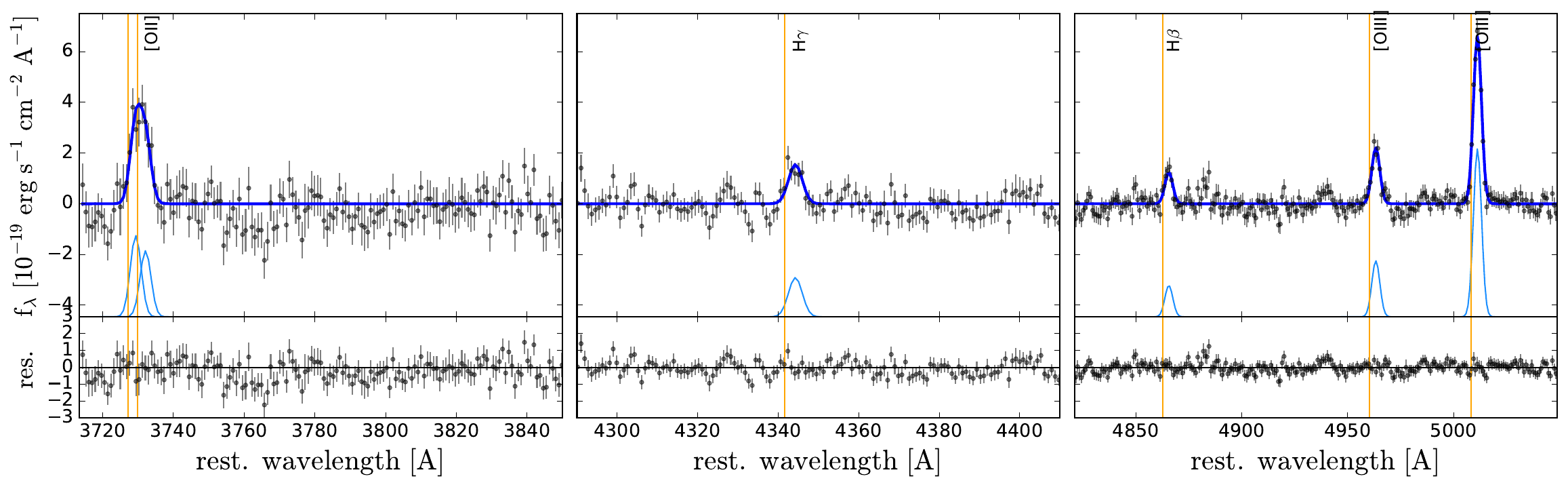}
\caption{Integrated spectrum of source `n3'.}
\end{figure*}

\begin{figure*}\ContinuedFloat 
\centering 
\includegraphics[width=0.9\textwidth]{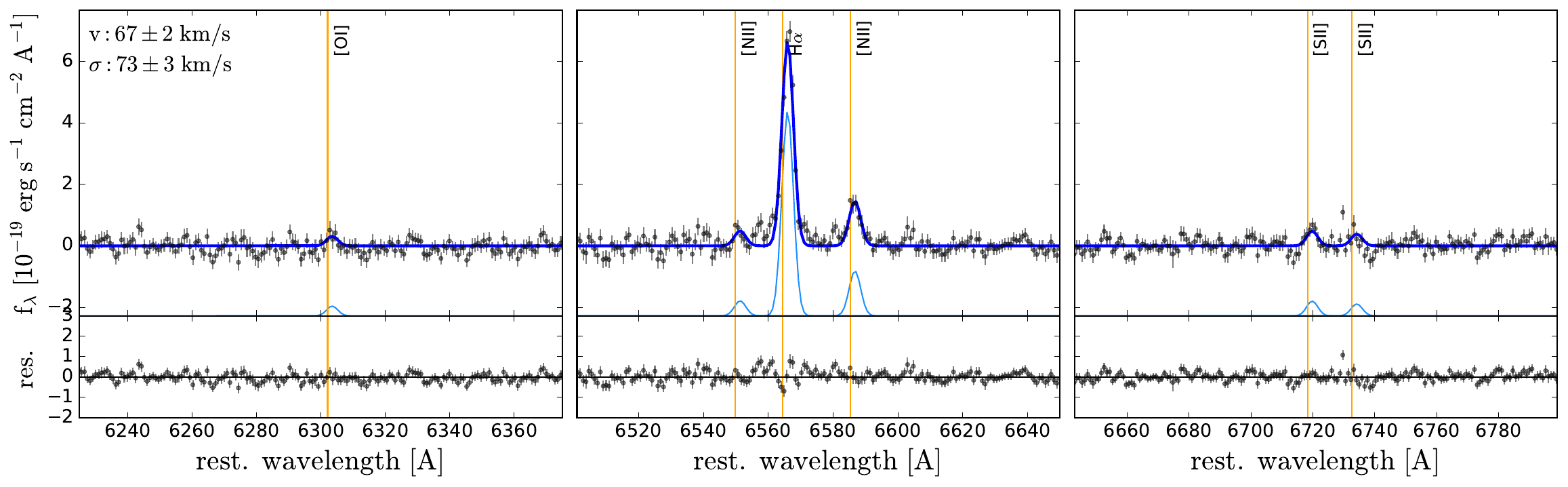}
\includegraphics[width=0.9\textwidth]{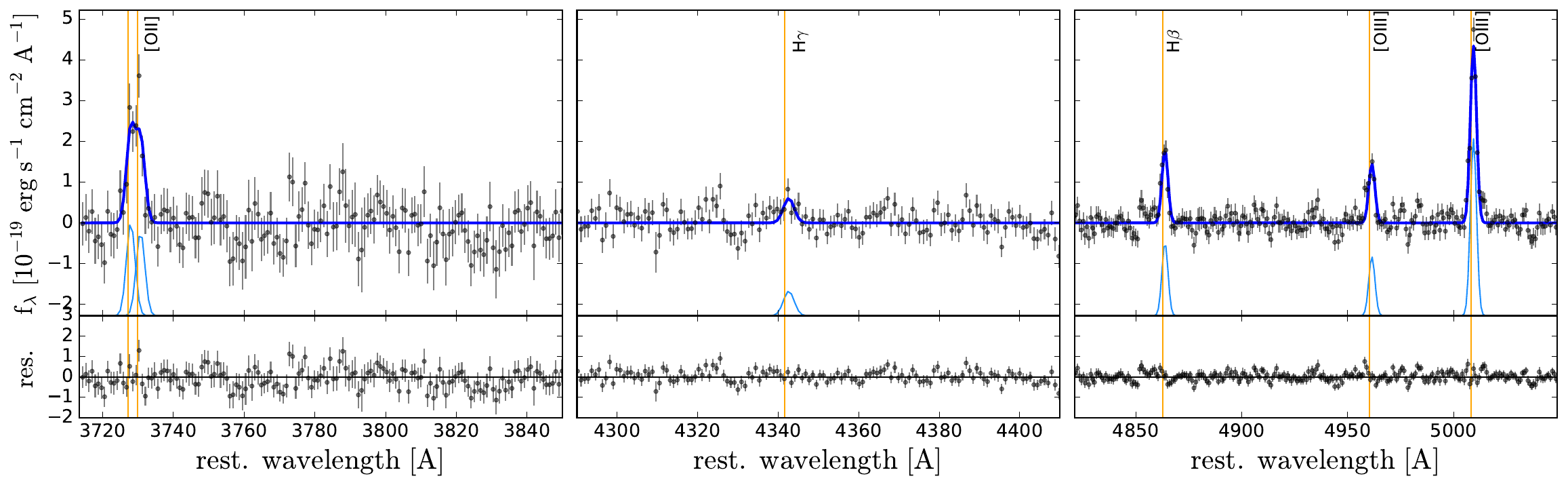}
\caption{Integrated spectrum of source `\clump'.}
\end{figure*}

\end{appendix}

\end{document}